\documentclass[12pt,aps,amsmath,latexsym,amsfonts]{JHEP3}
\usepackage{epsfig}
\usepackage{amsfonts,amssymb,amsmath}
\usepackage[hang]{subfigure}
\usepackage[numbers, square, sort&compress]{natbib}

\newcommand{\be}{\begin{equation}}
\newcommand{\ee}{\end{equation}}
\newcommand{\bea}{\begin{eqnarray}}
\newcommand{\eea}{\end{eqnarray}}

\title{Dp-brane dynamics and thermalization\,in\,type\\
IIB\,Ben\,Ami-Kuperstein-Sonnenschein\,models}
\author{
Dariush Kaviani\thanks{Email:dariush@ipm.ir} \\
School of Particles and Accelerators,
Institute for Research in Fundamental Sciences (IPM),
P.O. Box 19395-5531, Tehran, Iran
}

\abstract{We study the world volume Hawking temperature of all type IIB
rotating probe $Dp$-branes, dual to the temperature of different flavors
at finite R--charge, in the Ben Ami--Kuperstein--Sonnenschein holographic
models including the effects of spontaneous breakdown of the conformal
and chiral flavor symmetry. The model embeds type IIB probe flavor 
$Dp$-branes into the  Klebanov-Witten gravity dual of conformal gauge 
theory, with the embedding parameter, given by the minimal radial extension
of the probes, setting the IR scale of conformal and chiral flavor symmetry
breakdown. We show that when the minimal extension is positive definite 
and additional spin is switched on, the induced world volume metrics of only
certain type IIB probe branes admit thermal horizons and  Hawking temperatures
despite the absence of black holes in the bulk.  We find that the world volume
black hole nucleation is non-trivial and depends on the world volume dimension
and topology of the non-trivial internal cycle wrapped by the probe. We find, in 
addition, that by varying the minimal extension of the probe, large hierarchies of
temperature scales get produced, with the temperature  behavior varying 
dramatically, depending on the type of probe.  We also derive the  energy--stress
tensor of the thermal probes and study their backreaction and energy dissipation. 
We show that at the IR scale the backreaction is non-negligible and find the energy
can flow from the probes to the bulk, dual to the energy  dissipation from the flavor
sectors into the gauge theory.}


\preprint{IPM/PA-459}

\begin{document}

\section{Introduction}

Non-equilibrium steady states are of great interest in many branches of physics,
such as heavy ion physics (thermalization of the quark-gluon plasma), condensed
matter physics (quenches of cold atom systems), cosmology (non-equilibrium 
phase transitions and the  Kibble-Zurek mechanism) and more.  Studying such
states by standard field theory techniques is prohibitively difficult (except for 
special cases involving conformal symmetry or integrability),  however, they
can be easily constructed by using probe D-brane dynamics in gauge/gravity duality.

In gauge/gravity duality, \cite{Maldacena:1997re} (see also \cite{Ammon:2015wua}),
the pure gravity dual solution of conformal gauge theory is produced by taking the
near horizon limit of $N$ backreacting D3-branes at a smooth point of the transverse
space described by the $adS_5\times S^5$ background metric. The extension of such
holographic solution includes the addition of flavored quarks to the pure gauge theory, 
by embedding  additional $N_f$ flavor  branes into the pure gravity dual, taking the probe
limit, $N_f\ll N$,  \cite{Karch:2002sh} (see also  \cite{Ammon:2015wua})\footnote{In
gauge/gravity dual thermodynamics, \cite{Maldacena:1997re}  
(see also \cite{Ammon:2015wua}) the gravity dual solution of conformal gauge theory
at finite temperature is produced by taking the  near horizon limit of $N$ black D3-branes
described by known black hole background metric. In order to add flavored quarks to the
gauge theory and study their thermal equilibrium properties, one embeds additional $N_f$
flavor branes into the finite temperature gravity dual, taking the probe limit, $N_f\ll N$ 
\cite{Karch:2002sh} (see also \cite{Ammon:2015wua}). Such holographic setup including
probes,  \cite{Karch:2001cw,Karch:2000gx}, has been used to model flavor physics 
\cite{Karch:2002sh,Kruczenski:2003be}, and quantum critical phenomena,
\cite{Karch:2007pd}, complementary to other works on charged $adS$ black holes,
\cite{Herzog:2009xv}. Such phenomena have also been studied at zero temperature in
refs.\,\cite{O'Bannon:2008bz,PremKumar:2011ag}.}. In order to construct non-equilibrium
systems in such flavored holographic models, one then notes that when the flavored quark
sector (dual to the  probe flavor branes) attain finite temperature while the pure gauge 
theory itself  (dual to the pure gravity solution) is at zero temperature, the system describes
non-equilibrium steady states.  The prime examples of such non-equilibrium steady states
 have been constructed in ref.\,\cite{Das:2010yw} (see also \cite{Taghavi:2016nnp,
Russo:2008gb}), by  embedding type IIB probe flavor $Dp$-branes $(p=1,3,5,7)$ into the
$adS_5\times S^5$ gravity dual of conformal gauge theory ($\mathcal{N}=4$ SYM). 
(For $p=7$, the probe fills all the spacetime directions of the gravity dual and the corresponding
gauge theory is given by conformal field theory (CFT). For  $p<7$, the probes are localized in
some spacetime directions of the gravity dual and the  corresponding gauge theory is given by
\emph{defect} conformal field theory (dCFT)). In  ref.\,\cite{Das:2010yw} 
(see also \cite{Taghavi:2016nnp,Russo:2008gb}) rotating probe $Dp$-brane solutions have
been constructed and shown that when spin is turned on, the induced world volume metric on
\emph{all} type IIB rotating probe $Dp$-branes in $adS_5\times S^5$ admit thermal horizons
and Hawking temperatures despite the absence of  black holes in the bulk. By gauge/gravity 
duality, this means that when the flavor sector of the gauge theory  gets R--charged, it becomes
thermal. Since the gauge theory itself is at zero temperature, while its flavor sector, given, for 
instance, by a magnetic monopole $(p=1)$ or a  quark $(p=7)$, is at finite temperature, it was
concluded in ref.\,\cite{Das:2010yw}  such systems describe non-equilibrium steady states.  
However, by computing the energy-stress tensor, it has been shown in  ref.\,\cite{Das:2010yw}, 
that the energy of the probe (defect) flavor brane will eventually dissipate into the bulk and form,
with the large backreaction in the IR, a (localised) black hole in the bulk. By gauge/gravity duality,
this means that the energy from the (defect) flavor sector will eventually dissipate into the gauge
theory ($\mathcal{N}=4$ SYM).

This analysis has been extended in ref.\,\cite{Kaviani:2015rxa}, by studying the
induced world volume metrics on rotating probe $Dp$-branes ($p=1$ -- dual  to
magnetic monopoles), embedded in more general gauge/gravity dualities, including
 warped Calabi-Yau throats, refs.\,\cite{Klebanov:2000hb,Klebanov:2000nc,Klebanov:1998hh}
(see also refs.\,\cite{Buchel:2000ch} -- extending these gravity duals to finite temperature
systems).  The gravity dual solution, given by the throat background, is produced by placing
$N$ regular D3-branes and $M$ D5-branes (fractional D3-branes) at the generic Calabi-Yau
singularity -- the conifold tip point \cite{Candelas:1989js}. The solution then breaks some
supersymmetry and conformal invariance and is QCD-like -- admitting confinement and chiral symmetry
breakdown in the regular IR, with the gravity dual given by the Klebanov-Strassler (KS)
solution, \cite{Klebanov:2000hb}, and RG cascade in the singular UV, with the gravity
dual given by the Klebanov-Tseytlin (KT) solution, \cite{Klebanov:2000nc}. In the limit
where the fractional D3-branes are removed, the theory is conformal, with the gravity
dual given by the Klebanov-Witten (KW) solution, \cite{Klebanov:1998hh}. In 
ref.\,\cite{Kaviani:2015rxa} it has been shown that in such gravity duals of QCD-like, 
$\mathcal{N}=1$ gauge theories the induced world volume black hole on rotating probe
D1-branes (dual to magnetic monopoles) forms not in the regular IR of the KS solution, 
where the theory is confining and admits  conformal and chiral symmetry breakdown, but
in the singular UV  including  KT solution, where the theory is non-confining and admits
RG cascade and conformal symmetry breakdown. It has been shown in  
ref.\,\cite{Kaviani:2015rxa} that the world volume black hole on rotating probe D1-branes
forms about the KT conifold singularity with the world volume horizon and temperature 
changing dramatically with the scale of chiral symmetry breakdown. In the conformal limit
of the UV including the KW solution, it has been shown in ref.\,\cite{Kaviani:2015rxa} that
the induced world volume metric on rotating probe D1-branes coincides with the BTZ black
hole metric modulo the angular coordinate. It has been shown there that, in this case, the
world volume horizon and temperature increase linearly and much faster with the angular 
velocity (dual to R--charge), compared with the world volume  horizon and temperature in
the non-conformal KT solution. However, it has been found there that the functional form
and behavior of the world volume horizon and temperature get dramatically modified once
a  background gauge field from the $U(1)$ R--symmetry of the KW solution is activated. 
It has been shown there that, in this case, the world volume horizon equation describes
that of the  $adS$--Reissner--Nordstr\"om balck hole modulo the angular coordinate, and
that the world volume Hawking temperature admits two distinct branches: one where it 
decreases and another where it increases with growing horizon size, describing `small' and
`large' back holes, respectively. However, by computing the energy-stress tensor, it has
been shown in ref.\,\cite{Kaviani:2015rxa}, that the energy of the probe D1-brane will 
eventually dissipate into the bulk and  form, with the large backreaction in the IR, a localised
black hole in the bulk. By gauge/gravity duality, this means that the energy from the defect
flavor sector will eventually dissipate into the gauge theory ($\mathcal{N}=1$ SYM).

These analyses have been further extended in ref.\,\cite{Kaviani:2016fvo}, by studying 
thermalization of higher dimensional probe $Dp$-branes ($p=7$ -- dual to flavored quarks) 
in Kuperstein--Sonnenschein model, ref.\,\cite{Kuperstein:2008cq} 
(see also refs.\,\cite{Ihl:2012bm} -- extending the model to finite temperature and density, 
and refs.\,\cite{Sakai:2003wu,Dymarsky:2009cm} -- extending the model in the KS solution).
Motivated by the Sakai--Sugimoto model, ref.\,\cite{Sakai:2004cn}, the model embeds the
probe D7-brane(s) into the KW gravity dual, ref.\,\cite{Klebanov:1998hh},  
$adS_5\times T^{1,1}$ with $T^{1,1}\simeq S^2\times S^3$, of $\mathcal{N}=1$
conformal gauge theory\footnote{See also refs.\,\cite{Ouyang:2003df} -- considering 
holomorphic embeddings of D7-branes into the KS solution dual to supersymmetric gauge
theory without flavor chiral symmetry breaking, and refs.\,\cite{Benini:2006hh} --
studying the backreaction of D7-branes in the KS $\&$ KW solutions.}.  The probe wraps
$adS_5\times S^3$ in $adS_5\times T^{1,1}$ and the transverse space is the two-sphere
$S^2\subset T^{1,1}$. The probe D7-brane then starts from the UV boundary at infinity, 
bends at the minimum extension in the IR, and ends up back the boundary. The probe 
D7-brane therefore produces a U-shape. Since  the $D7$-brane and anti $D7$-brane differ
only by orientation, the probe describes a supersymmetry (SUSY) breaking $D7$-brane/anti
$D7$-brane pair, merged in the bulk at minimal extension. The SUSY breaking pair also 
guarantees tadpole cancellation on the transverse space by the annihilation of total $D7$
charge. When the minimal extension shrinks to zero at the conifold point, the embedding
appears as a disconnected   $D7$-brane/anti $D7$-brane pair. The $D7$-brane then 
produces a V-shape. In the V-shape configuration, the induced world volume metric on the
$D7$-brane is that of $adS_5\times S^3$ and the dual gauge theory describes the conformal
and chiral symmetric phase. On contrary, in the U-shape  configuration, the induced metric on
the $D7$-brane has no $adS$ factor by the embedding parameter, i.e., by  non-zero minimal
extension,--which sets the IR scale conformal and chiral flavor symmetry breakdown--, 
giving the VEV deformations in the  dual gauge theory\footnote{As discussed in $\S$2, this
setup cannot be realized in the well-used $adS_5\times S^5$ background.}. In the setup of
ref.\,\cite{Kaviani:2016fvo}, the probe has also been allowed, in addition, to rotate about the
transverse $S^2\subset T^{1,1}$ (dual to finite R--charge chemical potential). It has been
shown in ref.\,\cite{Kaviani:2016fvo} that when the embedding parameter,--the minimal radial
extension of the probe--, which sets the IR scale of conformal and chiral flavor symmetry
breakdown in the dual gauge theory, is positive definite and additional spin (dual to finite 
R-charge chemical potential) is turned on, the induced world volume metrics on rotating probe
D7-branes admit thermal horizons and Hawking  temperatures despite the absence of black
holes in the bulk KW. It has been shown there that the world volume horizon grows from the
minimal extension with the world volume temperature scale changing dramatically with the
minimal extension which sets the IR scale of conformal and chiral flavor symmetry breakdown.
It has also been shown there that when, in addition, the world volume gauge electric field
(dual to finite baryon chemical potential) is turned on, the behavior of the world volume 
temperature changes dramatically with growing horizon size, describing `small' and `large' 
black holes. However, it has been shown in  ref.\,\cite{Kaviani:2016fvo} that this behavior
in the world volume Hawking temperature depends, in turn, on the size of the minimal
extension. It has been shown there that when the size of the minimal extension gets 
increased,  the behavior of the temperature changes again dramatically,--describing only
`large' black holes. It has then been concluded in ref.\,\cite{Kaviani:2016fvo} that since
the bulk gravity dual of the $\mathcal{N}=1$ gauge theory is at zero temperature,  while
the (rotating) probe brane dual to the flavor sector of the  gauge theory is at finite 
temperature, such systems provide novel examples of non-equilibrium steady states in
the gauge theory conformal and chiral flavor symmetry breakdown. However, by computing
the energy--stress tensor, it has been shown in ref.\,\cite{Kaviani:2016fvo} that the energy
of the probe flavor brane will eventually dissipate into the bulk, forming, with the large 
backreaction in the IR, a black hole in the bulk. By gauge/gravity duality, it was thus found
that the energy of the flavor sector will finally dissipate into the gauge theory 
conformal and chiral flavor symmetry breakdown ($\mathcal{N}=1$ SYM).

The aim of this work is to extend and generalize such previous analysis and study
different non-equilibrium systems and their energy flow in holographic models dual to
\emph{dCFTs} with spontaneous breakdown of the conformal and chiral flavor 
symmetry. 

The model we consider consists of the Ben Ami--Kuperstein--Sonnenschein
(BKS) holographic model, \cite{Ben-Ami:2013lca} (see also ref.\,\cite{Filev:2014bna}  --
extending the model to finite temperature and density). Motivated by the 
Kuperstein--Sonnenschein model, ref.\,\cite{Kuperstein:2008cq}, the model embeds
type IIB  probe (defect) flavor $Dp$-branes $(p=3,5)$ into the KW gravity dual, 
ref.\,\cite{Klebanov:1998hh},  $adS_5\times T^{1,1}$ with 
$T^{1,1}\simeq S^2\times S^3$, of $\mathcal{N}=1$ conformal gauge theory.
(The model also embeds type IIA (defect) probe flavor $Dp$-branes (i.e. $p=4$) into
the ABJM gravity dual, \cite{Aharony:2008ug})\footnote{In this paper, we will not
consider such type IIA probe brane configurations.}.  In the example of probe D3-brane,
the probe wraps  $adS_2\times S^2$ in $adS_5\times T^{1,1}$ and the transverse
space is the three-sphere $S^3\subset T^{1,1}$. In the example of the probe 
D5-brane, there are two cases: The probe either wraps $adS_3\times S^3$ in
$adS_5\times T^{1,1}$ and the transverse space is the two-sphere 
$S^2\subset T^{1,1}$, or it wraps  $adS_4\times S^2$ in $adS_5 \times T^{1,1}$
and the transverse space is the three-sphere $S^3\subset T^{1,1}$. In all examples, 
the probe $Dp$-brane starts from the UV boundary at infinity, bends at minimal 
extension in the IR, and ends up at the boundary. The probe $Dp$-brane thus produces
a  U-shape. Since  the  $Dp$-brane and anti $Dp$-brane differ only by orientation, the
probe describes a supersymmetry (SUSY) breaking $Dp$-brane/anti $Dp$-brane pair, 
merged in the bulk at minimal extension. The SUSY breaking pair also guarantees tadpole
cancellation on the transverse space by the annihilation of total $Dp$ charge. When the
minimal extension shrinks to zero at the conifold point, the embedding appears as a 
disconnected $Dp$-brane/anti $Dp$-brane pair. The $Dp$-brane then produces a V-shape.
In the  V-shape configuration, the induced world volume metric on the $Dp$-brane is that 
of $adS_m\times S^n$ and the dual gauge theory describes the conformal and chiral 
symmetric phase. On contrary, in the U-shape  configuration, the induced metric on the
$Dp$-brane has no $adS$  factor by the embedding parameter, i.e., by  non-zero minimal
extension,--which sets the IR scale conformal and chiral flavor symmetry breakdown--, 
giving the VEV deformations in the  dual gauge theory\footnote{As discussed in $\S$2,
this setup cannot be realized in the well-used $adS_5\times S^5$ background.}. Here, in
our setup, we let, in addition, the probe(s) rotate about the transverse $S^2$ or 
$S^3\subset T^{1,1}$ and turn on, in addition, the world volume gauge electric field
 (dual to finite R--charge and density chemical potential). 

The motivation is the fact that in such model the induced world volume metrics 
on the rotating probe branes, \emph{if} given by black hole geometries, are then
expected to give the Hawking temperatures of the probes dual to the temperatures
of \emph{defect} flavors in the dCFT with spontaneous breakdown of the conformal 
and chiral flavor symmetry. Since the gauge field theory itself is at zero temperature
while its flavor sector is at finite temperature, such systems exemplify novel 
non-equilibrium steady states in the dCFT with conformal and chiral flavor symmetry
breakdown. However, interactions between different sectors are expected. The 
energy-stress tensor of the thermal probes is then expected to yield the energy
dissipation from the probes into the system, dual to the energy dissipation from the
flavor sectors into the gauge theory. We are also interested in modifying our analysis
by turning on world volume gauge fields on the probe branes,-- including finite baryon
density chemical potential--, corresponding to turning on external fields in the dual
gauge theory. The motivation is the fact that in the presence of such fields the  
R--symmetry of the gauge  theory gets broken\footnote{We note that earlier studies
in gauge/gravity, \cite{Filev:2007gb}, have shown this result.}  and the corresponding
modifications in the induced metrics and energy-stress tensors of the probes are 
expected to reveal new features of thermalization.

The main results we find are as follows. We find that when the probe embedding is
U-like, the induced world volume metrics of \emph{only certain} type IIB rotating
probe flavor $Dp$-branes embedded in the KW gravity dual of CFT 
($\mathcal{N}=1$ SYM)  with spontaneous breakdown of the conformal and chiral
flavor symmetry admit thermal horizons and Hawking temperatures of expected
features despite the absence of black holes in the bulk KW\footnote{This is in contrast
with the main result of ref.\,\cite{Das:2010yw} where it has been shown that in the
$adS_5\times S^5$ gravity dual of CFT ($\mathcal{N}=4$ SYM) the induced world
volume metrics of \emph{all} type IIB ratating probe flavor $Dp$-branes
admit thermal horizons and Hawking temperatures of expected features despite the
absence of black holes in the bulk. At this point, we note again that $adS_5\times S^5$
contains no non-trivial cycle and admits no U-like embeddings -- admitting spontaneous
breakdown of the conformal and chiral flavor symmetry.}. We find that when the 
embedding is U-like, world volume black hole formation on type IIB rotating probe
flavor $Dp$-branes embedded in the KW gravity dual of CFT, with spontaneous 
breakdown of the conformal and chiral flavor symmetry, depends on the world
volume dimension and topology of the non-trivial internal cycle wrapped by the
probe. We find that when the embedding parameter, the minimal extension of the
probe, changes, the world volume black hole temperature of both spacetime
filling ($p=7$) and defect flavor branes ($p=5$) changes dramatically, setting
large hierarchies of temperature scales, while leaving the temperature behaviors
unchanged. We find, however, that when the world volume electric field is turned
 on, the behavior of the world volume temperature of both probe flavor
branes ($p=5,7$) changes dramatically, while the temperature scales set moderate
hierarchies. We find that in the presence of the electric field the world volume 
temperatures of the U-like embedded probe flavor branes admit two distinct branches:
there is one branch where the temperatures increase and another where they decrease
with growing horizon size, corresponding to `large' and `small' black holes, respectively. 
However, by examining the parameter dependence of the black hole solutions, we find,
in the case of spacetime filling flavor brane, this temperature behavior parameter 
dependent, unlike in the case of defect flavor branes where we find the temperature 
behavior parameter independent. We find that when the minimal extension gets 
increased, the world volume temperature of the  spacetime filling flavor brane ($p=7$)
changes its behavior again dramatically, admitting only one class of black hole solution,
including `large' black holes,  whereas that of the of the defect flavor brane ($p=5$)
remains unchanged. By gauge/gravity duality, we thus find that when the IR scale of 
symmetry breakdown is positive definite and the flavor sector gets R--charged, flavor
thermalization and non-equilibrium steady state formation in the KW (d)CFT with
spontaneous conformal and chiral symmetry breakdown is non-trivial and depends on
the type of flavors inserted into the gauge theory. By gauge/gravity duality, we also
find that the temperature of both spacetime filling and defect flavored quarks varies
dramatically with the IR scale of conformal and chiral flavor symmetry breakdown, 
which sets large hierarchies of  temperature scales, while leaves the temperature 
behaviors unchanged.  By gauge/gravity duality, we find, however, that when the
external electric field is turned on, the  temperature behavior of both spacetime filling
and  defect flavored quarks changes dramatically. In this case, we find, however, that
when, in addition, the IR scale increases, the temperature behavior of spacetime filling
flavored quarks changes again dramatically, but that of defect flavors remains unchanged.  
However, by computing the energy--stress tensor of the U-like embedded rotating probe
flavor $Dp$-branes, we find that, in any case, the energy of both spacetime filling and 
defect probe flavor branes will eventually dissipate into the bulk and form, with the large
backreaction in the IR, a black hole in the bulk. By gauge/gravity duality, we thus find that
the energy from both spacetime filling and defect flavor sectors will eventually dissipate
into the $\mathcal{N}=1$ gauge theory conformal and chiral flavor symmetry breakdown.

The paper is organized as follows. In Sec.\,2, we review the basics of the type IIB BKS
holographic models. We first review very briefly the basics of the KW solution and write
down the specific form of the background metrics and probe $Dp$-brane action used in
the BKS models. We then review the BKS models constructing U-like embedded probe
$Dp$-brane ($p=3,5,7$) solutions in KW. In Secs.\,3 -- 6, we modify the BKS models
by turning on spin and world volume electric field on the probe $Dp$-branes. In Sec.\,3,
we first derive the induced world volume metric and Hawking temperature on the probe
$D7$-brane wrapping $adS_5\times S^3 \subset adS_5\times T^{1,1}$ and rotating
about $S^2\subset T^{1,1}$. We then derive the energy--stress tensors of the rotating
D7-branes and compute their backreaction and energy dissipation. In Sec.\,4, we do the
same computations for the probe $D5$-branes wrapping 
$adS_3\times S^3 \subset adS_5\times T^{1,1}$ and rotating about $S^2\subset T^{1,1}$.
In Sec.\,5, we do the same computations for the probe $D5$-branes wrapping wrapping 
$adS_4\times S^2 \subset adS_5\times T^{1,1}$ and rotating about $S^3\subset T^{1,1}$. 
In Sec.\,6, we do the same calculations for the probe $D3$-brane wrapping
$adS_2\times S^2 \subset adS_5\times T^{1,1}$ and  rotating about $S^3\subset T^{1,1}$. 
In Sec.\,7, we summarize and discuss our results with future outlook.

\section{Review of the BKS models}

\subsection{The KW solution}

The particular ten-dimensional holographic string solution we are interested
in is the KW solution, \cite{Klebanov:1998hh}, resulting from taking the 
near horizon limit of a stak of $N$ background D3-branes on the conifold
point. The conifold is a singular and non-compact Calabi-Yau three-fold 
described, in terms of complex coordinates $z_n\in \mathbb{C}^4$, by 
the constraint and three-form as:

\begin{equation}
\label{condef}
\sum_{n=1}^{4} z_n^2=0,\;\;\;\;\;\;\;\;\;\;\ \Omega=\frac{dz_2\wedge dz_3
\wedge dz_4}{z_1}.
\end{equation}
The constraint equation in (\ref{condef}) describes a singular non-compact
cone with symmetry $SO(4)\times U(1)\simeq SU(2)\times SU(2) \times U(1)$,
and the three-form in (\ref{condef}) is charged under the $U(1)$ R-symmetry.
The base of the cone is a five-dimensional Einstein manifold, $X_5=T^{1,1}$,
of topology $SO(4)/U(1)\simeq SU(2)\times SU(2)/ U(1)$, which can be identified
by intersecting the constraint in (\ref{condef}) with the unit three-sphere
$\sum_{n=1}^{4}|z_n|^2=1$.

The constraint equation in (\ref{condef}) can recast its form as:

\begin{equation}
\label{condef2}
\text{det}_{ i,j} z_{ij}=0,\;\;\;\;\ \text{with} \;\;\;\;\;\ z_{ij}=\sum_n\sigma_{ij}^n z_n.
\end{equation}
Here $\sigma^n$ denate the Pauli matrices with $n=1,2,3$ and $\sigma^4=i\mathbf{I}$.
To solve Eq.\,(\ref{condef2}), one may use unconstrained variables $z_{ij}=a_ib_j$, where
$a_i,b_j$ denote complex scalars with $i,j=1,2$. These scalars poses an additional 
$SU(2)\times SU(2)$ global symmetry, which is quotiented by the $U(1)$ symmetry
via $a_i\rightarrow \exp(i\alpha) a_i$ and $b_j\rightarrow \exp(-i\alpha) b_j$, so that
the global symmetry is $SU(2)\times SU(2)/U(1)$. 

To construct the gauge field theory of the solution, one may use $a_i, b_j$ and notes the
following points. First, since the conifold is a Calabi-Yau manifold, it preserves by the virtue
of Killing spinor equations, one quarter of the original supersymmetry. Thus the solution
has $\mathcal{N}=1$ supersymmetry and one may associate chiral $\mathcal{N}=1$ 
superfields $A_i,B_j$ with $i,j=1,2$ to the complex scalars $a_i,b_j$. Second, further to
the $SU(2)\times SU(2)/U(1)$ global symmetry discusseed above, the solution has a 
$U(1)_R$ symmetry. Third, the field theory exemplifies a quiver gauge theory. Namely,
the field theory of $N$ D3-branes placed on the conifold tip point has a $SU(N)\times SU(N)$
gauge symmetry where $A_i$ and $B_j$ transform in the ($\mathbf{N},\overline{\mathbf{N}}$)
and ($\overline{\mathbf{N}},\mathbf{N}$) representation, respectively. Fourth, anomaly
cancellation in $U(1)_R$ requires $A_i$ and $B_j$ have R--charge $1/2$, respectively.
Thus the gauge theory also includes a marginal superpotential 
$W=\epsilon^{ij}\epsilon^{kl}\text{Tr}A_iB_kA_jB_l$ which is uniquely fixed by the 
symmetries modulo an overall factor. 

To construct the gravity dual of this gauge theory, one notes that Eq.\,(\ref{condef2})
describes a cone over the base of coset space $T^{1,1}=SU(2)\times SU(2)/U(1)$.
One also notes that the $z_{ij}$ can be explicitly parameterized by five Euler angels 
($\psi, \phi_i$, $\theta_i$) with $i=1,2$ (not written out here) whereby
the metric on the conifold then takes the form \cite{Candelas:1989js}:

\begin{eqnarray}
\label{6DConmet1}
ds_{T^{1,1}}^2&=& \frac{1}{9}\left(d\psi+\sum_{i=1}^{2}\cos\theta_i d\phi_i\right)^2
+ \frac{1}{6}\sum_{i=1}^{2}\left(d\theta_i^2+\sin^2\theta_i d\phi_i^2\right).
\end{eqnarray}
Here $0\leq\psi\leq 4\pi$, $0\leq\phi_i\leq 2\pi$ and $0\leq\theta_i\leq \pi$. 
It is clear from (\ref{6DConmet1}) that the $T^{1,1}$ base is a $S^1$ bundle
over $S^2\times S^2$. It is also clear from (\ref{6DConmet1}) that the $T^{1,1}$
base has the topology $S^2\times S^3$ where the three-cycle $S^3$ can be
parameterized by $\theta_1=\phi_1=0$ and the two-cycle $S^2$ by $\psi=0$, 
$\theta_1=\theta_2$, $\phi_1=-\phi_2$, with both $S^2$ and $S^3$ shrinking
to zero size at the tip of the cone.

The alternative way of writing the metric on the conifold is \cite{Evslin:2007ux}:

\begin{eqnarray}
\label{6DConmet2}
ds_{T^{1,1}}^2&=&\frac{r^2}{3}\bigg[\frac{1}{4}(\Omega_1^2+\Omega_2^2)
+\frac{1}{3}\Omega_3^2+\Big(d\theta-\frac{1}{2}\Omega_2\Big)^2+
\Big(\sin\theta d\phi-\frac{1}{2}\Omega_1\Big)^2\bigg].
\end{eqnarray}
Here $r$ is the radial coordinate of the conifold (\ref{condef}), $\theta$
and $\phi$ parameterize the $S^2$, and $\Omega_i$ are one-forms parameterizing
the $S^3$. The $\Omega_i$s can be represented by Maurer-Cartan one-forms $w_i$
via two  $SO(3)$ matrices (not written out here) parameterized by $\theta$ and 
$\phi$, respectively,  which show that the $S^3$ is fibered trivially over the $S^2$.

Taking the near-horizon limit of $N$ D3-branes at the tip of the cone
over $T^{1,1}$, gives the gravity dual, $adS_5\times T^{1,1}$, of
the $\mathcal{N}=1$ superconformal gauge field theory described by
the superfields $A_i, B_i$ and the superpotential $W$, as above. The 
full ten-dimensional warped background metric takes the form 
\cite{Klebanov:1998hh}:

\begin{eqnarray}
\label{10DKWmet}
ds_{10}^2&=&\frac{r^2}{L^2}dx_{n}dx^{n}+
\frac{L^2}{r^2}(dr^2+r^2 ds_{T^{1,1}}^2).
\end{eqnarray}
Here the first term in (\ref{10DKWmet}) is the usual four-dimensional 
Minkowski spacetime metric, the second term is the metric on the 
conifold, given either by (\ref{6DConmet1}) or by (\ref{6DConmet2}),
and $L^4\equiv (27\pi/4) g_s N (\alpha^{\prime})^2$. In order to have
a valid supergravity solution, (\ref{10DKWmet}), the number of D3-branes,
$N$, has to large and the string coupling, $g_s$, has to be small, so that 
$g_sN\gg1$; $\alpha^{\prime}=l_s^2$ denotes the string scale. Here 
the dilaton is  constant, and the other non-trivial  background field is a 
self-dual R--R  five-form flux (not written out here).

To add flavored quarks to the pure conformal gauge field theory, one embeds
probe flavor $Dp$-brane(s) ($p=3,5,7$) into the gravity dual above. 
The action of a $Dp$-brane is the sum of the DBI and CS actions and takes the 
general form:
\begin{eqnarray}
 \label{DPACTION} S_{Dp}&=&-g_sT_p
\int{d^{p+1}\xi\,e^{-\Phi}\sqrt{-\det(\gamma_{ab}+\mathcal{F}_{ab})}}+g_s
T_p\int{\sum_{p}C_{p+1}\wedge e^{\mathcal{F}_{ab}}}.
\end{eqnarray}
Here $C_{p+1}$ are the IIB R--R background fields coupling to the $Dp$-brane
world volume; $\mathcal{F}_{ab}=\mathcal{B}_{ab}+2\pi\alpha^{\prime}F_{ab}$ is
the gauge invariant field strength with $F_2$ the world volume gauge field and
$\mathcal{B}_2$ the pullback of the NS--NS two-form background field, $B_2$, 
onto the world volume, $\mathcal{B}_{ab}=B_{MN}\partial_a X^M\,\partial_b X^N$, 
$\gamma_{ab}=g_{MN}\partial_a X^M\,\partial_b X^N$ the pullback of the 
ten-dimensional metric $g_{MN}$ in string frame, and $\Phi=0$. Lastly, $\xi^a$
denote the world volume coordinates and $T_p=[(2\pi)^p\, g_s (\alpha^{\prime})
^{(p+1)/2}]^{-1}$ is the $Dp$-brane tension.

In what follows, we first consider the KW gravity dual described with metric 
(\ref{10DKWmet}) and write down the explicit form of the probe $Dp$-brane
($p=3,5,7$) actions, (\ref{DPACTION}), and world volume fields in the U-like
embeddings of BKS models. We then modify the BKS models by turning on angular
motion and world volume fields, and compute the induced world volume metrics 
and Hawking temperatures on the rotating probes.

\subsection{U-like embedded probe D7-branes wrapping $adS_5\times S^3$ $\subset$
 $adS_5\times T^{1,1}$}

As first example, we consider the U-like brane embedding configuration of 
ref.\,\cite{Kuperstein:2008cq}, embedding U-like probe D7-branes into the
KW gravity dual, corresponding to adding flavored quarks to its dual gauge
theory with spontaneous breakdown of the conformal and chiral flavor
symmetry. In the KW gravity dual, the probe D7-brane wraps the entire dual
spacetime coordinates $\{t,x_i,r\}$ ($i=1,2,3$) of $adS_5$ in the 
01234-directions, and the three-sphere $S^3$ $\subset$ $T^{1,1}$ 
parameterized by the forms $\{\Omega_i\}$ in the 567-directions. Therefore
the transverse space is given by the two-sphere $S^2$ $\subset$ $T^{1,1}$
parameterized by the coordinates $\theta$ and $\phi$ in the 89-directions. 
This all can be summarized and represented by the array:

\[ \begin{array}{lcr}
\;\;\;\;\,\  0\,\,\, 1\,\,\, 2\,\,\; 3\,\,\, 4\,\,\, 5\,\,\, 6\,\,\, 7\,\,\, 8\,\, 9\\
\mbox{D3}  \times \times\times\times\\
\mbox{D7}  \times \times \times \times\times\times\times\times \\ 
\end{array}\]
Here we note that, as $w_i$ are left-invariant forms, the ansatz preserves one
of the $SU(2)$ factors of the global symmetry group of the conifold, 
$SU(2)\times SU(2)\times U(1)$. Thus, one may assume that the coordinates $\theta$
and $\phi$ are independent of the $S^3$ coordinates. The embedding breaks one of
$SU(2)$, but by expanding the action around the solution it can be shown that 
contribution from the nontrivial $S^3$ show up only at the second order fluctuations.
Thus, one can assume, in classical sense, that $\theta$ and $\phi$ depend only on the
radial coordinate, $r$, of the conifold geometry.

The Kuperstein--Sonnenschein embedding of the D7-branes includes two choices.
There is one choice where the D7-branes are placed on two separate points of the
$S^2$, with both of them stretched down to the tip of the conifold at $r=0$ where
both the $S^2$ and $S^3$ shrink to zero size. The configuration produced, in
this case, is \emph{V}-like. There is another choice where the D7-branes
are placed on the $S^2$ with both of them smoothly merged into a single stack
somewhere at  $r=r_0$ above the conifold tip point. The configuration produced,
in this case, is \emph{U}-like. In both of the \emph{U}-like and  \emph{V}-like 
configurations, the D7-brane(s) wrap the $adS_5\times S^3$ spacetime, as in the 
array above. However,  on the transversal $S^2$, here are two different pictures.
In the \emph{V}-like configuration, the D7-branes look like two separate fixed points
while the \emph{U}-like configuration results an arc along the equator. The position 
of the two points, setting the position of the D7-branes, depends on $r$ and with 
them smoothly connected in the midpoint arc at $r=r_0$. In the gravity dual, this
configuration produces a one-parameter family of D-brane profiles with the parameter
$r_0$ setting the minimal radial extension of the D7-brane.

By the choice of the world volume fields $\phi=\phi(r)$ and $\theta=\theta(r)$,
it is easy to derive the induced world volume metric from (\ref{10DKWmet}) in 
terms of (\ref{6DConmet2}) and obtain from (\ref{DPACTION})
the action of the form:
\begin{equation}
\label{KSA}
S_{D7}=-\tilde{T}_{D7}\int{drdt\,r^3\sqrt{1+\frac{r^2}{6}({\theta^{\prime}}^2+
\sin^2\theta{\phi^{\prime}}^2)}},
\end{equation}
where $\tilde{T}_{D7}=N_fV_{\mathbb{R}^3}V_{\mathbb{S}^3}T_{D7}$ 
with $T_{D7}=1/(2\pi)^7 g_s(\alpha^{\prime})^4$. The Lagrangian in 
(\ref{KSA}) is $SU(2)$ invariant and therefore one can restrict motion to 
the equator of the $S^2$ parameterized by $\phi$ and $\theta=\pi/2$.
In writing down the action (\ref{KSA}) from (\ref{DPACTION}), one has set $F_{ab}=0$
and noted that since the KW gravity dual contains only R--R four-form fluxes,
so there is no contribution from the CS part and therefore the D7-brane action
is just the DBI action.

The solution of the equation of motion from the action (\ref{KSA}) yields
a one-parameter family of D7-brane profiles of the form:

\begin{eqnarray}
\label{KUPSol.}
\phi(r)=\sqrt{6}r_0^4\int_{r_0}^{r}{\frac{dr}{r\sqrt{r^8-r_0^8}}}=
\frac{\sqrt{6}}{4}\cos^{-1}\bigg(\frac{r_0}{r}\bigg)^4.
\end{eqnarray}
The solution (\ref{KUPSol.}) describes two separate branches, a disconnected
D7 and an anti D7-brane pair, including the \emph{V}-like configuration, when
$r_0=0$. On contrary, the solution (\ref{KUPSol.}) describes a single branch,
including the \emph{U}-like configuration,  when the two branches merge at
$r=r_0>0$. Furthermore, when the configuration is \emph{U}-like, taking the 
limit $r\rightarrow r_0$ implies $\phi^{\prime}(r)\rightarrow\infty$.

The solution (\ref{KUPSol.}) contains several important features. First, when 
the configuration is \emph{V}-like, it is clear from  $d\theta=d\phi=0$ that the 
induced world volume metric is that of $adS_5\times S^3$ and the configuration
describes the conformal and chiral symmetric phase. On contrary, when the 
configuration is \emph{U}-like, the induced world volume metric has no $adS$
factor and the conformal and chiral flavor symmetry of the dual gauge theory
must be broken spontaneously. Second, by taking the asymptotic UV limit, 
$r\rightarrow\infty$,  (\ref{KUPSol.}) describes two constant solutions 
$\phi_{\pm}=\pm\sqrt{6}\pi/8$. This gives an asymptotic UV separation
between the branes as $\Delta\phi=\phi_{+}-\phi_{-}=\sqrt{6}\pi/4$, and an 
asymptotic expansion as $\phi\simeq\pm\sqrt{6}\pi/8\pm\frac{\sqrt{6}}{4}
\left(r_0/r\right)^4+\cdots$. The asymptotic UV separation between the branes
is $r_0$ independent. In the dual gauge field theory $r_0$ corresponds to a 
normalizable mode, a vacuum expectation value (VEV). The fact that $r_0$ is
a modulus, or a flat direction, implies the spontaneous breaking of the conformal
symmetry. The expansion reveals that a $\Delta=4$ marginal operator has a VEV
fixed by $r_0$ as $\left< O\right>\sim r_0^4/(\alpha^{\prime})^2$, the 
fluctuations of which gives the Goldstone boson associated with the conformal 
symmetry breakdown. Third, the solutions $\phi_{\pm}$, giving an $r_0$-independent
UV separation, make the brane anti-brane interpretation natural. This is because
the brane world volume admits two opposite orientations once the asymptotic points 
$\phi_{\pm}$ are approached. Fourth, the presence of both the D7 and anti D7-brane
guarantees tadpole cancellation and annihilation of total charge on the transverse
$S^2$, and it breaks supersymmetry explicitly with the embedding being non-holomorphic.

To this end, one also notes that the above setup cannot be embedded in
the $adS_5\times S^5$ solution. This because the $S^5$ contains no 
nontrivial cycle and therefore the D7-brane will shrink to a point on
the $S^5$. This problem may be fixed by a specific choice of boundary
conditions at infinity, but this turns out to be incompatible with the
\emph{U}-shape configuration of interest. In addition, tadpole cancellation
by an anti-D7-brane is not required, since one has no 2-cycle as in the
conifold framework.

\subsection{U-like embedded probe D5-branes wrapping $adS_3\times S^3$
$\subset$ $adS_5\times T^{1,1}$}

As second example, we consider the U-like brane embedding configuration of 
ref.\,\cite{Ben-Ami:2013lca}, embedding U-like probe D5-branes into the
KW gravity dual, corresponding to adding defect flavored quarks to its dual 
gauge theory with spontaneous breakdown of the conformal and chiral flavor
symmetry. In the KW gravity dual, the probe D5-brane wraps some of the
dual spacetime coordinates, including $adS_3$ $\subset$ $adS_5$, given by
$\{t,x_i,r\}$ ($i=1$) in the  014-directions, and the three-sphere
$S^3$ $\subset$ $T^{1,1}$  parameterized by the forms $\{\Omega_i\}$ in
the 567-directions. Thus, in this eample, the probe D5-brane wraps the 
$adS_3\times S^3$ spacetime and the transverse space is given by the 
two-sphere  $S^2$ $\subset$ $T^{1,1}$ parameterized by the coordinates
$\theta$ and $\phi$ in the 89-directions.  This all can be summarized and 
represented by the array:

\[ \begin{array}{lcr}
\;\;\;\;\,\  0\,\,\, 1\,\,\, 2\,\,\; 3\,\,\, 4\,\,\, 5\,\,\, 6\,\,\, 7\,\,\, 8\,\, 9\\
\mbox{D3}  \times \times\times\times\\
\mbox{D5}  \times \times \;\;\;\;\;\;\;\ \times\times\times\times \\ 
\end{array}\]

By the choice of the world volume fields similar to the D7-brane above, $\phi=\phi(r)$ 
and $\theta=\text{const.}$, it is easy to derive the induced world volume metric from
(\ref{10DKWmet}) in terms of (\ref{6DConmet2}) and obtain from (\ref{DPACTION})
 the action of the form:
\begin{equation}
\label{BKSAD51}
S_{D5}=-\tilde{T}_{D5}\int{drdt\,r\sqrt{1+\frac{r^2}{6}{\phi^{\prime}}^2}},
\end{equation}
where $\tilde{T}_{D5}\equiv N_fV_{\mathbb{S}^3}T_{D5}$. Here we note that
the probe D5-brane wraps the same $S^3$ as the probe D7-brane above and the 
Lagrangian density in (\ref{BKSAD51}) is the same as the Lagrangian density in 
(\ref{KSA}) modulo $r^2$.

The solution of the equation of motion from the action (\ref{BKSAD51}) yields
a one-parameter family of D5-brane profiles of the form:

\begin{eqnarray}
\label{BKSSol.D51}
\phi(r)=\frac{\sqrt{6}}{2}\left(\pm\frac{\pi}{2}\mp\arctan\left(\frac{r_0^2}
{\sqrt{r^4-r_0^4}}\right)\right).
\end{eqnarray}
The solution (\ref{BKSSol.D51}) has the same general features, and satisfies the
same boundary conditions, as before. However, in this example, the UV separation
is $\Delta\phi=\sqrt{6}\pi/2$, instead of $\Delta\phi=\sqrt{6}\pi/4$, in the previous
example. Thus (\ref{BKSSol.D51}) represents another U-shape solution, though this
time, dual to dCFT, with the additional quark fields residing in just two out of the four 
spacetime dimensions. The expected  near boundary behavior of (\ref{BKSSol.D51})
is that of a VEV deformation of a marginal ($\Delta=2$) operator in one dimension.

\subsection{U-like embedded probe D5-branes wrapping $adS_4\times S^2$ $\subset$
$adS_5\times T^{1,1}$}

As third example, we consider the U-like brane embedding configuration of 
ref.\,\cite{Ben-Ami:2013lca}, embedding U-like probe D5-branes alternatively,
\cite{Filev:2013vka,Arean:2004mm}, into the KW gravity dual, corresponding
to adding defect flavored quarks to its dual  gauge theory with spontaneous
breakdown of the conformal and chiral flavor symmetry. In the KW gravity dual,
the probe D5-brane wraps some of the dual spacetime coordinates, including
$adS_4$ $\subset$ $adS_5$, given by $\{t,x_i,r\}$ ($i=1,2$) in the 
0124-directions, and the two-sphere  $S^2$ $\subset$ $T^{1,1}$  parameterized
by $\theta_1$, $\phi_1$ in the 56-directions. Thus, in this example, the probe
D5-brane wraps the $adS_4\times S^2$ spacetime and the transverse space is
given by the  three-sphere $S^3$ $\subset$ $T^{1,1}$ parameterized by the
coordinates  $\theta_2$, $\phi_2$ and $\psi$ in the  789-directions.  This all
can be summarized and represented by the array:

\[ \begin{array}{lcr}
\;\;\;\;\,\  0\,\,\, 1\,\,\, 2\,\,\; 3\,\,\, 4\,\,\, 5\,\,\, 6\,\,\, 7\,\,\, 8\,\, 9\\
\mbox{D3}  \times \times\times\times\\
\mbox{D5}  \times \times \times \;\;\;\ \times\times\times \\ 
\end{array}\]

By the choice of the world volume fields $\theta_2=\pi-\theta_1$, 
$\phi_1=\phi_2$, and $\psi=\psi(r)$  it is easy to derive the induced
world volume metric from the metric (\ref{10DKWmet}) in terms of
(\ref{6DConmet1}) and obtain from (\ref{DPACTION}) the action of
the form:

\begin{equation}
\label{BKSAD52}
S_{D5}=-\tilde{T}_{D5}\int{drdt\,r^2\sqrt{1+\frac{r^2}{9}{\psi^{\prime}}^2}},
\end{equation}
where $\tilde{T}_{D5}\equiv N_fV_{\mathbb{S}^2}T_{D5}$. Here we note that
the probe D5-brane wraps this time the $S^2$, unlike in the previous example, and
the Lagrangian density in (\ref{BKSAD52}) is not the same as the Lagrangian 
density in (\ref{BKSAD51}).

The solution of the equation of motion from the action (\ref{BKSAD52}) yields
a one-parameter family of D5-brane profiles of the form:

\begin{eqnarray}
\label{BKSSol.D52}
\psi(r)=\pm\frac{\pi}{2}\mp\arcsin\left(\frac{r_0^3}{r^3}\right).
\end{eqnarray}
The solution (\ref{BKSSol.D52}) has the same of the general features, and meets
the same boundary conditions, as before. However, in this example, the UV separation
is given by $\Delta\psi=\psi_{+}-\psi_{-}=\pi/2-(-\pi/2)=\pi$, with both $\psi_{\pm}$ 
corresponding to supersymmetric D5-brane embedding and preserving different
supersymmetries. Thus (\ref{BKSSol.D52}) represents another U-shape solution, 
though this time, supersymmetric and dual to dCFT, with the additional quark fields
residing in three out of the four spacetime dimensions. The expected  near boundary
behavior of (\ref{BKSSol.D52}) is that of a VEV deformation of a marginal $\Delta=2$ 
operator.

\subsection{U-like embedded probe D3-branes wrapping $adS_2\times S^2$ $\subset$
 $adS_5\times T^{1,1}$}

As final example, we consider the U-like brane embedding configuration of 
ref.\,\cite{Ben-Ami:2013lca}, embedding U-like probe D3-branes into the
KW gravity dual, corresponding to adding defect flavored quarks to its dual
gauge theory with spontaneous breakdown of the conformal and chiral flavor
symmetry. In the KW gravity dual, the probe D3-brane wraps some of the
dual spacetime coordinates, including $adS_2$ $\subset$ $adS_5$, given by
 $\{t,r\}$ in the 04-directions, and the two-sphere $S^2$ $\subset$ $T^{1,1}$
parameterized by $\theta_1$, $\phi_1$ in the 56-directions. Thus, the probe
D3-brane wraps the $adS_2\times S^2$ spacetime and the transverse space is
given by the  three-sphere $S^3$ $\subset$ $T^{1,1}$ parameterized by the
coordinates  $\theta_2$, $\phi_2$ and $\psi$ in the  789-directions.  This all
can be summarized and represented by the array:

\[ \begin{array}{lcr}
\;\;\;\;\,\  0\,\,\, 1\,\,\, 2\,\,\; 3\,\,\, 4\,\,\, 5\,\,\, 6\,\,\, 7\,\,\, 8\,\, 9\\
\mbox{D3}  \times \times\times\times\\
\mbox{D3}  \times \;\;\;\;\;\;\;\;\;\;\ \times\times\times \\ 
\end{array}\]

By the choice of the world volume fields similar to the probe D5-brane
above, $\phi=\theta=\text{const.}$ and $\psi=\psi(r)$, it is easy to
derive the induced world volume metric from (\ref{10DKWmet}) in terms
of (\ref{6DConmet1}) and obtain from (\ref{DPACTION}) the
action of the form:
\begin{eqnarray}
\label{dbiacd03}
S_{D3}=-\tilde{T}_{D3}\int{dr dt\sqrt{1+\frac{r^2(\psi^{\prime})^2}{9}}},
\end{eqnarray} 
where $\tilde{T}_{D3}=N_fV_{\mathbb{S}^2}T_{D3}$. We also note that
the Lagrangian density in (\ref{dbiacd03}) is, up to a the radial factor,  $r^2$,
the same as the density of the probe D5-brane in (\ref{BKSAD52}).

The solution of the equation of motion from the action (\ref{dbiacd03}) yields
a one-parameter family of D3-brane profiles of the form:

\begin{eqnarray}
\label{BKSSol.D3}
\psi(r)=\mp\frac{3\pi}{2}\pm 3\arctan\left(\frac{r_0}{\sqrt{r^2-r_0^2}}\right).
\end{eqnarray}
The solution (\ref{BKSSol.D3}) has the same general features, and satisfies the
same boundary conditions, as before. However, in this example, the UV separation
is $\Delta\psi=3\pi$ whereby the brane never wraps the internal cycle more than
once, as $\psi$ $\in$ $[0,4\pi]$. Thus (\ref{BKSSol.D3}) represents another
U-shape solution dual to dCFT, with the additional quark fields being
``quantum mechanical",-- depending only on time and not on space. The expected 
near boundary behavior of (\ref{BKSSol.D3}) is that of a VEV deformation of a 
marginal ($\Delta=1$) operator in one dimension.

\section{Induced metric and temperature on U-like embedded
probe D7-brane wrapping $adS_5\times S^3\subset adS_5\times T^{1,1}$
and spinning about $S^2\subset T^{1,1}$ with world volume electric field
turned on}

As first example in our study, we consider the U-like probe D7-brane model
wrapping $adS_5\times S^3$ in $adS_5\times T^{1,1}$ reviewed in Sec.\,2,
and turn on, in addition, spin as well as world-volume gauge electric field on 
the probe, following closely \cite{Kaviani:2016fvo}. We include additional 
spin degrees of freedom, dual to finite R--charge chemical potential, by 
allowing in our system the D7-brane rotate in the $\phi$ direction of the 
transverse $S^2$ with conserved angular momentum. Thus, in our setup,
 we allow $\phi$ to have time-dependence as well, such that 
$\dot{\phi}(r,t)=\omega=const.$, with $\omega$ denoting the angular
velocity of the probe. This way, we construct rotating solutions. Hence,
the world-volume field is given by $\phi(r,t)$, with other directions fixed.
We also include the additional contribution from world-volume fields
strengths, $F_{ab}$, in (\ref{DPACTION}), corresponding to world-volume
gauge  fields, dual to finite baryon density chemical potential, by noting
the following:  In the presence of $N_f$ flavors, the gauge theory posses
a global  $U(N_f)\simeq SU(N_c)\times U(1)_q$ symmetry. The $U(1)_q$
counts the net number of quarks, that is, the number of baryons times $N_c$. 
In the gravity dual, this global symmetry corresponds to the $U(N_f)$ gauge
symmetry on the world volume of the $N_f$ D7-brane probes. The conserved
currents associated with the $U(N_f)$ symmetry of the gauge theory are dual
to the gauge fields, $A_{\mu}$, on the D7-branes. Hence the introduction of a
chemical potential $\mu$ or a non-vanishing $n_B$ for the baryon number the
gauge theory corresponds to turning on the diagonal $U(1)\subset U(N_f)$ 
gauge field, $A_{\mu}$ on the world volume of the D7-branes. We may describe
external fields in the field theory, coupled to anything having $U(1)$ charge, by
introducing non-normalizable modes for $A_{\mu}$ in the gravity dual
 (e.g. see ref.\,\cite{Kruczenski:2003be,O'Bannon:2008bz}).

In this section therefore we will study U-like embedded probe D7-branes spinning
with an angular  frequency $\omega$, and with a $U(1)$ world volume gauge field
$A_{\mu}$.  We note that in order to have the gauge theory at finite chemical 
potential or baryon number density, it suffices to turn on the time component 
of the gauge field, $A_t$. By symmetry considerations, one may take $A_t=A_t(r)$. 
As we shall briefly discuss below, a potential of this form will support an electric field.

Therefore, we will consider the ansatz for the D7-brane world-volume
field of the form $\phi(r,t)=\omega t+f(r)$, with other directions
fixed, and $F_{ab}=F_{rt}=\partial_rA_t(r)$. Using this ansatz and 
the metric (\ref{10DKWmet}) in terms of (\ref{6DConmet2}) , it is
easy to derive the components of the induced world-volume metric
on the D7-brane, $g_{ab}^{D7}$, and compute the determinant, 
$\det g_{ab}^{D7}$, resulting the DBI action (\ref{DPACTION}), as:

\begin{eqnarray}
\label{dbiac2}
S_{D7}=-\tilde{T}_{D7}\int{dr dt\,r^3\sqrt{1-\frac{L^4\dot{\phi}^2}{6r^2}+
\frac{r^2\,(\phi^{\prime})^2}{6}-(A^{\prime}_t(r))^2}}.
\end{eqnarray}
Here we note that by setting $\dot{\phi}=\omega=A_t(r)=0$, our action
(\ref{dbiac2}) reduces to that of the probe D7-brane action in the
Kuperstein--Sonnenschein model, (\ref{KSA}). As in the
Kuperstein--Sonnenschein model reviewed in Sec.\,2, we restrict
brane motion to the $\phi$-direction of the transverse $S^2$ sphere and
fix other directions constant. Thus, in our set-up we let, in addition,
the probe rotate about the $S^2$, and furthermore turn on a non-constant
 world-volume electric on the probe.

The D7-brane equations of motion from the action (\ref{dbiac2}) read:

\begin{eqnarray}
\label{D7eq}
\frac{\partial}{\partial r}\Bigg[\frac{r^5\,\phi^{\prime}}{\sqrt{1+
\frac{r^2\,(\phi^{\prime})^2}{6}-\frac{L^4\, \dot{\phi}^2}{6r^2}-
(A^{\prime}_t(r))^2}}\Bigg]&=&\frac{\partial}{\partial t}\Bigg[\frac{L^4
\,r\,\dot{\phi}}{\sqrt{1+\frac{r^2\,(\phi^{\prime})^2}{6}-\frac{L^4\,
\dot{\phi}^2}{6r^2}-(A^{\prime}_t(r))^2}}\Bigg],\;\;\ \\ \label{D7eqA}
 \frac{\partial}{\partial r}\Bigg[\frac{r^3\, A^{\prime}_t(r)}{\sqrt{1+
\frac{r^2\,(\phi^{\prime})^2}{6}-\frac{L^4\, \dot{\phi}^2}{6r^2}
-(A^{\prime}_t(r))^2}}\Bigg]&=& 0.
\end{eqnarray}
By taking the large radii limit, $r\rightarrow\infty$, it
is straightforward to see that Eq.\,(\ref{D7eqA}) solves to
$A_t(r)\simeq \mu-a_B/r^2$, with $\mu$  being the chemical
potential and $a_B$ the VEV of baryon density number. We note
that this solution for $A_t$ is of expected form, since an 
electric field will be supported by a potential of the from
$A_t\simeq r^{-2}$. Since this is a rank one massless field
in $adS$, it must correspond to a dimension four operator
or current in the gauge theory. This is just what one would
expect from an R--current, to which gauge fields correspond. 

Take rotating solutions to (\ref{D7eq}) of the form:

\begin{eqnarray}
\label{rotsol2}
\phi(r,t)&=&\omega t+f(r),\;\;\;\;\ f(r)=\sqrt{6}r_0^4
\int_{r0}^{r}{\frac{dr}{r}\sqrt{\frac{1-L^4\,\overline{\omega}^2
/r^2-(A^{\prime}_t(r))^2}{r^8-r_0^8}}}.
\end{eqnarray}
Here we set $\overline{\omega}=\omega/\sqrt{6}$ and note that when
$\omega=A_t(r)=0$, our solution (\ref{rotsol2}) integrates to that
of probe D7-brane in the Kuperstein--Sonnenschein model, 
\cite{Kuperstein:2008cq}, reviewed in Sec.\,2, with the probe wrapping
$adS_3\times S^3$ in $adS_5\times T^{1,1}$ (see Eq.\,(\ref{KUPSol.})).
The solution (\ref{rotsol2}) is parameterized by $(r_0,\omega, a_B)$
and describes probe D7-brane motion, with non-constant world-volume
gauge field $A_t(r)$ and angular velocity $\omega$ about the transverse
$S^2\subset T^{1,1}$, starting and ending up at the boundary. The probe
descends from the UV boundary at infinity to the  minimal extension 
$r_0$ in the IR where it bends back up the boundary. We also note that
inspection of (\ref{rotsol2}) illustrates, in the limit $r\rightarrow\text{large}$,
that the behavior of $df(r)/dr=f_r(r)$ does not depend on $\omega$
and $a_B$ (see Fig.\,\ref{fig:KWgp2}). This illustrates that in the
large radii limit the solution $f(r)$ in (\ref{rotsol2}) gives the
world-volume field $\phi(r)$ of the Kuperstein--Sonnenschein model
(see Sec.\,2) with the boundary values $\phi_{\pm}$ in the asymptotic
UV limit,  $r\rightarrow\infty$ (see also Fig.\,\ref{fig:KWgp2}).
But, we note that in the (other) IR limit, i.e., when $r\rightarrow\text{small}$,
the behavior of $f_r(r)$ does depend on $\omega$. Furthermore, we note here
that by turning on, in addition, the world volume electric field merely
changes the scale but leaves the behavior of $f_r(r)$ in the IR unchanged. 
Inspection of (\ref{rotsol2}) illustrates that in the IR the behavior of
$f_r(r)$ with $\omega>0$ is comparable to that of with $\omega=0$, only
if certain $\omega>0$ are chosen (see Fig.\,\ref{fig:KWgp2}). This 
illustrates that in the small radii limit in the IR the behavior of $f_r(r)$
(here) does compare to that of $\phi^{\prime}(r)=d\phi/dr$ in the 
Kuperstein--Sonnenschein model (see Sec.\,2),--with 
$\phi^{\prime}(r)\rightarrow\infty$ in the IR limit $r\rightarrow r_0$--,
consistent with U-like embedding, only if certain $\omega>0$ are chosen.

To derive the induced metric on the D5-brane, we put the rotating solution
(\ref{rotsol2}) into the background metric (\ref{10DKWmet}) in terms of
(\ref{6DConmet2}) and obtain:

\begin{eqnarray}
\label{ind2}
ds_{D7}^2&=&-\frac{1}{3L^2}(3r^2-L^4\omega^2)dt^2\notag\\ && +\frac{L^2}{r^2}
\left[\frac{3r^2(r^8-r_0^8)+r_0^8(6r^2(1-(A^{\prime}_t(r))^2)-L^4\omega^2)}{3r^2
(r^8-r_0^8)}\right]dr^2\notag\\ &&+\frac{2L^2 \omega r_0^4}{3r^2}\sqrt{\frac{6r^2
(1-(A^{\prime}_t(r))^2)-L^4\,\omega^2}{r^8-r_0^8}}drdt\notag\\ &&+\frac{r^2}{L^2}
(dx^2+dy^2+dz^2)+\frac{L^2}{3}\left[\frac{1}{2}(\Omega_1^2+\Omega_2^2)+\frac{1}{3}
\Omega_3^2\right]\notag\\ &&-\frac{L^2}{3}\left[\omega\Omega_1 dt+ \frac{r_0^4}{r^2}
\sqrt{\frac{6r^2(1-(A^{\prime}_t(r))^2)-L^4\,\omega^2}{r^8-r_0^8}}\Omega_1 dr\right].
\end{eqnarray}
Here we note that by setting $\omega=A_t=0$, our induced world-volume
metric (\ref{ind2}) reduces to that of the Kuperstein--Sonnenschein
model, \cite{Kuperstein:2008cq}, reviewed in Sec.\,2. In this case, for $r_0=0$
the induced world-volume metric is that of $adS_5\times S^3$ and the dual gauge
theory describes the conformal and chiral symmetric phase. On contrary, for 
$r_0>0$ the induced world volume metric has no $adS$ factor and the conformal
invariance of the dual gauge theory must be broken in such case. In order to 
find the world-volume horizon and Hawking temperature, we first eliminate the
relevant cross term. To eliminate the relevant cross-term in (\ref{ind2}),
we consider a coordinate transformation:

\begin{equation}
\tau=t-\omega\,L^4 r_0^4 \int{\frac{dr\,(6r^2(1-(A^{\prime}_t(r))^2)-L^4
\,\omega^2)^{1/2}}{r^2(3r^2-L^4 \omega^2)(r^8-r_0^8)^{1/2}}}.
\end{equation}
The induced metric on the rotating D7-brane(s) then takes the form:

\begin{eqnarray}
\label{indAD7}
ds_{D7}^2 &=&-\frac{(3r^2-L^4 \omega^2)}{3L^2}d\tau^2\notag\\ &&+
\frac{L^2}{r^2}\left[\frac{(3r^2-L^4 \omega^2)(r^8-r_0^8)+r_0^8(6r^2
(1-(A^{\prime}_t(r))^2)-L^4 \omega^2)}{(3r^2-L^4 \omega^2)(r^8-r_0^8)}
\right]dr^2\notag\\ && +\frac{r^2}{L^2}(dx^2+dy^2+dz^2)+\frac{L^2}{3}
\left[\frac{1}{2}(\Omega_1^2+\Omega_2^2)+\frac{1}{3}\Omega_3^2\right]
\notag\\ &&-\frac{L^2}{3}\left[\omega\Omega_1d\tau+\frac{3\,r_0^4}
{3r^2-L^4 \omega^2} \sqrt{\frac{6r^2(1-(A^{\prime}_t(r))^2)-L^4\,\omega^2}
{r^8-r_0^8}}\Omega_1 dr\right].
\end{eqnarray}
Here we note that for $r_0=0$ the induced world-volume metric (\ref{indAD7})
has no horizon solving $-g_{\tau\tau}=g^{rr}=0$, and thus not given by the
black hole geometry. On contrary, for $r_0>0$ the induced world-volume 
horizon solves the equation:

\begin{figure}[t]
\begin{center}
\epsfig{file=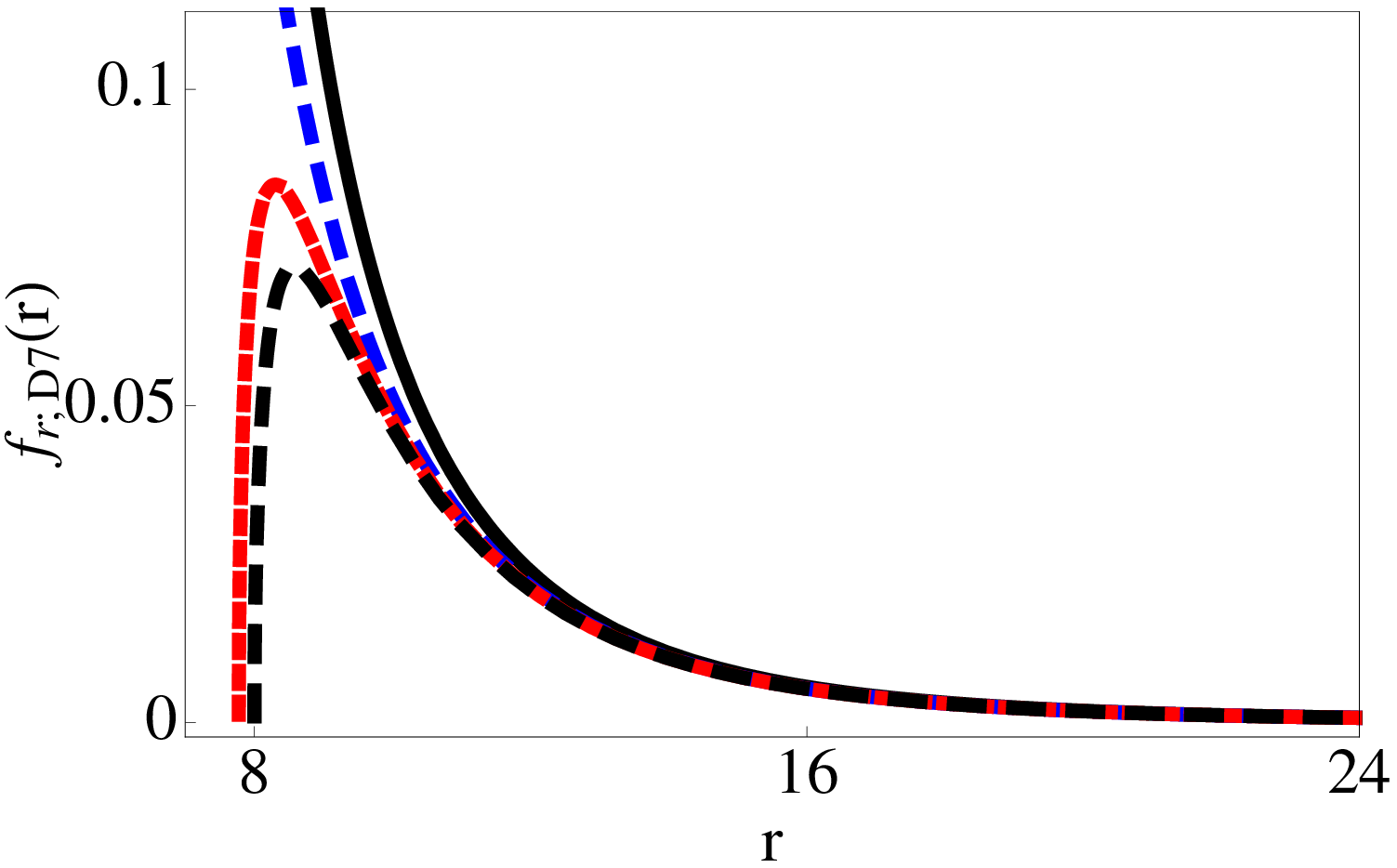,width=.50\textwidth}~~\nobreak
\epsfig{file=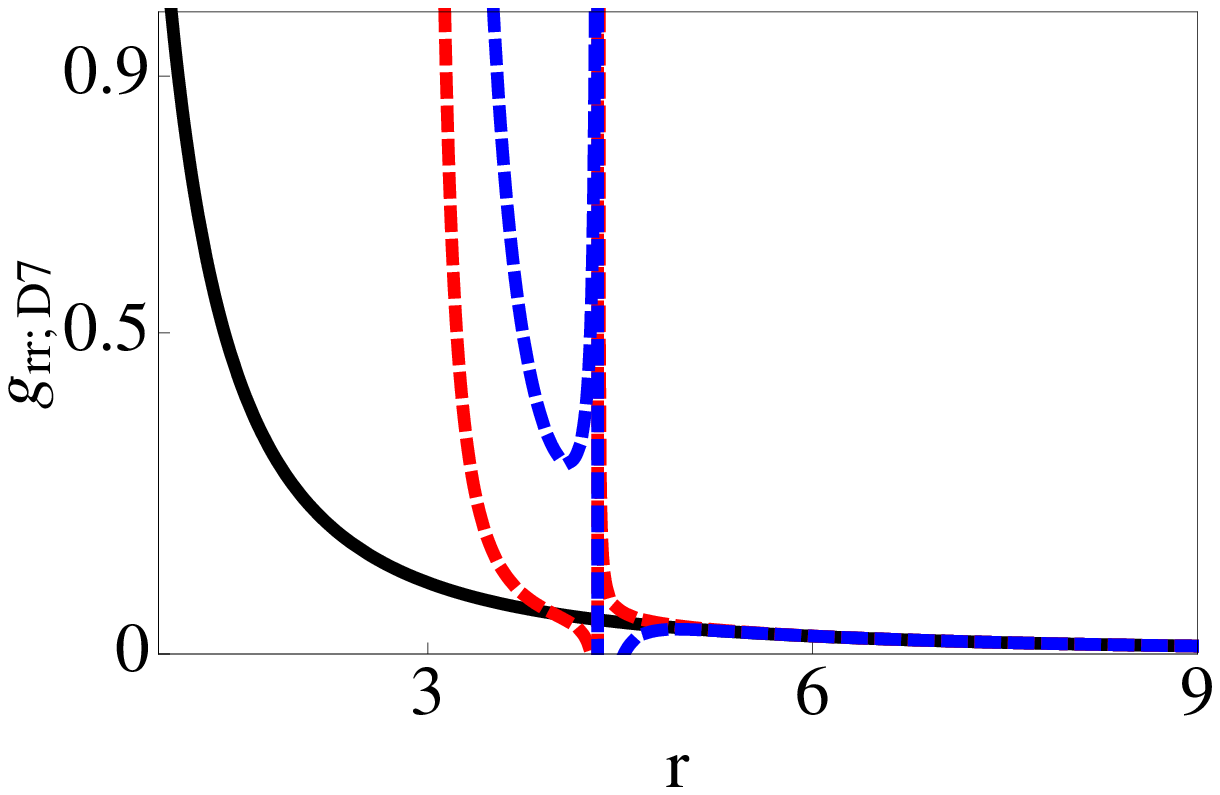,width=.50\textwidth}
\caption{[Left] The behavior of the derivative of the world
volume field with respect to $r$ with $L=1, r_0=7$, $\omega=0$,
$a_B=0$ (black-solid), $\omega=7.5$, $a_B=100$ (blue-dashed),
$a_B=180$ (red-dashed), and $a_B=200$ (black-dashed). [Right]
The behavior of the $g_{rr}$ component of the induced world volume
metric with $L=1,r_0=\omega=a_B=0$ (black-solid), $r_0=3,
\omega=7.5, a_B=20$ (red-dashed), and $a_B=60$ (blue-dashed).}
\label{fig:KWgp2}
\end{center}
\end{figure}

\begin{equation}
\label{HorEq2}
H(r)=r^2(r^8-r_0^8)(3r^2-L^4\omega^2)=0.
\end{equation}
Eq.\,(\ref{HorEq2}) contains two real positive definite zeros 
(see also Fig.\,\ref{fig:KWgp2}). The thermal horizon of the induced
world-volume black hole geometry is given by the zero $3r_H^2-L^4\omega^2=0$,
with the horizon varying continuously with varying the angular velocity
$\omega$, as expected. Eq.\,(\ref{HorEq2}) also shows that the horizon must
grow from the minimal extension $r_0\neq0$ with increasing the angular
velocity  $\omega$. We thus conclude at this point that when $r_0$ is positive
definite ($r_0>0$) and spin is turned on  ($\omega>0$), the induced world-volume
metric on the rotating probe D7-brane wrapping $adS_5\times S^3$ in 
$adS_5\times T^{1,1}$ admits a thermal horizon growing continuously with increasing
the angular velocity, regardless of the  presence of the world-volume gauge electric
field.

The Hawking temperature on the probe D7-brane can be found from the induced
world-volume metric (\ref{indAD7}) in the form:

\begin{equation}
\label{TH2}
T_{H;D7}=\frac{(g^{rr})^{\prime}}{4\pi}\bigg|_{r=r_H}=\frac{3r_H^3(r_H^8-r_0^8)}
{2\pi L^2 r_0^8[6r_H^2(1-(A^{\prime}_t(r))^2)-L^4\omega^2]}=\frac{r_H^7
(r_H^8-r_0^8)}{2\pi L^2 r_0^8(r_H^6-8a_B^2)}.
\end{equation}
Here, we note  from (\ref{TH2}) that when the world volume horizon
is at the minimal radial extension, $r_H=r_0$, the Hawking temperature of
the world volume black hole geometry is identically zero, $T_{H; D7}=0$.
The Hawking temperature (\ref{TH2}) of the world-volume black hole solution
(\ref{indAD7}) also has a number of interesting features, which depend on the 
choice of parameters $\{a_B,r_0, \omega\}$. Thus we note the following points:
$\\$
$\bullet$ For $a_B=0$, $r_0=4$ and $\omega\geq 0$, the world-volume 
Hawking temperature $T_{H;D7}$ (\ref{TH2}) increases monotonically with
increasing the angular velocity $\omega$, by which the world volume horizon
size $r_H$ grows (see Fig.\,\ref{fig:T0TAD7}). In this case, changing $r_0$
changes the scale, but not the behavior of the temperature (\ref{TH2}). In 
particular, inspection of (\ref{TH2}) shows that decreasing/increasing $r_0$ 
increases/decreases the temperature and produces a large hierarchy between
temperature scales, $T_{H;D7}(r_0<1)/T_{H;D7}(r_0>1)\simeq 10^8$, 
while the temperature behavior remains unchanged (see Fig.\,\ref{fig:T0TAD7}). 
$\\$
$\bullet$ For $0<a_B\leq 180,r_0=4$ and $\omega\geq 0$, the world-volume
Hawking temperature $T_{H;D7}$ (\ref{TH2}) no longer behaves monotonically
with growing the horizon size $r_H$. Instead, the temperature of the solution
(\ref{TH2}) admits three distinct branches, including two obvious classes of 
black hole solutions (see Fig.\,\ref{fig:T0TAD7}). The first branch appears once
the horizon starts to grow from the minimal extension $r_0$. In this case, the 
temperature $T_{H;D7}$ (\ref{TH2}) decreases almost immediately to negative
values\footnote{Here we note that in other interesting study in the literature, 
ref.\,\cite{Nakamura:2013yqa}, using similar D-brane systems, it has been shown
that when just an electric field is turned on (in place of rotation, or R--charge, 
considered here), in certain codimensions, the induced world-volume temperature
of the probe is given by a decreasing function of the electric field (see Eq.\,(24) in 
ref.\,\cite{Nakamura:2013yqa})}, before peeking off very sharply, whereby producing
a divergent-like behavior, hitting zero, and then growing into positive values 
(see Fig.\,\ref{fig:T0TAD7}). The second branch appears once the horizon 
continues to grow, away from the zero point. In this case, the temperature 
$T_{H;D7}$ (\ref{TH2}) decreases to positive values, in an `inverse-like' manner
with the horizon, before reaching its non-zero minimum (see Fig.\,\ref{fig:T0TAD7}). 
Therefore in the neighborhood of the zero point the world-volume temperature of
the solution decreases with increasing horizon size. These `small' black holes have
the familiar behavior of five-dimensional black holes in asymptotically flat spacetime,
since their temperature decreases with increasing horizon size $r_H$. The third branch
appears once the horizon continues to grow from its radius setting the
non-zero minimum of $T_{H;D7}$. In this case, the other class of black hole
solution appears with its temperature only increasing with increasing horizon
size $r_H$, like `large' black holes (see Fig.\,\ref{fig:T0TAD7}). This behavior
is like that of the temperature with $a_B=0$, however, in the latter two 
branches, the temperature of the black hole solution scales higher than that
of with $a_B=0$ (see Fig.\,\ref{fig:T0TAD7}). It is thus remarkable that by
varying $a_B$, the scale and behavior of the temperature changes 
continuously with growing horizon size $r_H$ (see Fig.\,\ref{fig:T0TAD7}). 
It is clear that by increasing/deceasing $a_B$ the temperature behavior
changes dramatically, while the temperature scales change very moderately,
producing only small hierarchies, $T_{H;D7}(a_B<)/T_{H;D7}(a_B>)
\simeq 0.5$ (see Fig.\,\ref{fig:T0TAD7}).
$\\$
$\bullet$ By varying $r_0$, while keeping the other parameters within
the above range, the scale and behavior of the temperature of the black
hole solution, $T_{H;D7}$ (\ref{TH2}), change dramatically once again.
For $r_0<1$, only  the temperature scale changes, but for $r_0>1$, i.e.,
$r_0\simeq 8$, the scale and, in particular, the behavior of the temperature
chage dramatically. In this latter case, the temperature scales much less 
than in previous cases ($r_0>1$), increases with increasing $a_B$ as before,
and behaves similarly to the temperature with $a_B=0$, increasing
continuously with growing $r_H$, despite now $a_B\neq 0$ 
(see Fig.\,\ref{fig:T0TAD7}). In this latter case, therefore there is only
one branch, corresponding to one class of black hole solution, including
`large' black holes only, with temperature scales setting relatively large
hierarchies, $T_{H;D7}(r_0<)/T_{H;D7}(r_0>)\simeq 10^3$.

We therefore conclude at this point that when the minimal extension
changes, the temperature scale of the world volume black hole solution
changes dramatically and sets large hierarchies, while the temperature
behavior remains unchanged. We conclude, however, that when the 
world wolume electric field is turned on, the temperature behavior 
changes dramatically and includes two distinct branches corresponding
to `large' and `small' black hole solutions, respectively, while the 
temperature scale changes moderately and thereby sets only moderate
hierarchies. In this case, we conclude, however, that when, in addition,
the minimal extension gets increased, the temperature behavior changes
once again dramatically and includes only one branch corresponding to
`large' black holes, with the temperature scale setting relatively large 
hierarchies.

We further notice that by taking into account the backreaction of
the above solution to the KW gravity dual, $adS_5\times T^{1,1}$, 
one accordingly awaits the D7-brane to form a mini black hole in KW, 
corresponding to a locally thermal gauge field theory in the probe limit.
Thereby, the rotating D7-brane describes a thermal object in the dual
gauge field theory. In the KW example here, the configuration is dual
to  $\mathcal{N}=1$ gauge theory coupled to a flavor
subject to an external electric field. As the gauge theory itself is at zero
temperature, while its flavor sector is at finite temperature, given by
(\ref{TH2}), we conclude that the configuration describes non-equilibrium
steady state. However, as we shall show below, the energy from the flavor
sector will in the end dissipate into the gauge theory conformal and chiral
flavor symmetry breakdown.

So far, in this example, we neglected the backreaction of the D7-brane
to the KW gravity dual, as we worked in the probe limit. It is useful to 
observe the extend this is justifiable. To this end, we derive the components
of the energy--stress tensor of the rotating D7-brane, which take the form:

\begin{eqnarray}
\label{TA1}
\sqrt{-g}J_{t;D7}^{t}&\equiv&\frac{\tilde{T}_{D7}\,r^3(1+r^2(\phi^{\prime})^2/6)}
{\sqrt{1+r^2(\phi^{\prime})^2/6-L^4\dot{\phi}^2/6r^2-(A^{\prime})^2}}\notag\\
&=&\frac{\tilde{T}_{D7}(r^4(r^{10}-L^4\overline{\omega}^2r_0^8)-4r_0^8a_B^2)}
{r^4\sqrt{(r^8-r_0^8)(r^6-L^4\overline{\omega}^2 r^4-4a_B^2)}},\;\;\;\;\;\;\;\;\
\\ \label{TA2}\sqrt{-g}J_{r;D7}^{r}&\equiv&-\frac{\tilde{T}_{D7}\,
r^3(1-L^4\dot{\phi}^2/6r^2)}{\sqrt{1+r^2(\phi^{\prime})^2/6-L^4\dot{\phi}^2/6r^2
-(A^{\prime})^2}}\notag\\ &=&-\tilde{T}_{D7}(r^2-L^4\overline{\omega}^2)
\sqrt{\frac{r^8-r_0^8}{r^6-L^4\overline{\omega}^2 r^4-4a_B^2}}, \\  \label{TA3}
\sqrt{-g}J_{t;D7}^{r}&\equiv&\frac{\tilde{T}_{D7}r^5\dot{\phi}\phi^{\prime}}
{\sqrt{1+r^2(\phi^{\prime})^2/6-L^4\dot{\phi}^2/6r^2-(A^{\prime})^2}}=
\tilde{T}_{D7}r_0^4\omega.
\end{eqnarray}
Using (\ref{TA1})--(\ref{TA3}), we can derive the total energy and energy
flux of the D-brane system. The total energy of the D7-brane in the above
configuration is given by:

\begin{figure}[t]
\begin{center}
\epsfig{file=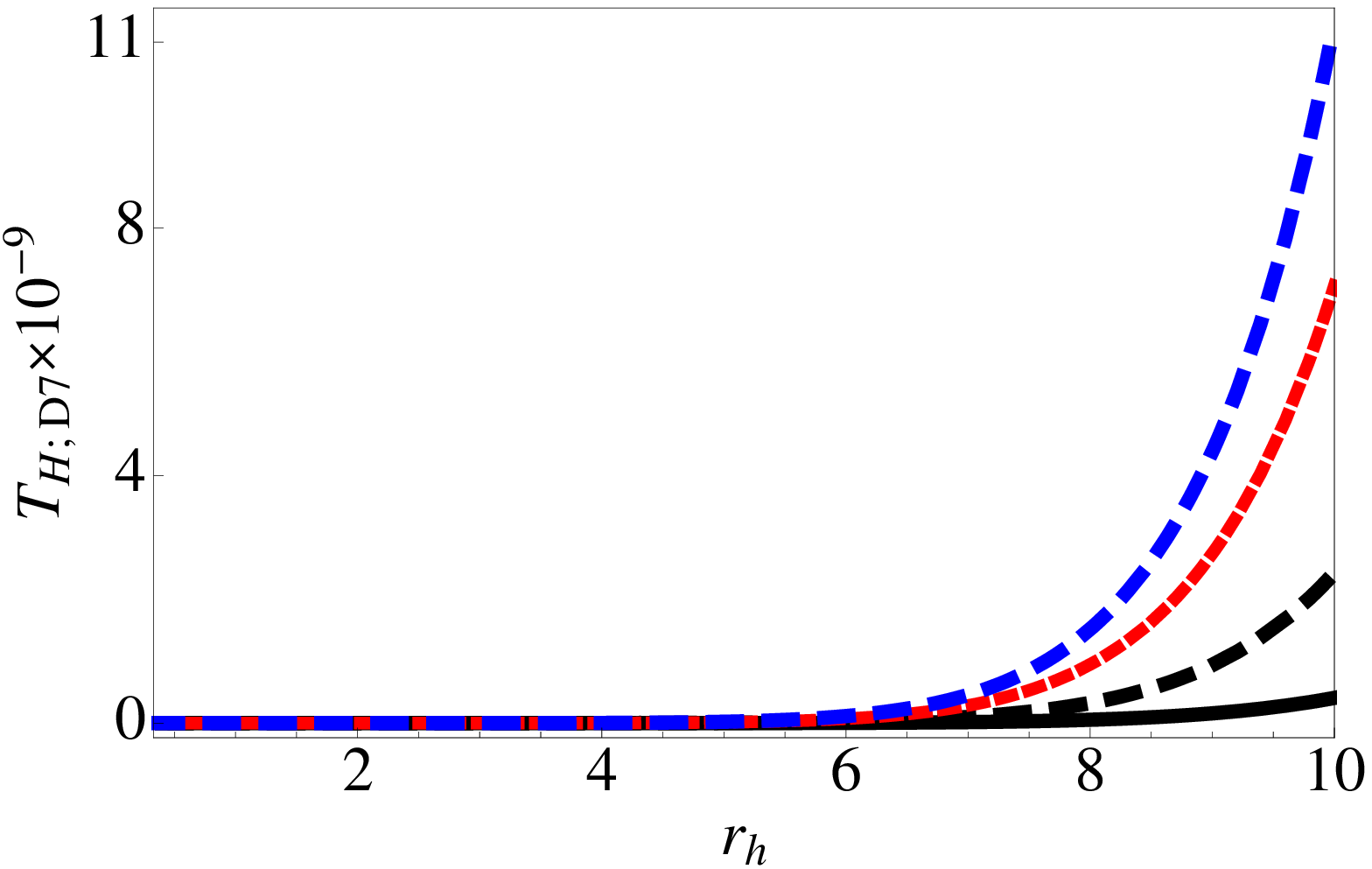,width=.45\textwidth}~~\nobreak
\epsfig{file=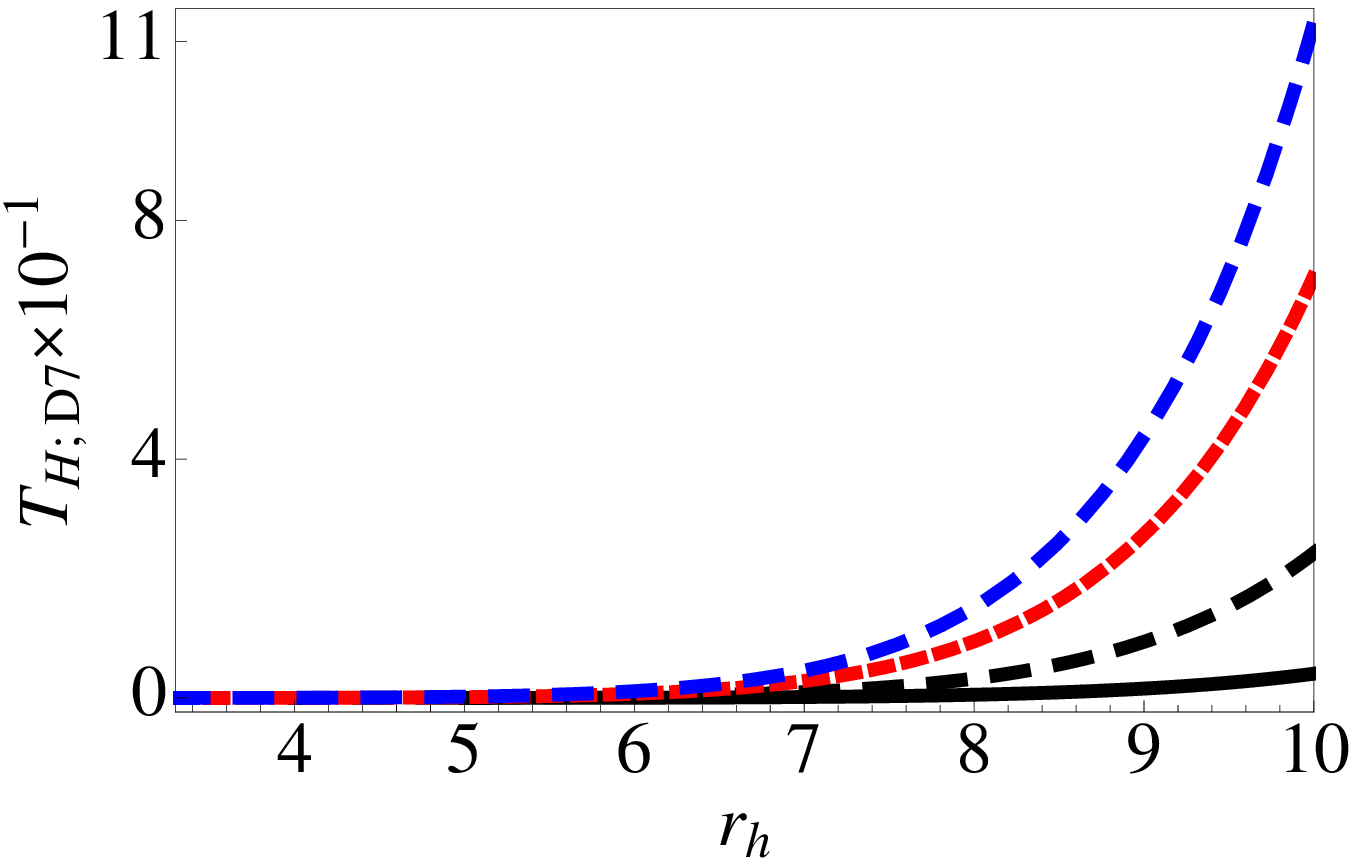,width=.45\textwidth}
\epsfig{file=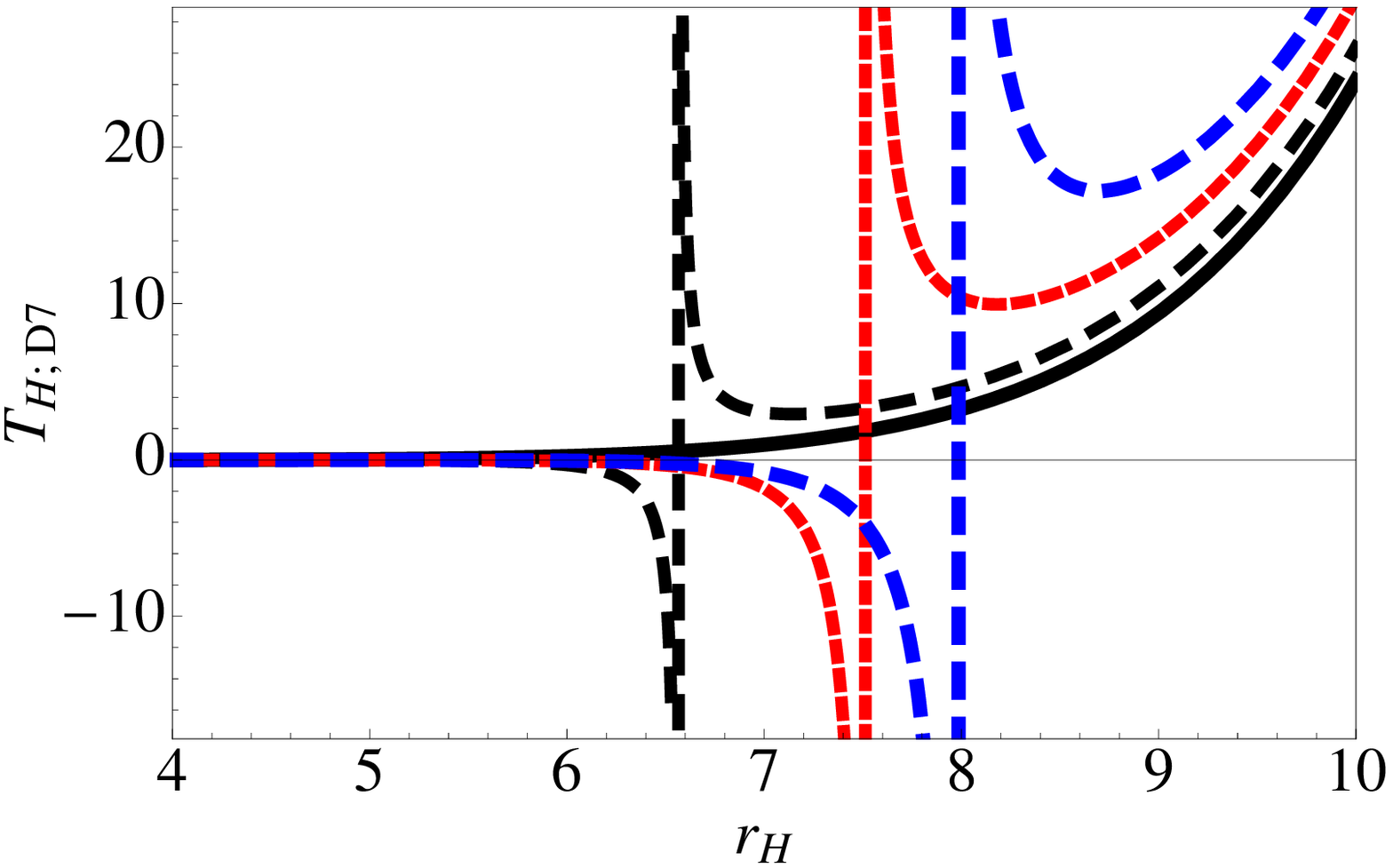,width=.45\textwidth}~~\nobreak
\epsfig{file=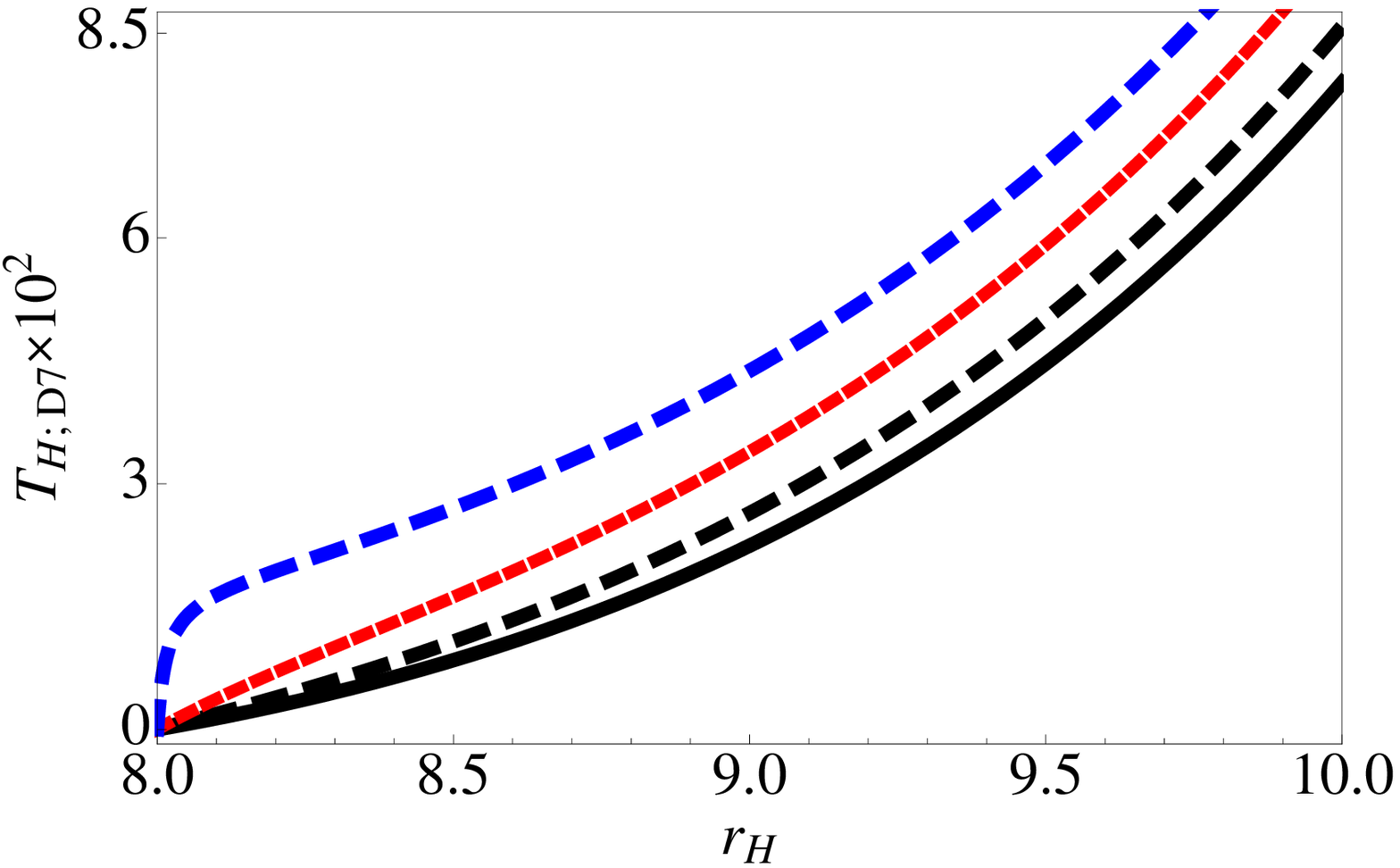,width=.45\textwidth}
\epsfig{file=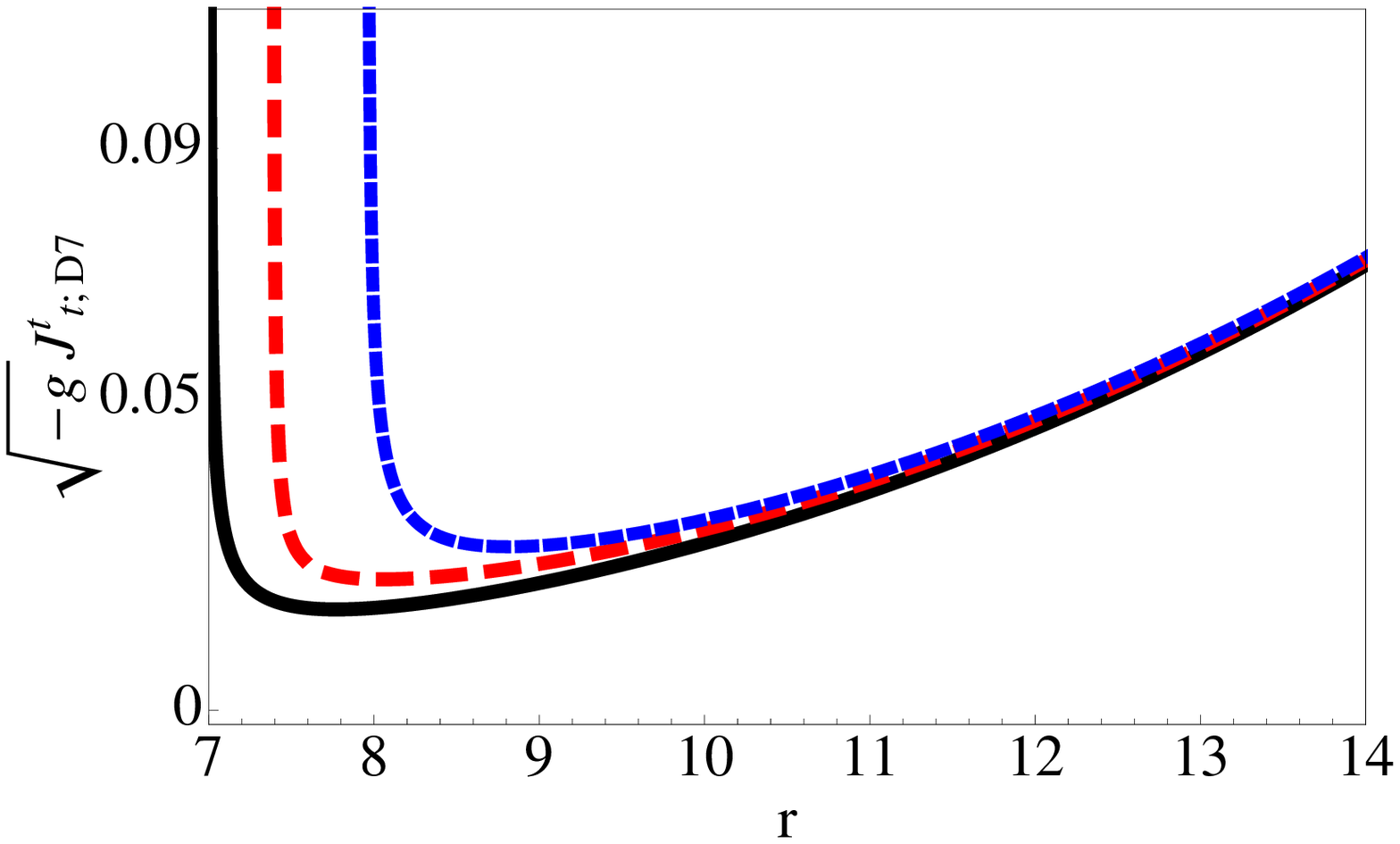,width=.45\textwidth}~~\nobreak
\epsfig{file=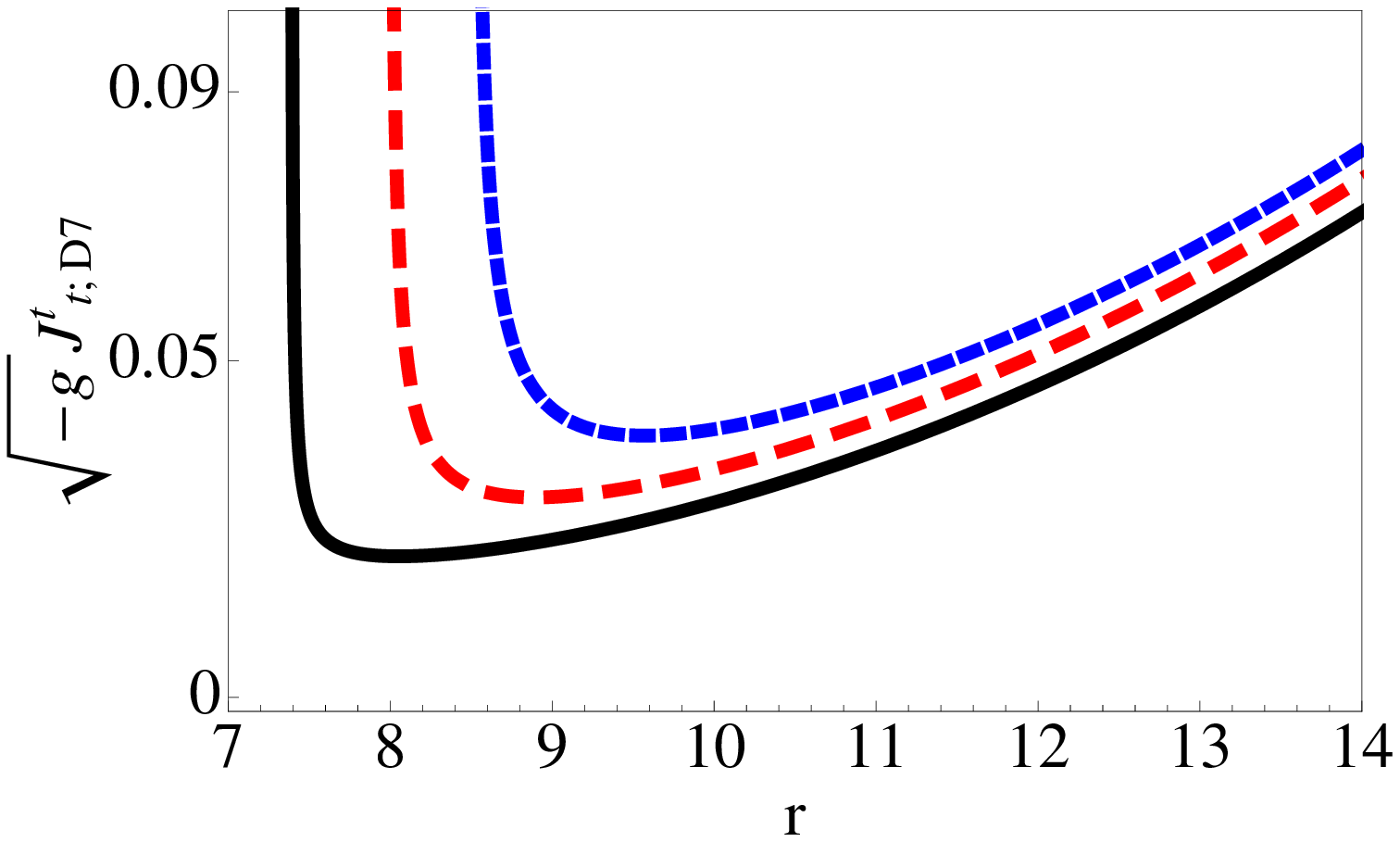,width=.45\textwidth}
\caption{[Up-Left] The behavior of the world volume Hawking
temperature with $r_0=5$ (solid), $r_0=4$ (black-dashed), 
$r_0=3.5$ (gray-dashed), $r_0=3.3$ (blue-dashed), and $L=10$.
[Up-Right] The behavior of the temperature with $r_0=0.5$ (black-solid),
$r_0=0.4$ (black-dashed), $r_0=0.35$ (red-dashed), $r_0=0.33$
(blue-dashed), and $L=10$. 
[Middle-Left] The behavior of the temperature in the presence of the
world volume electric field with $a_B=0$ (black-solid), $a_B=100$ 
(black-dashed), $a_B=150$ (red-dashed), $a_B=180$ (blue-dashed),
$r_0=4$, and $L=10$.
[Middle-Right] The behavior of the temperature in the presence of the
world volume electric field with $a_B=0$ (black-solid), $a_B=100$ 
(black-dashed), $a_B=150$ (red-dashed), $a_B=180$ (blue-dashed),
$r_0=8$, and $L=10$.
[Down-Left] The behavior of the energy-density, $\sqrt{-g}J_t^t$, with
$a_B=0$ (black-solid), $a_B=200$ (red-dashed) $a_B=250$ (blue-dashed),
and $r_0=7$ and $\omega=1$. 
[Down-Right] The behavior of the energy-density with $\omega=1$ (black-solid),
$\omega=5$ (red-dashed) $\omega=6.5$ (blue-dashed),
and $r_0=7$ and $a_B=200$. 
In all cases, $L=\alpha^{\prime}=1$ and $g_s=0.1$.}
\label{fig:T0TAD7}
\end{center}
\end{figure}

\begin{equation}
\label{Eden2}
E=\tilde{T}_{D7}\int_{r_0}^{\infty}{\frac{dr\,(r^4(r^{10}-L^4\overline{\omega}^2
r_0^8)-4r_0^8a_B^2)} {r^4\sqrt{(r^8-r_0^8)(r^6-L^4\overline{\omega}^2 r^4-4a_B^2)}}}.
\end{equation}
It is straightforward to see from (\ref{Eden2}) that when $r\rightarrow r_0$, 
the energy density of the flavor brane, given by (\ref{TA1}), becomes very 
large and blows up precisely at the minimal extension, $r=r_0$, at the IR scale
of  conformal and chiral flavor symmetry breakdown, independent from the
presence of the world volume electric field. Thus we  conclude that at the IR
scale, $r=r_0$, the backreaction of the D7-brane to the gravity dual metric is
non-negligibly large and forms a black hole centered at the IR scale $r=r_0$ in
the bulk. The black hole size should grow as the energy gets pumped into it 
from the D7-brane steadily. In order to obtain this energy flux, we use the 
components of the energy--stress tensor (\ref{TA1})--(\ref{TA3}).  We note
 that when the minimal radial extension is positive definite ($r_0>0$) and spin
is turned on ($\omega>0$), the component (\ref{TA3}) is non-vanishing and 
hence we compute the time evolution of the total energy as:
\begin{equation}
\label{dEAD7}
\dot{E}_{D7}=\frac{d}{dt}\int{dr\sqrt{-g}J_{t;D7}^{t}}=\int{dr\partial_r(\sqrt{-g}J_{t;D7}^{r})}
=\sqrt{-g}J_{t;D7}^{r}|_{r=r_0}^{\infty}=\tilde{T}_{D7}r_0^4\omega^2-\tilde{T}_{D7}r_0^4
\omega^2=0.
\end{equation}
Here we note that when the electric field is turned on, $J_{t;D7}^{r}$, given by
(\ref{TA3}), and the energy dissipation $\dot{E}_{D7}$, given by (\ref{dEAD7}),
remain unchanged. Therefore, independent from the electric field, the energy 
dissipation from the brane into the bulk will form a black hole, centered at the IR
scale $r=r_0$. Thus, by this flow of energy form the probe to the bulk, we conclude
by duality that the energy from the flavor sector will eventually dissipate into the
gauge theory conformal and chiral flavor symmetry breakdown, independent from
the electric field. To see this external injection of energy in our stationary rotating
solution, we may introduce UV and IR cut offs to the D7-brane solution, such that
it extends from $r=r_{\text{IR}}=r_0$ to $r=r_{\text{UV}}\gg r_{\text{IR}}$. 
It is clear from  (\ref{TA3}) that at both $r_{\text{IR}}$ and $r=r_{\text{UV}}$
we have $J_t^r>0$. This shows the presence of equal energy flux incoming from
$r=r_{\text{UV}}$ and outgoing at $r=r_{\text{IR}}$ at which the energy gets
not reflected back but its backreaction will nucleate a black hole intaking the
injected energy, independent from the electric field.

The energy--density (\ref{TA1}) of the world-volume black hole solution
(\ref{indAD7}) has a number of interesting features. These depend on the choice
of parameters $\{r_0, a_B, \omega\}$, and therefore we note the following points.
$\\$
$\bullet$ Increasing $\omega$, while keeping the other parameters fixed,
increases the scale of the minimum energy-density near the IR scale, and
also increases the amount of energy density away from the IR scale
(see Fig.\,\ref{fig:T0TAD7}).  However, the increase in the energy-density
is so that the backreaction remains negligibly small away from the blowing
up point, located at the IR scale $r_0$ (see Fig.\,\ref{fig:T0TAD7}).
$\\$
$\bullet$ Increasing $a_B$ by relatively large values, while keeping the
other parameters fixed, only shifts the minimum of the energy-density 
and leaves the scale and behavior of the energy-density unaltered
(see Fig.\,\ref{fig:T0TAD7}). 
$\\$
$\bullet$ Increasing/decreasing $r_0$, while keeping the other
parameters fixed, leaves the scale and behavior of the energy-density
unaltered (as in Fig.\,\ref{fig:T0TAD7}).

In all cases, the energy-density falls sharply from its infinite value
at the IR scale to its minimum value, whereupon it increases to finitely
small values away from the IR scale. We thus conclude that altering the
parameters of the solution results almost no alteration in the scale and
behavior of the energy-density, meaning that the density blows up at
the IR scale, whereat the  backreaction is non-negligible, and stays
finitely small away from the IR scale, where the backreaction can be
neglected.

\section{Induced metric and temperature on U-like embedded
probe D5-brane wrapping $adS_3\times S^3\subset adS_5\times T^{1,1}$
and spinning about $S^2$ $\subset T^{1,1}$ with world volume
electric field turned on}

As second example in our study, we consider the U-like probe D5-brane model
wrapping $adS_3\times S^3$ in $adS_5\times T^{1,1}$ reviewed in Sec.\,2,
and turn on, in addition, spin as well as world-volume gauge electric field on 
the probe. We include additional  spin degrees of freedom, dual to finite 
R--charge chemical potential, by allowing in our system the D5-brane rotate
in the $\phi$ direction of the transverse $S^2$ with conserved angular 
momentum. Thus, in our setup, we allow $\phi$ to have time-dependence
as well, such that $\dot{\phi}(r,t)=\omega=const.$, with $\omega$ denoting
the angular velocity of the probe. This way, we construct rotating solutions. 
Hence, the world-volume field is given by $\phi(r,t)$, with other directions fixed.
We also include the additional contribution from world-volume fields
strengths, $F_{ab}$, in (\ref{DPACTION}), corresponding to world-volume
gauge fields, dual to finite baryon density chemical potential, by noting
the following:  In the presence of $N_f$ flavors, the gauge theory posses
a global  $U(N_f)\simeq SU(N_c)\times U(1)_q$ symmetry. The $U(1)_q$
counts the net number of quarks, that is, the number of baryons times $N_c$. 
In the gravity dual, this global symmetry corresponds to the $U(N_f)$ gauge
symmetry on the world volume of the $N_f$ D5-brane probes. The conserved
currents associated with the $U(N_f)$ symmetry of the gauge theory are dual
to the gauge fields, $A_{\mu}$, on the D5-branes. Hence the introduction of a
chemical potential $\mu$ or a non-vanishing $n_B$ for the baryon number the
gauge theory corresponds to turning on the diagonal $U(1)\subset U(N_f)$ 
gauge field, $A_{\mu}$ on the world volume of the D5-branes. We may describe
external fields in the field theory, coupled to anything having $U(1)$ charge, by
introducing non-normalizable modes for $A_{\mu}$ in the gravity dual
 (e.g. see ref.\,\cite{Kruczenski:2003be,O'Bannon:2008bz}).

In this section, we will study U-like embedded probe D5-branes spinning with
angular frequency $\omega$, and with a $U(1)$ world volume gauge field
$A_{\mu}$. To have the gauge theory at finite chemical potential or baryon
number density, we note that it is sufficient to turn on the time component
of the gauge field, $A_t$. By symmetry considerations, we then take
$A_t=A_t(r)$, corresponding to the world volume electric field.

Therefore, we will consider the ansatz for the D5-brane world volume
field of the form $\phi(r,t)=\omega t+f(r)$, with other directions
fixed, but now  $F_{ab}=F_{rt}=\partial_rA_t(r)$. Using this ansatz
and the metric (\ref{10DKWmet}) in terms of (\ref{6DConmet2}),
it is easy to derive the components of the induced world volume metric
on the D5-brane, $g_{ab}^{D5}$, and compute the determinant, 
$\det g_{ab}^{D5}$, resulting the DBI action (\ref{DPACTION}), as:

\begin{eqnarray}
\label{dbiacd5A}
S_{D5}\simeq-T_{D5}\int{drdt\,r\sqrt{1+\frac{r^2(\phi^{\prime})^2}{6}
-\frac{L^4\dot{\phi}^2}{6r^2}-(A_t^{\prime})^2}}.
\end{eqnarray}
Here we note that by setting $\dot{\phi}=\omega=A_t(r)=0$, our action
(\ref{dbiacd5A}) reduces to that of the probe D5-brane action in the BKS
 model, (\ref{KSA}). As in the BKS model reviewed in Sec.\,2, we restrict
brane motion to the $\phi$-direction of the transverse $S^2$ sphere and
fix other directions constant. Thus, in our set-up we let, in addition, the 
probe rotate about the $S^2$, as before, and furthermore turn on a 
non-constant world-volume electric on the probe. 

The D5-brane equation of motion from the action (\ref{dbiacd5A}) is:

\begin{eqnarray}
\label{D5eqA}
\frac{\partial}{\partial r}\Bigg[\frac{r^3\phi^{\prime}}{\sqrt{1+\frac{r^2(\phi^{\prime})^2}{6}
-\frac{L^4\dot{\phi}^2}{6r^2}-(A_t^{\prime})^2}}\Bigg]&=&
\frac{\partial}{\partial t}\Bigg[\frac{L^4\dot{\phi}}{r\sqrt{1+\frac{r^2(\phi^{\prime})^2}{6}
-\frac{L^4\dot{\phi}^2}{6r^2}-(A_t^{\prime})^2}}\Bigg],\\ \label{D5eqA2} 
\frac{\partial}{\partial r}\Bigg[\frac{rA_t^{\prime}}{\sqrt{1+\frac{r^2(\phi^{\prime})^2}{6}
-\frac{L^4\dot{\phi}^2}{6r^2}-(A_t^{\prime})^2}}\Bigg]&=&0.
\end{eqnarray}
By taking the large radii limit, $r\rightarrow\infty$, it is straightforward
to see that Eq.\,(\ref{D5eqA2}) implies $A_t^{\prime}\simeq a_B/r$, with $a_B$
denoting the VEV of baryon density number.

Take rotating solutions to (\ref{D5eqA}) of the form:

\begin{eqnarray}
\label{rotsold5A}
\phi(r,t)&=&\omega t+f(r),\;\;\;\;\ f(r)=\sqrt{6}r_0^2\int_{r_0}^{r}{\frac{dr}{r}
\sqrt{\frac{1-L^4\,\overline{\omega}^2/r^2-(A_t^{\prime})^2}{r^4-r_0^4}}}.
\end{eqnarray}
Here we set $\overline{\omega}=\omega/\sqrt{6}$ and note that when
$\omega=A_t(r)=0$, our solution (\ref{rotsold5A}) integrates to that
of probe D5-brane in the BKS model, \cite{Ben-Ami:2013lca}, reviewed
 in Sec.\,2, with the probe wrapping $adS_3\times S^3$ in $adS_5\times T^{1,1}$
(see Eq.\,(\ref{BKSSol.D51})). The solution (\ref{rotsold5A}) is
parameterized by $(r_0,\omega, a_B)$ and describes probe D5-brane
motion, with non-constant world volume gauge field $A_t(r)$ and
angular velocity $\omega$ about the transverse $S^2\subset T^{1,1}$,
starting and ending up at the boundary. The probe descends from the
UV boundary at infinity to the  minimal extension $r_0$ in the IR
where it bends back up the boundary. We also note that inspection
of (\ref{rotsold5A}) illustrates, in the limit $r\rightarrow\text{large}$,
that  the behavior of $df(r)/dr=f_r(r)$ does not depend on $\omega$
and $a_B$ (see Fig.\,\ref{fig:fpgrrD5}). This illustrates that in the
large radii limit the solution $f(r)$ in (\ref{rotsold5A}) gives the
world volume field $\phi(r)$ of the BKS model (see Sec.\,2) with the
boundary values $\phi_{\pm}$ in the asymptotic UV limit,  
$r\rightarrow\infty$ (see also Fig.\,\ref{fig:fpgrrD5}). But, we note
that in the (other) IR limit, i.e., when $r\rightarrow\text{small}$, the
behavior of $f_r(r)$ does depend on $\omega$. Furthermore, we note
here that by turning on, in addition, the world volume electric field merely
changes the scale but leaves the behavior of $f_r(r)$ in the IR unchanged. 
Inspection of (\ref{rotsold5A}) illustrates that in the IR the behavior of
$f_r(r)$ with $\omega>0$ is comparable to that of with $\omega=0$, only
if certain $\omega>0$ are chosen.  (see Fig.\,\ref{fig:fpgrrD5}). This 
illustrates that in the small radii limit in the IR the behavior of $f_r(r)$
(here) does compare to that of $\phi^{\prime}(r)=d\phi/dr$ in the 
BKS (see Sec.\,2),--with 
$\phi^{\prime}(r)\rightarrow\infty$ in the IR limit $r\rightarrow r_0$--,
consistent with U-like embedding, only if certain $\omega>0$ are chosen.

To derive the induced metric on the D5-brane, we put the rotating solution
(\ref{rotsold5A}) into the background metric (\ref{10DKWmet}) in terms
of (\ref{6DConmet2}) and obtain:

\begin{eqnarray}
\label{indD51A}
ds_{D5}^2&=&-\frac{(3r^2-L^4\omega^2)}{3L^2}dt^2+\frac{L^2}{r^2}
\left[\frac{3r^2(r^4-r_0^4)+r_0^4(6r^2(1-(A_t^{\prime})^2)-L^4\omega^2)}
{3r^2(r^4-r_0^4)}\right]dr^2\notag\\ &&+\frac{2L^2 \omega\, r_0^2}
{3r^2}\sqrt{\frac{6r^2(1-(A_t^{\prime})^2)-L^4\,\omega^2}{r^4-r_0^4}}drd
t+\frac{r^2}{L^2}dx^2\notag\\&&+\frac{L^2}{3}\left[\frac{1}{2}(\Omega_1^2
+\Omega_2^2)+\frac{1}{3}\Omega_3^2+ \omega\Omega_1dt- \frac{r_0^2}{r^2}
\sqrt{\frac{6r^2(1-(A_t^{\prime})^2)-L^4\,\omega^2}{r^4-r_0^4}}\Omega_1 
dr\right].\notag\\
\end{eqnarray}
Here we note that by setting $\omega=A_t=0$, our induced world volume
metric (\ref{indD51A}) reduces to that of the BKS model, \cite{Ben-Ami:2013lca}, 
reviewed in Sec.\,2. In this case, for $r_0=0$ the induced world volume metric is 
that of $adS_3\times S^3$ and the dual gauge theory describes the conformal 
and chiral symmetric phase. On contrary, for $r_0>0$ the induced world volume
metric has no $adS$ factor and the conformal invariance of the dual gauge theory
must be broken in such case. In order to find the world volume horizon and Hawking
temperature, we first eliminate the relevant cross term. To eliminate the relevant
cross-term in (\ref{indD51A}), we consider a coordinate transformation:

\begin{equation}
\tau=t-\omega\,L^4 r_0^2 \int{\frac{dr\,(6r^2(1-(A^{\prime}_t(r))^2)-L^4
\,\omega^2)^{1/2}}{r^2(3r^2-L^4 \omega^2)(r^4-r_0^4)^{1/2}}}.
\end{equation}
The induced metric on the rotating D5-brane, (\ref{indD51A}), then takes the form:

\begin{eqnarray}
\label{indD52}
ds_{D5}^2 &=&-\frac{(3r^2-L^4 \omega^2)}{3L^2}d\tau^2\notag\\ &&+\frac{L^2}{r^2}
\left[\frac{(3r^2-L^4 \omega^2)(r^4-r_0^4)+r_0^8(6r^2(1-(A_t^{\prime})^2)-L^4
\omega^2)}{(3r^2-L^4 \omega^2)(r^4-r_0^4)}\right]dr^2\notag\\ && +\frac{r^2}{L^2}dx^2
+ -\frac{L^2}{3}\left[\frac{1}{2}(\Omega_1^2+\Omega_2^2)+\frac{1}{3}\Omega_3^2\right]
\notag\\ && -\frac{L^2}{3}\left[\omega\Omega_1d\tau+\frac{3\,r_0^2}{3r^2-L^4 \omega^2}
\sqrt{\frac{6r^2(1-(A_t^{\prime})^2)-L^4\,\omega^2}{r^4-r_0^4}}\Omega_1 dr\right].
\end{eqnarray}
Here we note that for $r_0=0$ the induced world volume metric (\ref{indD52}) has
no horizon solving $-g_{\tau\tau}=g^{rr}=0$, and thus not given by the black hole
geometry. On contrary, for $r_0>0$ the induced world volume horizon solves the 
equation:

\begin{figure}[t]
\begin{center}
\epsfig{file=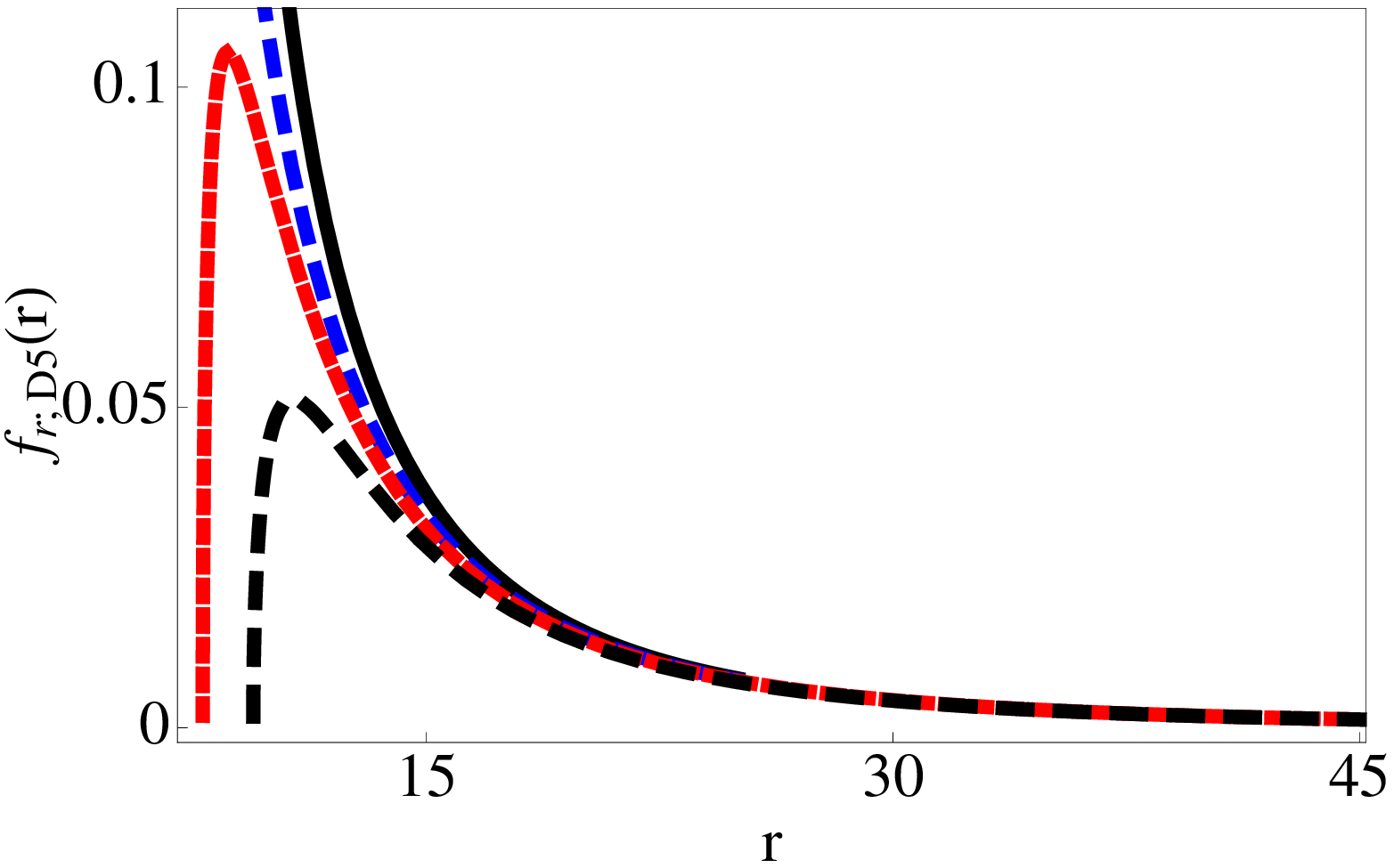,width=.50\textwidth}~~\nobreak
\epsfig{file=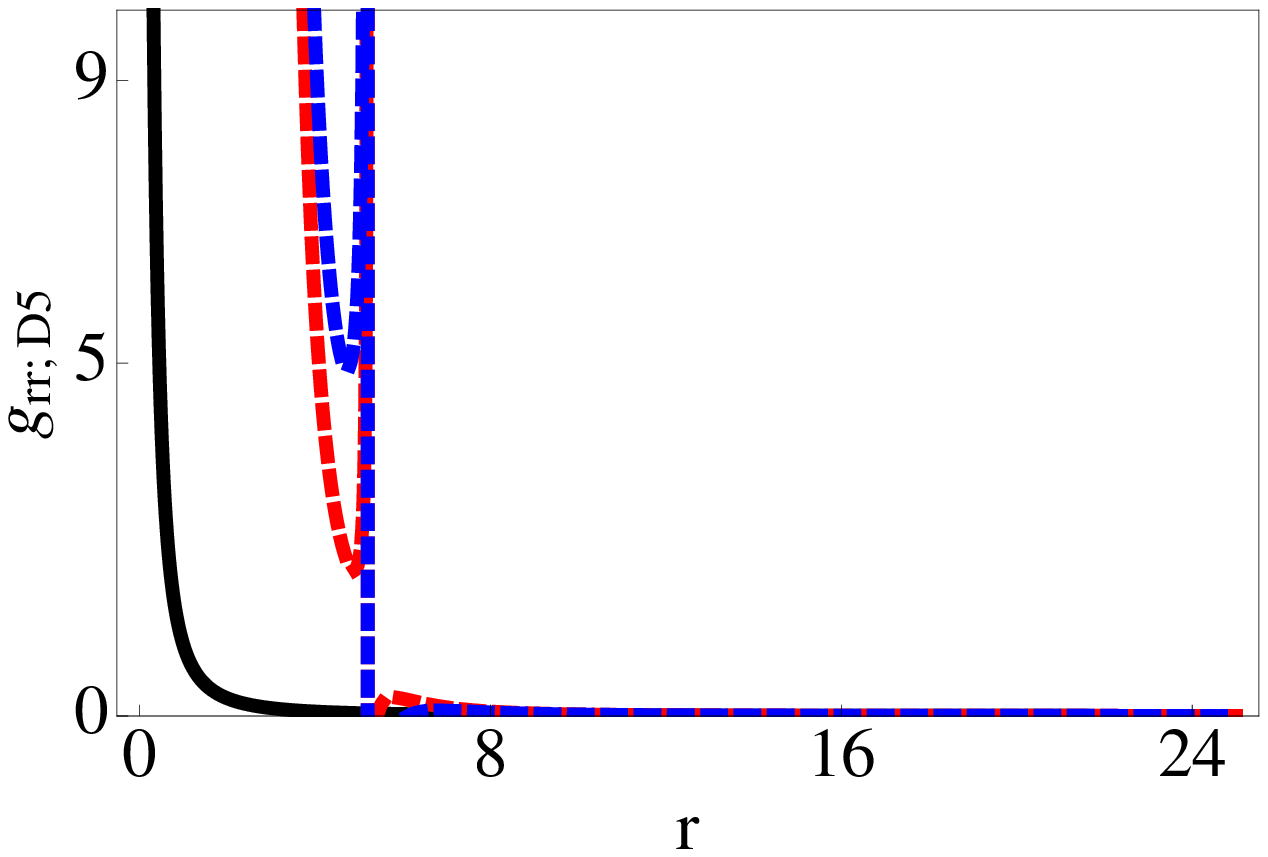,width=.50\textwidth}
\caption{[Left] The behavior of the derivative of the world
volume field with respect to $r$ with $L=1, r_0=7$, $\omega=0$,
$a_B=0$ (black-solid), $\omega=5$, $a_B=4$ (blue-dashed),
$a_B=6$ (red-dashed), and $a_B=8$ (black-dashed). [Right]
The behavior of the $g_{rr}$ component of the induced world volume
metric with $L=1,r_0=\omega=a_B=0$ (black-solid), $r_0=3,
\omega=9, a_B=4$ (red-dashed), and $a_B=5$ (blue-dashed).}
\label{fig:fpgrrD5}
\end{center}
\end{figure}

\begin{equation}
\label{HorEqD5A}
H(r)=r^2(r^4-r_0^4)(3r^2-L^4\omega^2)=0.
\end{equation}
Eq.\,(\ref{HorEqD5A}) contains two real positive definite zeros 
(see also Fig.\,\ref{fig:fpgrrD5}). The thermal horizon of the induced world
volume black hole geometry is given by the zero $3r_H^2-L^4\omega^2=0$,
with the horizon varying continuously with varying the angular velocity
$\omega$, as expected. Eq.\,(\ref{HorEqD5A}) also shows that the horizon 
must grow from the minimal extension $r_0\neq0$ with increasing the angular
velocity  $\omega$, as before. We thus conclude at this point that when $r_0$
is positive definite ($r_0>0$) and spin is turned on ($\omega>0$), the induced
world volume metric on the rotating probe D5-brane wrapping $adS_3\times S^3$
in $adS_5\times T^{1,1}$ admits a thermal horizon growing continuously with
increasing the angular velocity, regardless of the presence of the world volume
gauge electric field.

The Hawking temperature on the probe D5-brane can be found from the induced
world-volume metric (\ref{indD52}) in the form:

\begin{equation}
\label{THD5A}
T_{H;D5}=\frac{(g^{rr})^{\prime}}{4\pi}\bigg|_{r=r_H}=\frac{3r_H^3(r_H^4-r_0^4)}
{2\pi L^2 r_0^4[6r_H^2(1-(A^{\prime}_t)^2)-L^4\omega^2]}=\frac{r_H^3
(r_H^4-r_0^4)}{2\pi L^2 r_0^4(r_H^2-2a_B^2)}.
\end{equation}
Here, we note that from (\ref{THD5A}) that when the world volume horizon
is at the minimal radial extension, $r_H=r_0$, the Hawking temperature of
the world volume black hole geometry is identically zero, $T_{H; D5}=0$.
The Hawking temperature (\ref{THD5A}) of the world-volume black hole solution
(\ref{indD52}) also has a number of interesting features, which depend on the 
choice of parameters $\{a_B,r_0, \omega\}$. Therefore we note the following points:
$\\$
$\bullet$ For $a_B=0$, $r_0=4$ and $\omega\geq 0$, the world-volume Hawking
temperature $T_{H;D5}$ (\ref{indD52}) increases monotonically with increasing
the angular velocity $\omega$, by which the world-volume horizon size $r_H$
grows (see Fig.\,\ref{fig:TJD5}). In this case, changing $r_0$ changes the scale, 
but not the behavior of the temperature (\ref{THD5A}). In particular, inspection of
(\ref{THD5A}) shows that decreasing/increasing $r_0$ increases/decreases the
temperature and produces relatively large hierarchies between temperature scales, 
$T_{H;D5}(r_0<1)/T_{H;D5}(r_0>1)\simeq 10^4$, while the temperature behavior
remains unchanged (see Fig.\,\ref{fig:TJD5}). 
$\\$
$\bullet$ For $0<a_B\leq 15, r_0=4$ and $\omega\geq 0$, the world-volume
Hawking temperature $T_{H;D5}$ (\ref{THD5A}) no longer behaves monotonically
with growing the horizon size $r_H$ (see Fig.\,\ref{fig:TJD5}). Instead, the
temperature of the solution (\ref{THD5A}) admits three distinct branches, including
two obvious classes of black hole solutions. The first branch appears once the horizon
starts to grow from the minimal extension. In this case, the temperature $T_{H;D5}$
(\ref{THD5A}) decreases almost immediately to negative values before peeking off
very sharply, producing a divergent-like behavior, hitting zero, and then growing into
positive values (see Fig.\,\ref{fig:TJD5}). The second branch appears once the horizon
continues to grow, away from the zero point. In this case, the temperature $T_{H;D5}$
(\ref{THD5A}) decreases to positive values, in an `inverse-like' manner with the horizon,
before reaching its non-zero minimum  (see Fig.\,\ref{fig:TJD5}). Therefore in the 
neighborhood of the zero point the world-volume temperature of the solution decreases
with increasing horizon size. These `small' black holes have the familiar behavior
of five-dimensional black holes in asymptotically flat spacetime, since their temperature
decreases with increasing horizon size $r_H$. The third branch appears once the horizon
continues to grow from its radius setting the non-zero minimum of $T_{H;D5}$. In this 
case, the other class of black hole solution appears with its temperature only increasing
with increasing horizon size $r_H$, like `large' black holes (see Fig.\,\ref{fig:TJD5}). 
This behavior is like that of the temperature with $a_B=0$, however, in the latter two 
branches, the temperature of the black hole solution scales higher than that of with 
$a_B=0$ (see Fig.\,\ref{fig:TJD5}). It is thus remarkable that by varying $a_B$, the
scale and behavior of the temperature changes  continuously with growing horizon size
$r_H$ (see Fig.\,\ref{fig:TJD5}). It is clear that by increasing/deceasing $a_B$ the
temperature behavior changes dramatically, while the temperature scales change very
moderately, producing only small hierarchies, $T_{H;D5}(a_B<)/T_{H;D5}(a_B>)
\simeq 0.5$ (see Fig.\,\ref{fig:TJD5}).
$\\$
$\bullet$ By varying $r_0$, while keeping the other parameters within
the above range, only the scale but not the behavior of the temperature
of the black hole solution, $T_{H;D5}$ (\ref{THD5A}), changes.
For both $r_0<1$ and $r_0>1$, i.e., $r_0\simeq 8$, only the temperature
scale changes, and the temperature behavior remains unchanged. In this 
latter case, the temperature scales less than in previous cases ($r_0>1$), 
increases with increasing $a_B$ as before, and behaves as before, increasing
and decreasing continuously with growing $r_H$ (see Fig.\,\ref{fig:TJD5}).
In this latter case, therefore there are still two branches, corresponding to
two classes of black hole solutions, including both `large' and `small' black 
holes, with temperature scales setting relatively  small hierarchies, 
$T_{H;D5}(r_0<)/T_{H;D5}(r_0>)\simeq 10$.

We therefore conclude at this point that when the minimal extension
changes, the temperature scale of the world volume black hole solution
changes dramatically and sets large hierarchies, while the temperature
behavior remains unchanged. We conclude, however, that when the 
world wolume electric field is turned on, the temperature behavior 
changes dramatically and includes two distinct branches corresponding
to`large' and `small' black hole solutions, respectively, while the 
temperature scale changes moderately and thereby sets only moderate
hierarchies. In this case, we also conclude that when, in addition, the 
minimal extension gets inccreased, the temperature behavior remains
unchanged and includes the same two branchs corresponding to `large'
and `small' black hole solutions, with the temperature scales setting relatively
small hierarchies.

We further notice that by taking into account the backreaction of the
above solution to the KW gravity dual, $adS_5\times T^{1,1}$, one
accordinglly awaits the D5-brane to form a mini black hole in KW, 
corresponding to a locally thermal gauge field theory in the probe limit.
Thereby, the rotating D5-brane describes a thermal object in the dual
gauge field theory. In the KW example here, the configuration is dual
to  $\mathcal{N}=1$ gauge theory coupled to a defect flavor subject to
an external electric field. As the gauge theory itself is at zero temperature
as the defect flavor is at finite temperature, given by (\ref{THD5A}), we
conclude that the configuration describes non-equilibrium steady state.
However, as we shall show below, the energy from the flavor sector will in
the end dissipate to the gauge theory.

So far, in this example, we neglected the backreaction of the D5-brane
to the KW gravity dual, as we worked in the probe limit. It is useful
to observe the extend this is justifiable. To this end, we derive the
components of the energy--stress tensor of the D5-brane, which take
the form:

\begin{eqnarray}
\label{T1D5A}
\sqrt{-g}J_{t;D5}^{t}&\equiv&\frac{\tilde{T}_{D5}\,r(1+r^2(\phi^{\prime})^2/6)}
{\sqrt{1+r^2(\phi^{\prime})^2/6-L^4\dot{\phi}^2/6r^2-(A^{\prime})^2}}\notag\\ &&=
\frac{\tilde{T}_{D5}(r^{6}-r_0^4(L^4\overline{\omega}^2+a_B^2))}{r^2\sqrt{(r^4-r_0^4)
(r^2-L^4\overline{\omega}^2-a_B^2)}},\\ \label{T2A}\sqrt{-g}J_{r;D5}^{r}&\equiv&
-\frac{\tilde{T}_{D5}\,r(1-L^4\dot{\phi}^2/6r^2)}{\sqrt{1+r^2(\phi^{\prime})^2/6
-L^4\dot{\phi}^2/6r^2-(A^{\prime})^2}}\notag\\ &&=-\tilde{T}_{D5}(r^2-L^4
\overline{\omega}^2)\sqrt{\frac{r^4-r_0^4}{r^2-L^4\overline{\omega}^2-a_B^2}},
\;\;\ \\ \label{T3D5A} \sqrt{-g}J_{t;D5}^{r}&\equiv&\frac{\tilde{T}
_{D5}r^3\dot{\phi}\phi^{\prime}}{\sqrt{1+r^2(\phi^{\prime})^2/6-L^4\dot{\phi}^2
/6r^2-(A^{\prime})^2}}=\tilde{T}_{D5}r_0^2\overline{\omega}.
\end{eqnarray}
Using (\ref{T1D5A})--(\ref{T3D5A}), we can derive the total energy and energy flux
of the D-brane system. The total energy of the D5-brane in the above configuration
is given by:

\begin{figure}[t]
\begin{center}
\epsfig{file=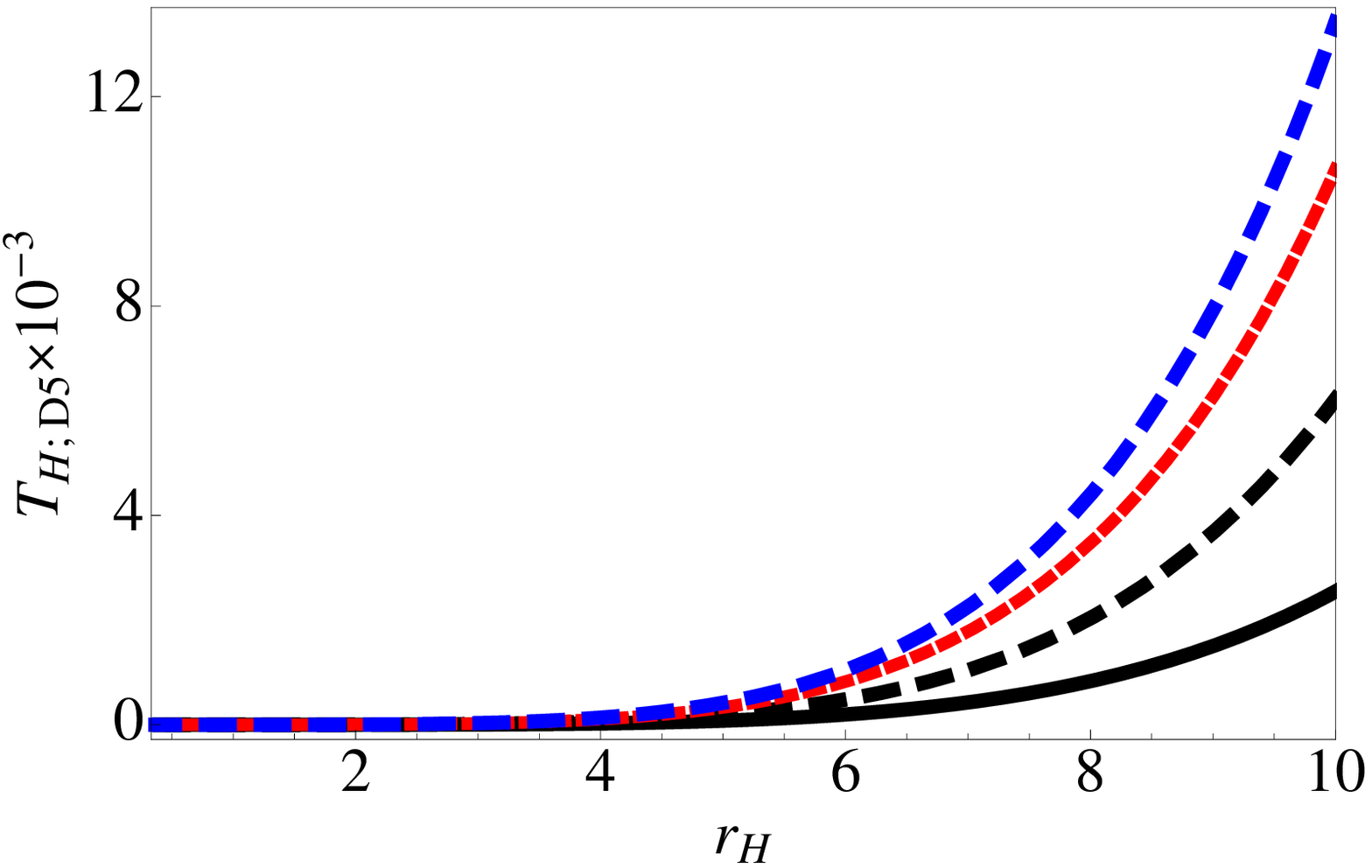,width=.48\textwidth}~~\nobreak
\epsfig{file=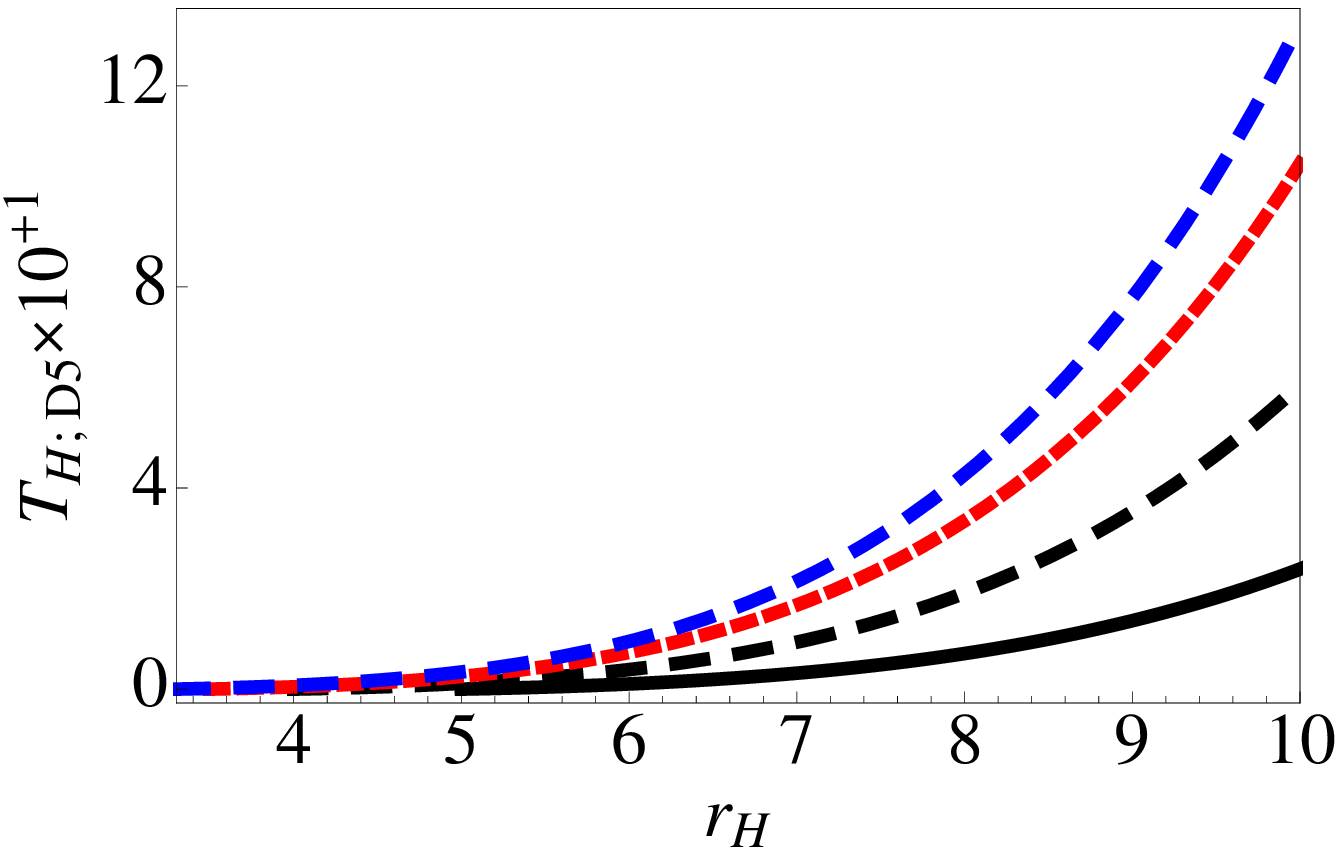,width=.48\textwidth}
\epsfig{file=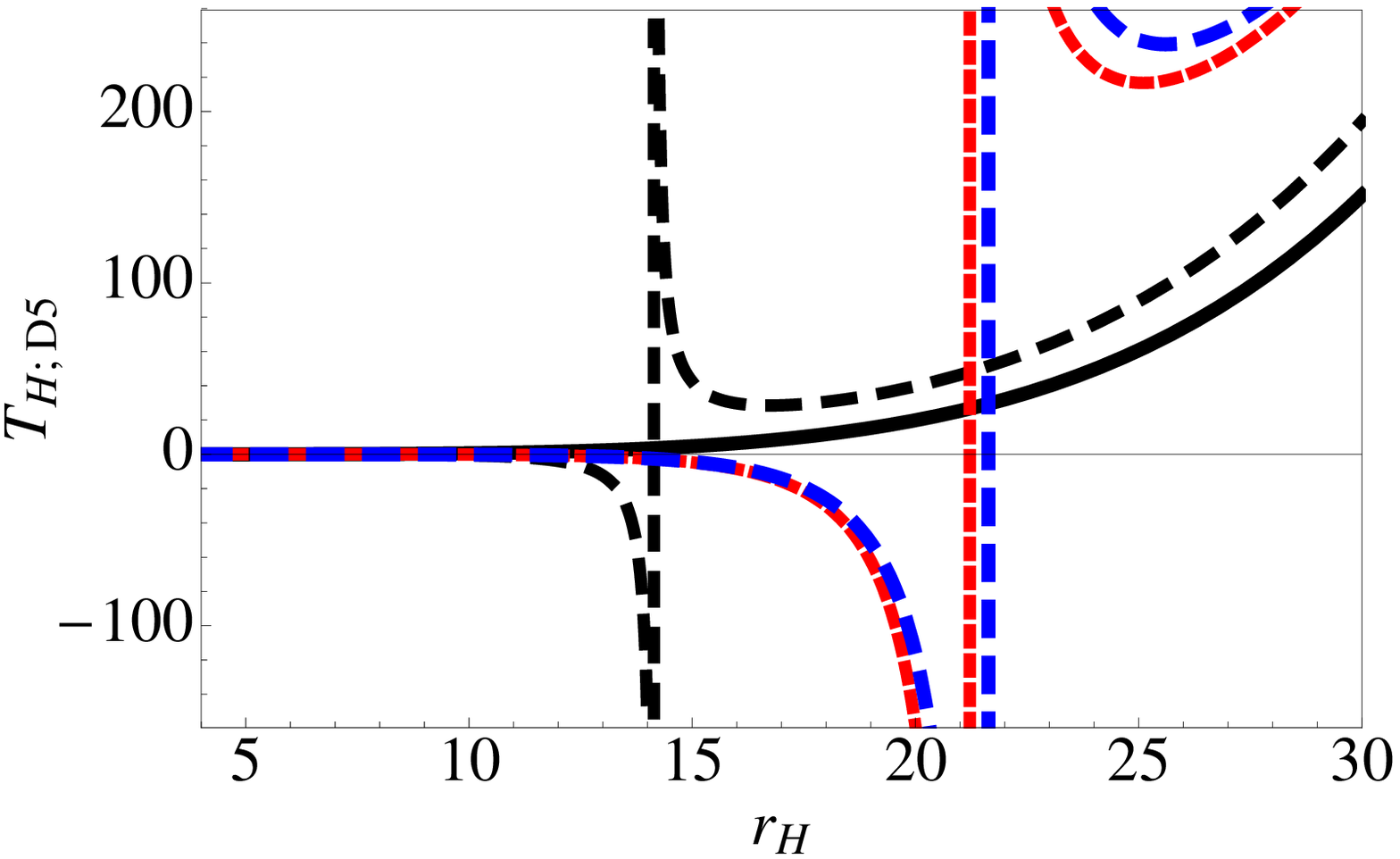,width=.48\textwidth}~~\nobreak
\epsfig{file=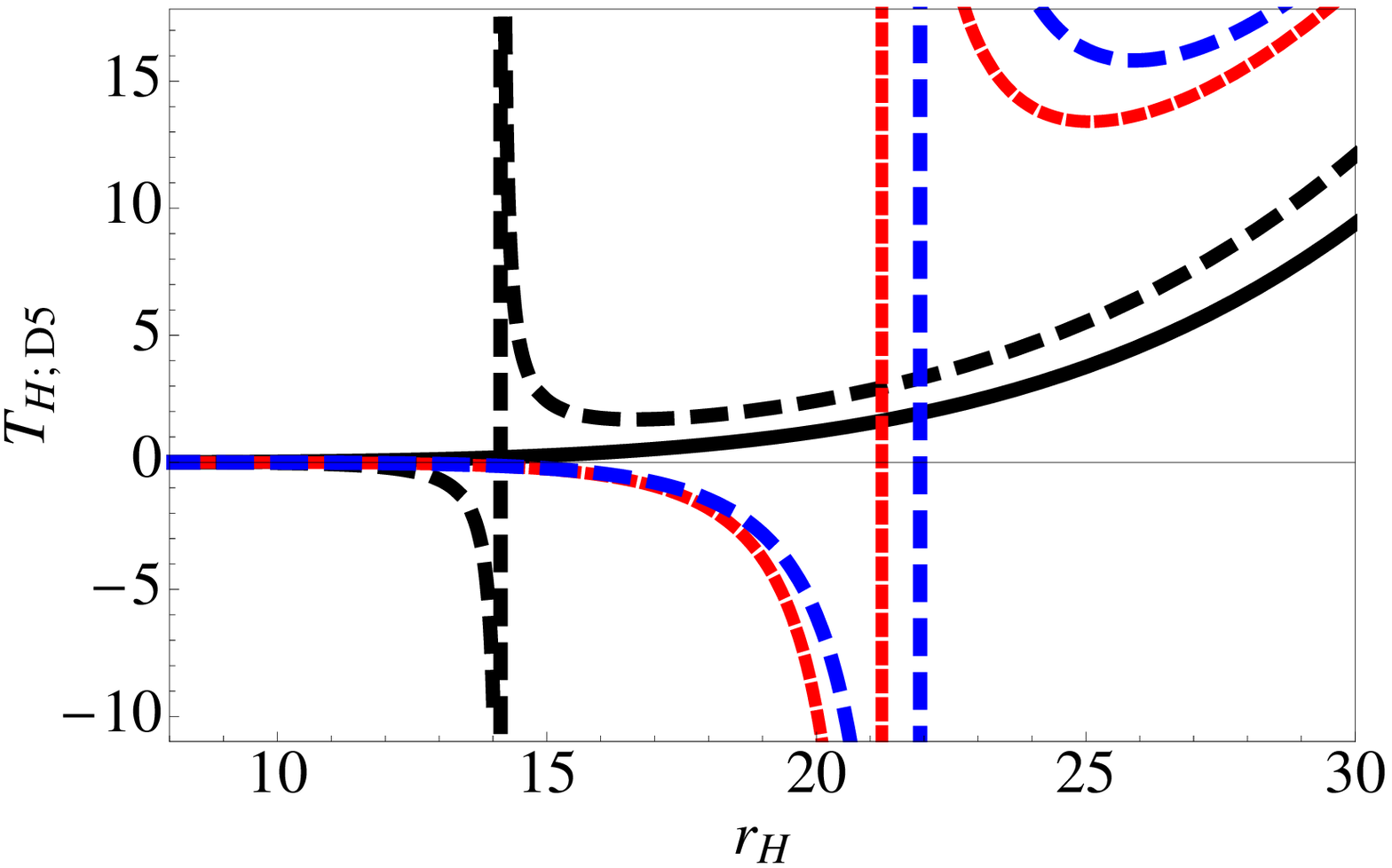,width=.48\textwidth}
\epsfig{file=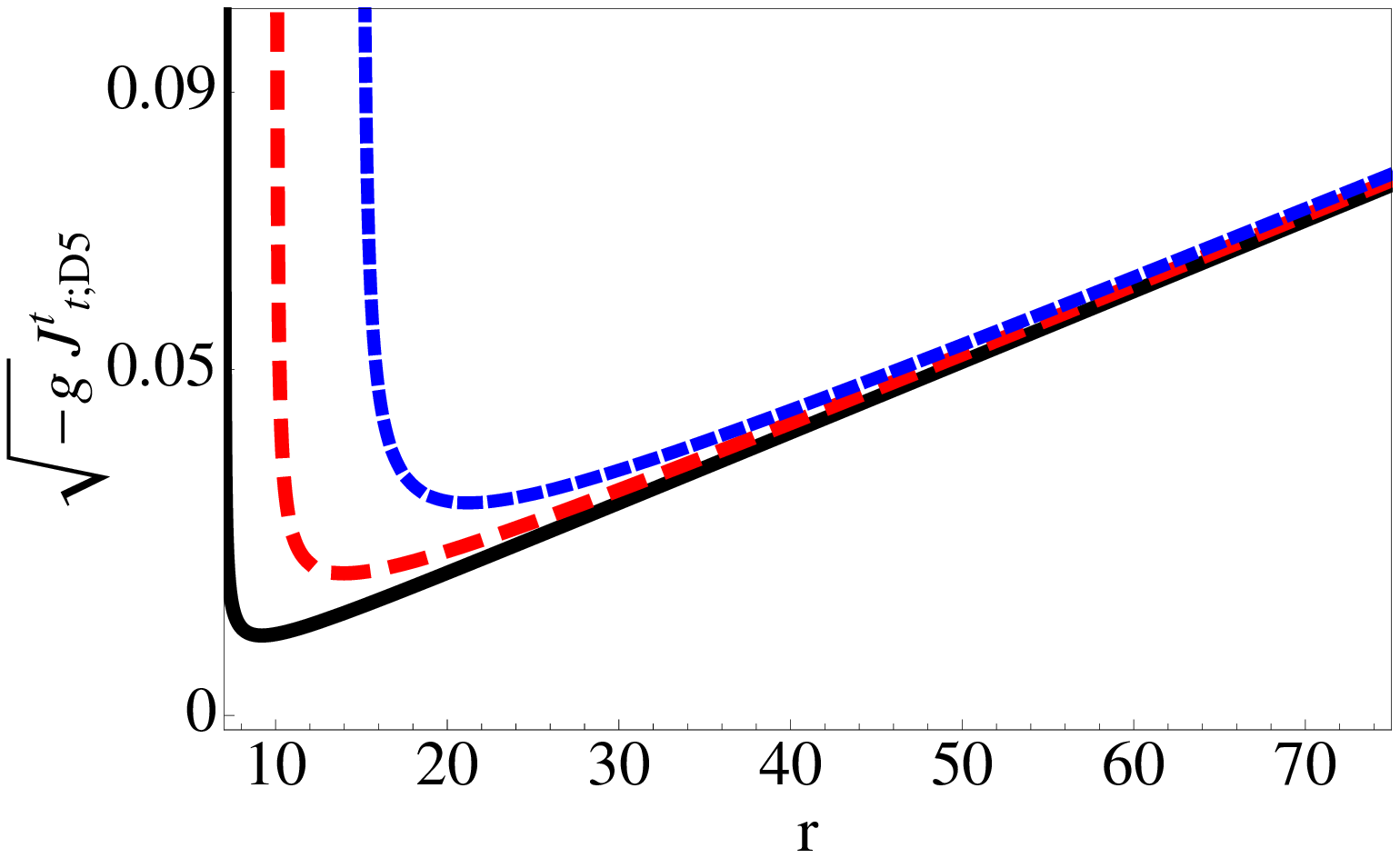,width=.48\textwidth}~~\nobreak
\epsfig{file=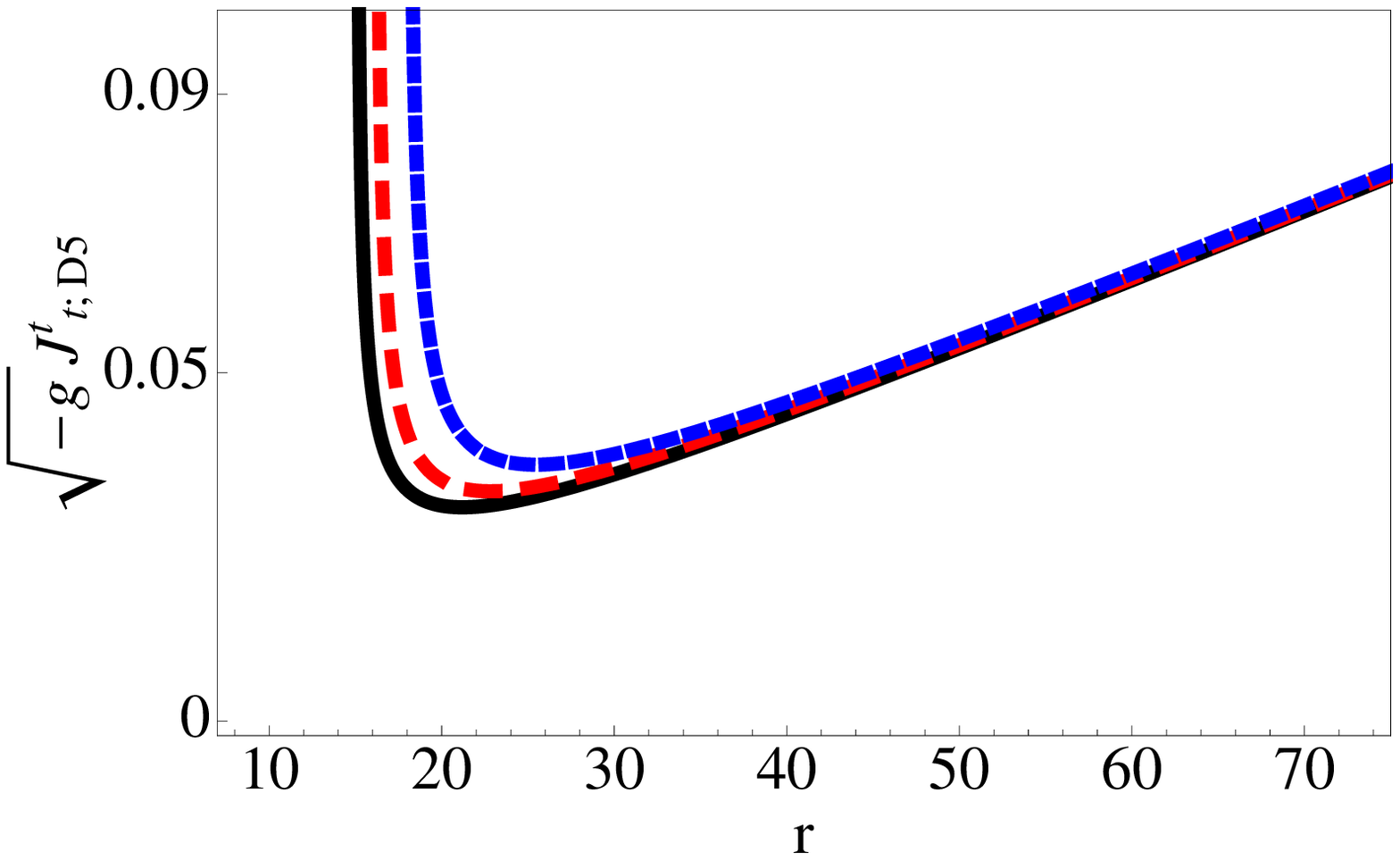,width=.48\textwidth}
\caption{[Up-Left] The behavior of the world volume Hawking
temperature with $r_0=0.5$ (black-solid), $r_0=0.4$ (black-dashed),
$r_0=35$ (red-dashed), $r_0=0.33$ (blue-dashed), $a_B=0$, and $L=10$.
[Up-Right] The behavior of the temperature with $r_0=5$ (black-solid),
$r_0=4$ (black-dashed), $r_0=3.5$ (red-dashed), $r_0=3.3$ 
(blue-dashed), $a_B=0$, and $L=10$. [Middle-Left] The behavior of the
temperature with $a_B=0$ (black-solid), $a_B=10$ (black-dashed), 
$a_B=15$ (red-dashed), $r_0=15.5$ (blue-dashed), $r_04=0$, and $L=10$.
 [Middle-Right] The behavior of the temperature with $a_B=0$ (black-solid), 
$a_B=10$ (black-dashed), $a_B=15$ (red-dashed), $r_0=15.5$ (blue-dashed),
$r_0=8$, and $L=10$. [Down-Left] The behavior of the energy-density, 
$\sqrt{-g}J_t^t$, with $\omega=1$ (black-solid), $\omega=6$ (blue-dashed),
$\omega=10$ (red-dashed), $a_B=15$ and $r_0=7$. [Down-Right] The behavior
of the energy-density with $a_B=0$ (black-solid), $a_B=10$ (blue-dashed), 
$a_B=15$ (red-dashed), $\omega=1$, and $r_0=7$. In all cases, 
$L=\alpha^{\prime}=1$ and $g_s=0.1$.}
\label{fig:TJD5}
\end{center}
\end{figure}
 
\begin{equation}
\label{EdenD5A}
E_{D5}=\tilde{T}_{D5}\int_{r_0}^{\infty}{\frac{dr\,(r^{6}-r_0^4(L^4\overline{\omega}^2
+a_B^2))}{r^2\sqrt{(r^4-r_0^4)(r^2-L^4\overline{\omega}^2-a_B^2)}}}.
\end{equation}
Taking the limit $r\rightarrow r_0$, Eq.\,(\ref{T1D5A}) shows that the energy density
of the (defect) flavor brane, the integrand in (\ref{EdenD5A}), enlarges dramatically,
and  blows up exactly at the minimal extension, $r=r_0$, at the IR scale of conformal
and chiral flavor symmetry breakdown. Therefore, we  conclude that at the IR scale
of symmetries breakdown, $r=r_0$, the backreaction of the D5-brane to the gravity
dual metric is large and cannot be neglected and forms a localized black hole centered
at the IR scale $r=r_0$ in the bulk. The size of the black hole ought to increase due to
the energy injected into it from the D5-brane continuously. By using the  components
of the energy--stress tensor (\ref{T1D5A})--(\ref{T3D5A}), we can find this energy flux.
We notice that in the brane configuration of interest here, with $r_0>0$ and $\omega>0$, 
the component (\ref{T3D5A}) is positive definite and so we derive the time evolution of the
total energy in the form:

\begin{equation}
\label{dEA}
\dot{E}_{D5}=\frac{d}{dt}\int{dr\sqrt{-g}J_{t;D5}^{t}}=\int{dr\partial_r(\sqrt{-g}J_{t;D5}^{r})}
=\sqrt{-g}J_{t;D5}^{r}|_{r=r_0}^{\infty}=\tilde{T}_{D5}r_0^2\overline{\omega}-\tilde{T}_{D5}r_0^2
\overline{\omega}=0.
\end{equation}
Relation (\ref{dEA}) shows that, independent from the presence of the world volume
electric field, the energy per unit time injected at the boundary $r=\infty$ by some
external system equals the energy dissipated from the IR into the bulk, despite the
total energy is time-independent, as before. This dissipation of energy from the
D5-brane to the bulk will create a localized black hole in the bulk. Given this flow of
energy form the probe to the bulk, we thereby conclude by duality that the energy
from the defect flavor sector will in the end dissipate into the gauge theory, 
independent from the electric field. The injection of energy can also be seen clearly
in our stationary solution by introducing cut offs, as before. We can, again, introduce
UV and IR cut offs to the D5-brane solution, such that it extends from 
$r=r_{\text{IR}}=r_0$ to $r=r_{\text{UV}}\gg r_{\text{IR}}$. It is clear from 
(\ref{T3D5A}) that, again, at both $r_{\text{IR}}$ and $r=r_{\text{UV}}$
we have $J_t^r>0$. This shows the presence of equal energy flux incoming from
$r=r_{\text{UV}}$ and outgoing at $r=r_{\text{IR}}$ at which the energy gets
not reflected back but its backreaction will nucleate a localized black hole 
intaking the injected energy, independent from the electric field, as before.

The energy--density (\ref{T1D5A}) of the world-volume black hole solution
(\ref{indD52}) has a number of interesting features. These depend on the choice
of parameters $\{r_0,a_B, \omega\}$, and therefore we note the following points.
$\\$
$\bullet$ Increasing $\omega$, while keeping the other parameters fixed,
leaves the behavior of the energy density unaltered, increases the scale of
the minimum energy-density near the IR scale, but leaves the scale of energy
density more or less unaltered away from the IR scale (see Fig.\,\ref{fig:TJD5}).
However, the increase in the energy-density is, so that the backreaction remains 
negligibly small away from the blowing up point, located at the IR scale $r_0$ 
(see Fig.\,\ref{fig:TJD5}).
$\\$
$\bullet$ Increasing $a_B$, while keeping other parameters fixed, only
shifts the minimum of the density and leaves the scale and behavior
of the density unaltered (see Fig.\,\ref{fig:TJD5}). 
$\\$
$\bullet$ Increasing/decreasing $r_0$, while keeping the other
parameters fixed, leaves the scale and behavior of the energy-density
unaltered (as in Fig.\,\ref{fig:TJD5}).

In all cases, the energy-density falls sharply from its infinite value
at the IR scale to its minimum value, whereupon it increases to finitely
small values away from the IR scale. We thus conclude that altering the
parameters of the theory yields almost no alteration in the scale and
behavior of the energy-density, meaning that the density blows up at
the IR scale, whereat the  backreaction is non-negligible, and staying
finitely small away from the IR scale, where the backreaction can be
neglected.

\section{Induced metric and temperature on U-like embedded probe 
D5-brane wrapping $adS_4\times S^2\subset adS_5\times T^{1,1}$
and spinning about $S^3\subset T^{1,1}$}

As third example in our study, in this section, we continue with our analysis
of the U-like embedded probe D5-brane and consider, instead, the probe
wrapping   $adS_4\times S^2$ in  $adS_5\times T^{1,1}$ reviewed in 
Sec.\,2, and turn on, in addition, spin degrees of freedom. Using spherical
symmetry, we allow in our setup, the probe to rotate about the 
$\psi$-direction of the transverse $S^3$ with conserved angular momentum. 
Thus, in our setup, we allow $\psi$ to have, in addition, time-dependence, so
that $\dot{\psi}(r,t)=\omega=const.$, with $\omega$ denoting the angular
velocity of the probe. This way, we construct rotating solutions. Hence, the
world-volume field is given by $\psi(r,t)$, with other directions fixed.

Therefore, we may consider an ansatz for the D5-brane world volume
fields $\theta$, $\psi(r,t)=\omega t+f(r)$, and $F_{ab}=0$.
Using this ansatz and the metric (\ref{10DKWmet}) in terms of 
(\ref{6DConmet1}, it is easy to derive the components of the induced
world volume metric on the D5-brane, $g_{ab}^{D5}$, and compute the
determinant, $\det g_{ab}^{D5}$, resulting the DBI action (\ref{DPACTION})
of the form:

\begin{eqnarray}
\label{dbiacd5}
S_{D5}\simeq-T_{D5}\int{drdt\,r^2\sqrt{1+\frac{r^2(\psi^{\prime})^2}{9}
-\frac{L^4\dot{\psi}^2}{9r^2}}}.
\end{eqnarray} 
Here we note that by setting $\dot{\psi}=\omega=0$, our action
(\ref{dbiacd5}) reduces to that of the probe D5-brane action in
the BKS model, (\ref{BKSAD52}). As in the BKS model reviewed
in Sec.\,2, restrict brane motion to the $\psi$-direction of the 
transverse $S^3$ sphere and fix other directions constant. Thus,
in our set up we let, in addition, the probe rotate about the $S^3$. 
The equation of motion from the action (\ref{dbiacd5}) is:

\begin{equation}
\label{D5eqF}
\frac{\partial}{\partial r}\Bigg[\frac{r^4\psi^{\prime}}
{\sqrt{1+\frac{r^2(\psi^{\prime})^2}{9}-\frac{L^4\dot{\psi}^2}
{9r^2}}}\Bigg]=\frac{\partial}{\partial t}\Bigg[\frac{L^4
\dot{\psi}}{\sqrt{1+\frac{r^2(\psi^{\prime})^2}{9}
-\frac{L^4\dot{\psi}^2}{9r^2}}}\Bigg].
\end{equation}

Take rotating solutions to (\ref{D5eqF}) of the form:

\begin{eqnarray}
\label{rotsold5F}
\psi(r,t)&=&\omega t+f(r),\;\;\;\;\ f(r)=r_0^3\int_{r_0}^{r}{\frac{dr}{r^2}
\sqrt{\frac{r^2-L^4\,\overline{\omega}^2}{r^6-r_0^6}}}.
\end{eqnarray}
Here we set $\overline{\omega}=\omega/3$ and note that
when $\omega=0$, our solution (\ref{rotsold5F}) integrates
to that of probe D5-brane in the BKS model,  
\cite{Ben-Ami:2013lca}, reviewed in Sec.\,2, with the probe
wrapping  $adS_4\times S^2$ in $adS_5\times T^{1,1}$
(see Eq.\,(\ref{BKSSol.D52})). The solution (\ref{rotsold5F}) is
parameterized by $(\omega,r_0)$ and describes probe D5-brane
motion, with angular velocity $\omega$ about the transverse
$S^3\subset T^{1,1}$, starting and ending up at the boundary.
The probe descends from the UV boundary at infinity to the 
minimal extension $r_0$ in the IR where it bends back up the
boundary. We also note that inspection of (\ref{rotsold5F})
illustrates, in the limit $r\rightarrow\text{large}$, that 
the behavior of $df(r)/dr=f_r(r)$ does not depend on $\omega$
(see Fig.\,\ref{fig:D5}). This illustrates that in the large
radii limit the solution $f(r)$ in (\ref{rotsold5F}) gives the
world volume field $\psi(r)$ of the BKS model  (see Sec.\,2)
with the boundary values $\psi_{\pm}$ in the asymptotic UV limit,
$r\rightarrow\infty$ (see also Fig.\,\ref{fig:D5}). But, we note that
in the (other) IR limit, i.e., when $r\rightarrow\text{small}$, the 
behavior of $f_r(r)$ does depend on $\omega$. Inspection of 
(\ref{rotsold5F}) illustrates that in the IR the behavior of $f_r(r)$
with $\omega>0$ is comparable to that of with $\omega=0$, only
if certain $\omega>0$ are chosen. (see Fig.\,\ref{fig:D5}). This 
illustrates that in the small radii limit in the IR the behavior of $f_r(r)$
(here) does compare to that of $\psi^{\prime}(r)=d\phi/dr$ in the BKS
model (see Sec.\,2),--with $\psi^{\prime}(r)\rightarrow\infty$ in the IR
limit $r\rightarrow r_0$--, consistent with U-like embedding, only for 
certain $\omega>0$.

To derive the induced metric on the D5-brane, we put the rotating solution
(\ref{rotsold5F}) into the background metric (\ref{10DKWmet}) and obtain:

\begin{eqnarray}
\label{indD5F}
dS_{D5}^2&=&-\frac{1}{L^2}(r^2-L^4\overline{\omega}^2)dt^2+\frac{L^2}{r^2}
\bigg[\frac{r^8-r_0^6 L^4\overline{\omega}^2}{r^2(r^6-r_0^6)}\bigg]dr^2
+\frac{2L^2r_0^3\overline{\omega}}{r^2}\sqrt{\frac{r^2-L^4\overline{\omega}^2}
{r^6-r_0^6}}drdt\notag\\ &&+\frac{r^2}{L^2}(dx_1^2+dx_2^2).
\end{eqnarray}
Here we note that by setting $\omega=0$, our induced world volume
metric (\ref{indD5F}) reduces to that of the BKS model, 
\cite{Ben-Ami:2013lca}, reviewed in Sec.\,2. In this case, for $r_0=0$ the
induced world volume metric is that of $adS_4\times S^2$ and the dual
gauge theory describes the conformal and chiral symmetric phase. On 
contrary, for  $r_0>0$ the induced world volume metric has no $adS$
factor and the conformal invariance of the dual gauge theory must be 
broken in such case.

In order to  find the world volume horizon and Hawking temperature, 
we first eliminate the relevant cross term. To eliminate the relevant 
cross-term in (\ref{indD5F}), we consider a coordinate transformation:

\begin{equation}
\tau=t-L^4\overline{\omega} r_0^3\int{\frac{dr}{r^2 \sqrt{(r^2-L^4
\overline{\omega}^2)
(r^6-r_0^6)}}}.
\end{equation}
The induced metric on the rotating D5-brane, (\ref{indD5F}), then takes the form:

\begin{figure}[t]
\begin{center}
\epsfig{file=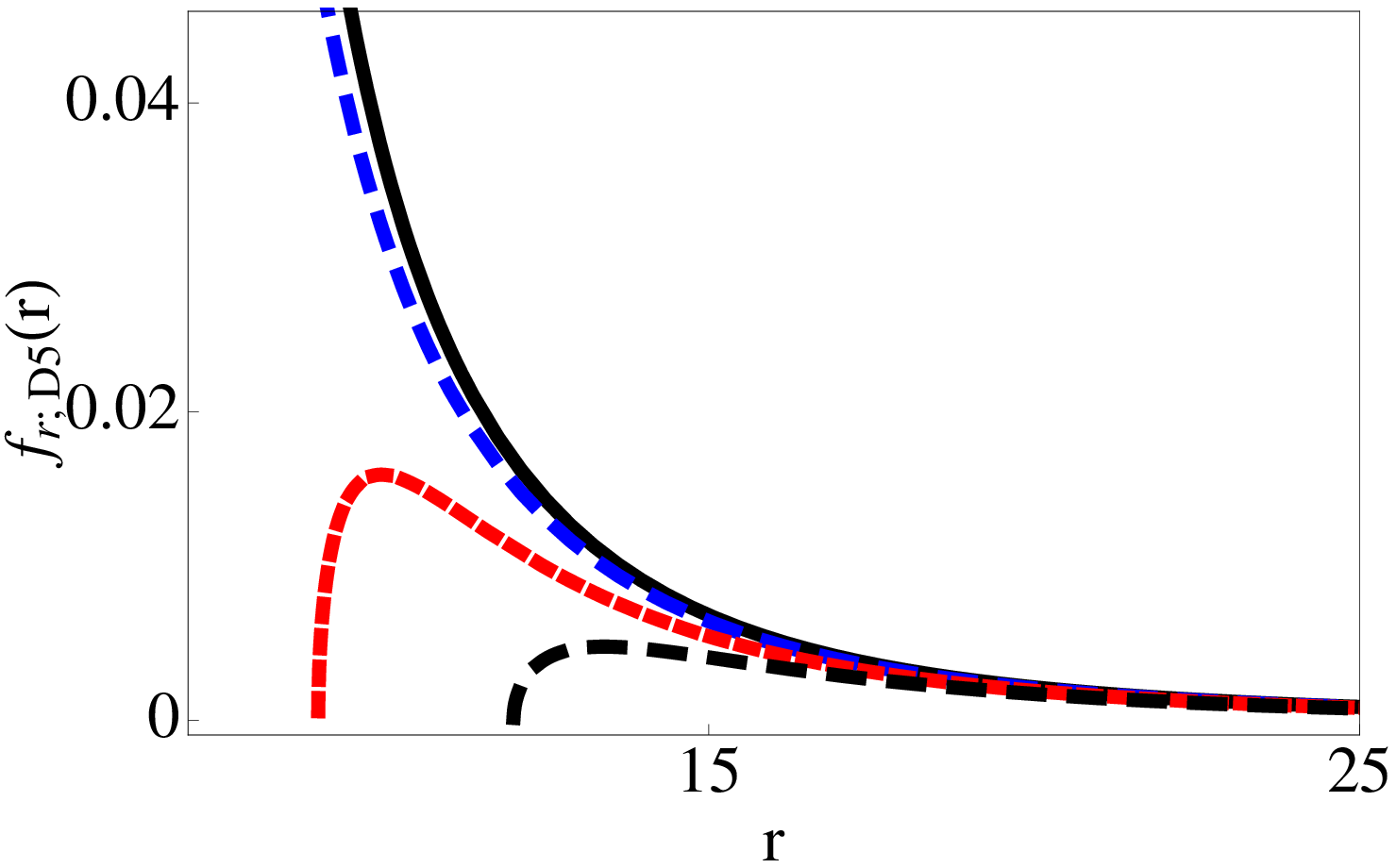,width=.50\textwidth}~~\nobreak
\epsfig{file=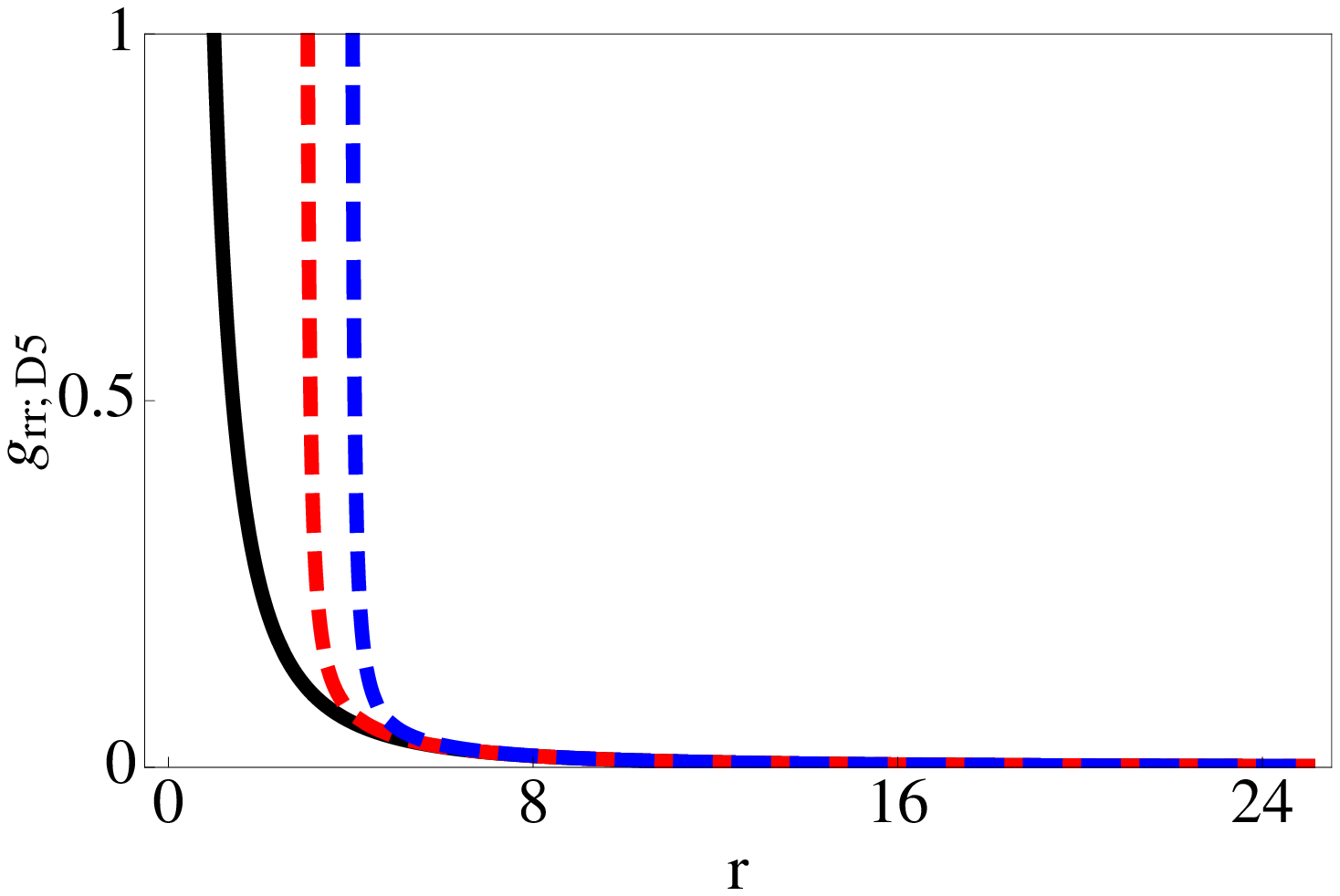,width=.50\textwidth}
\caption{[Left] The behavior of the derivative of the world
volume field with respect to $r$ with $L=1, r_0=7$, $\omega=0$
(black-solid), $\omega=5$ (blue-dashed), $\omega=9$ (red-dashed),
and $\omega=12$ (black-dashed). [Right] The behavior of the $g_{rr}$
component of the induced world volume metric with $L=1,r_0=0$ 
(black-solid), $r_0=3$ (red-dashed), and $r_0=4$ (blue-dashed).}
\label{fig:D5}
\end{center}
\end{figure}

\begin{eqnarray}
\label{indD5F2}
dS_{D5}^2&=&-\frac{1}{L^2}(r^2-L^4\overline{\omega}^2)d\tau^2
+\frac{L^2 r^4}{r^6-r_0^6}dr^2
+\frac{r^2}{L^2}(dx_1^2+dx_2^2).
\end{eqnarray}
The induced world volume metric (\ref{indD5F2}) is not given by the black hole
geometry with thermal horizon solving $-g_{\tau\tau}=g^{rr}\big|_{r=r_H=0}=0$.
Inspection of (\ref{indD5F2}) may seem to indicate though that for a very specific,
degenerate choice of parameters, $r_0=L^{2/3}\overline{\omega}$, the induced
world volume metric (\ref{indD5F2}) is given by the black hole geometry. But, when
the  embedding is fixed U-like ($r_0>0$), this degeneracy conflicts the continuity
of the `would be' world-volume thermal horizon: The `would be' world volume 
thermal horizon, $r_H=L^2\overline{\omega}$, grows continuously with increasing
the angular velocity continuously from zero, $\omega=0$, to positive definite
values, $\omega>0$. This means that, in this limit, when $\omega=0$, by degeneracy,
the embedding is rather V-like, and only then U-like, when $\omega>0$. Thus, in this 
example, there is no continuous world volume horizon consistent with the U-like
embedding of the probe. We therefore conclude that the induced world volume 
metric on the U-like embedded probe D5-brane wrapping the $adS_4\times S^2$
spacetime (in the BKS model) and rotating about the transverse $S^3\subset T^{1,1}$
does not describe the world volume black hole geometry with thermal horizon and 
Hawking temperature of expected features.

\section{Induced metric and temperature on U-like embedded probe D3-brane
wrapping $adS_2\times S^2 \subset adS_5\times T^{1,1}$
and spinning about $S^3\subset T^{1,1}$}

As fourth and final example in our study, in this section, we consider another
type of probe, namely, the U-like embedded probe D3-brane wrapping  
$adS_2\times S^2$ in $adS_5\times T^{1,1}$ reviewed in Sec.\,2, taking
into account additional spin degrees of freedom. Using spherical symmetry,
we allow in our setup, the probe to rotate about the  $\psi$-direction of the
transverse $S^3$ with conserved angular momentum. Thus, in our setup, we
allow $\psi$ to have, in addition, time-dependence, so that 
$\dot{\psi}(r,t)=\omega=const.$, with $\omega$ denoting the angular 
velocity of the probe. This way, we construct rotating solutions. Hence, the
world-volume field is given by $\psi(r,t)$, with other directions fixed.

Thus, we consider an ansatz for the D3-brane world volume
$\psi(r,t)=\omega t+f(r)$, with other directions fixed, and $F_{ab}=0$.
Using this ansatz and the metric (\ref{10DKWmet}) in terms of (\ref{6DConmet1}), 
it is easy to derive the components of the induced world volume metric on the
D3-brane, $g_{ab}^{D3}$, and compute the determinant, $\det g_{ab}^{D3}$, 
resulting the DBI action (\ref{DPACTION}) of the form:

\begin{eqnarray}
\label{dbiacd3}
S_{D3}\simeq-T_{D3}\int{dr dt\sqrt{1+\frac{r^2(\psi^{\prime})^2}{9}
-\frac{L^4\dot{\psi}^2}{9r^2}}}.
\end{eqnarray} 
Here we note that by setting $\dot{\psi}=\omega=0$, our action
(\ref{dbiacd3}) reduces to that of the probe D3-brane action in
the BKS model, (\ref{KSA}). As in the BKS model reviewed in Sec.\,2,
restrict brane motion to the $\psi$-direction of the transverse
$S^3$ sphere and fix other directions constant. Thus, in our set
up, we let, in addition, the probe rotate about the $S^3$. The
equation of motion from the action (\ref{dbiacd3}) is:

\begin{equation}
\label{D3eq}
\frac{\partial}{\partial r}\Bigg[\frac{r^2\psi^{\prime}}
{\sqrt{1+\frac{r^2(\psi^{\prime})^2}{9}-\frac{L^4\dot{\psi}^2}
{9r^2}}}\Bigg]=\frac{\partial}{\partial t}\Bigg[\frac{L^4\dot{\psi}}
{r^2\sqrt{1+\frac{r^2(\psi^{\prime})^2}{9}-\frac{L^4\dot{\psi}^2}
{9r^2}}}\Bigg].
\end{equation}

\begin{figure}[t]
\begin{center}
\epsfig{file=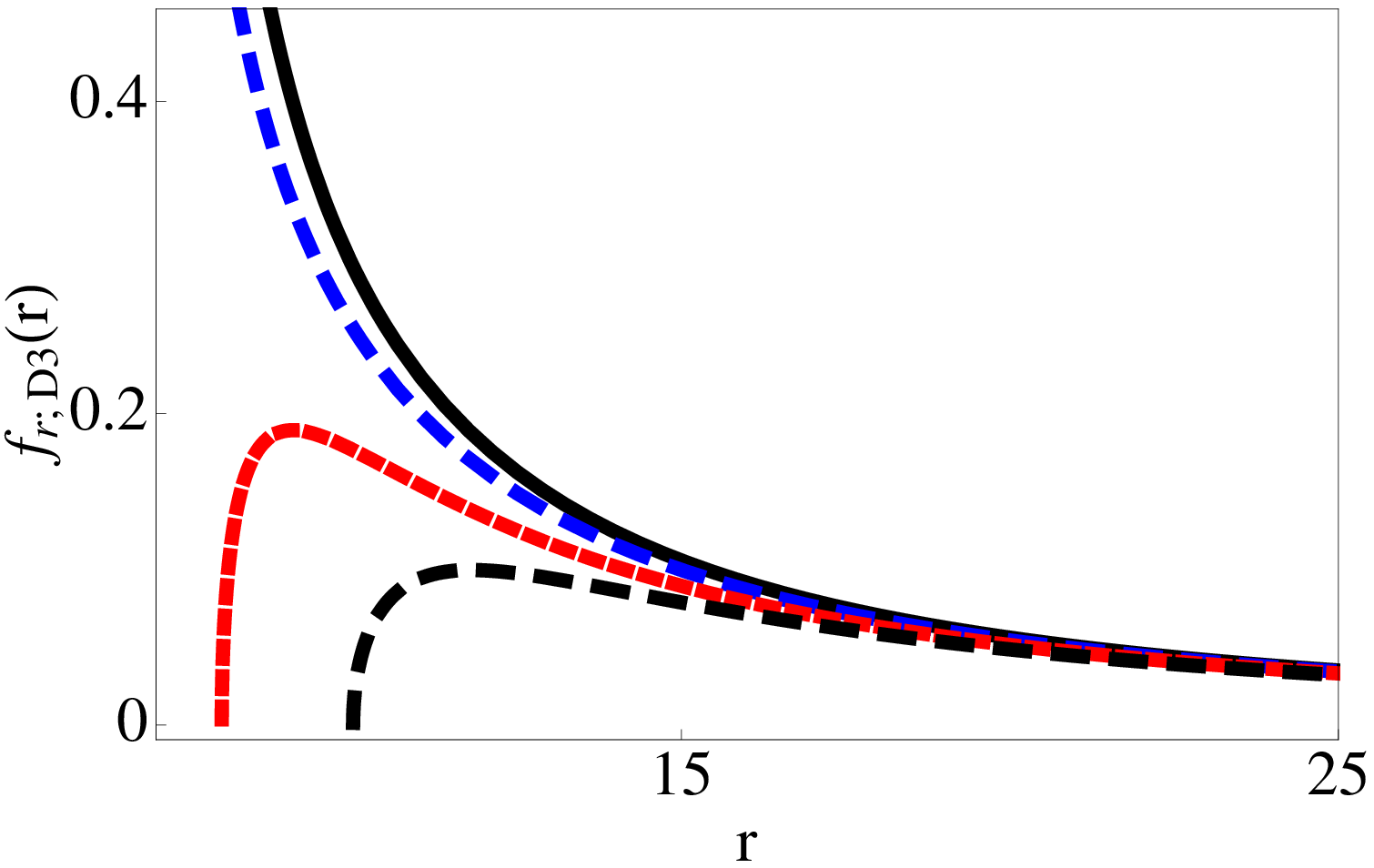,width=.50\textwidth}~~\nobreak
\epsfig{file=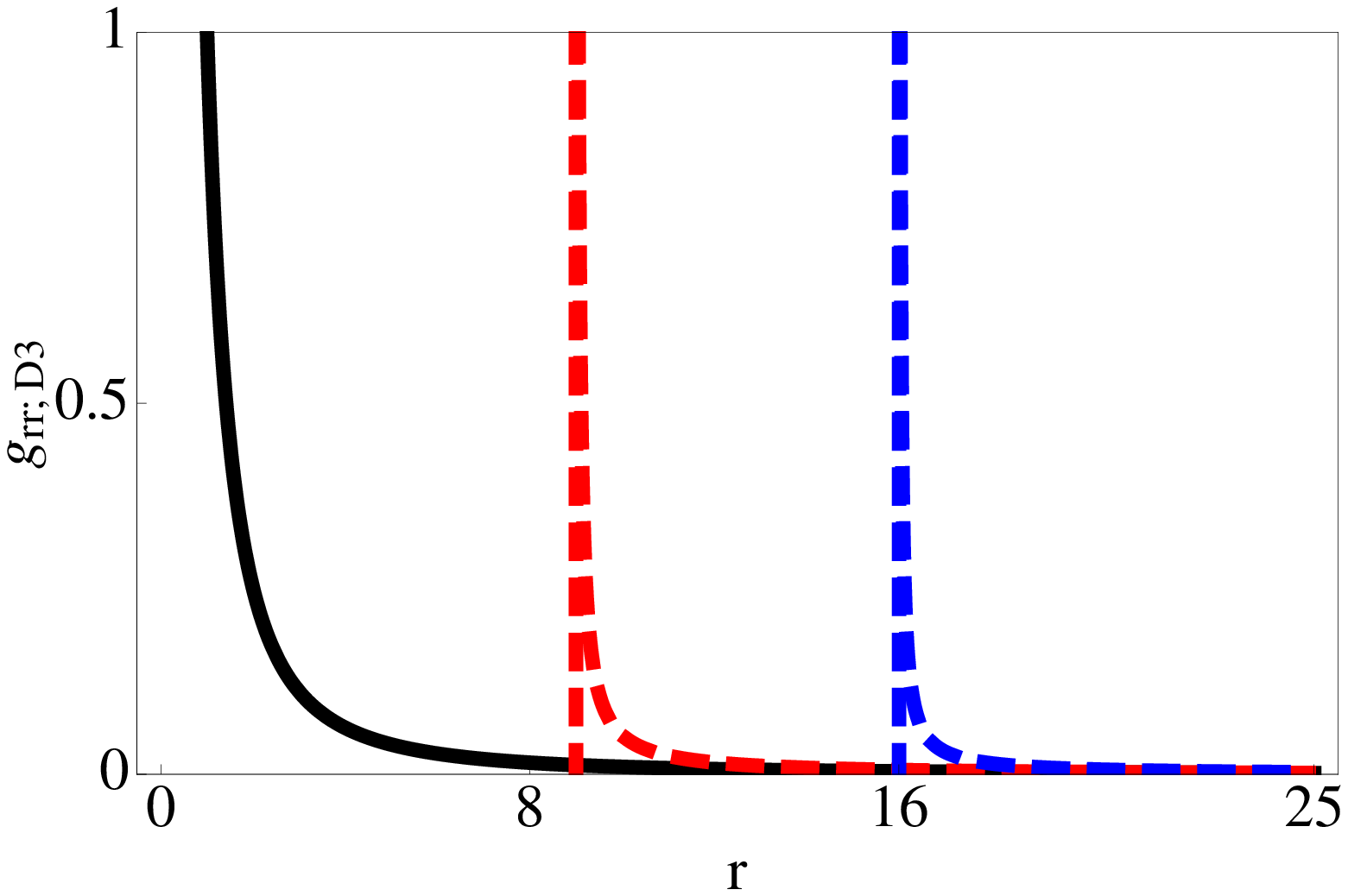,width=.50\textwidth}
\caption{[Left] The behavior of the derivative of the world
volume field with respect to $r$ with $L=1, r_0=7$, $\omega=0$
(black-solid), $\omega=5$ (blue-dashed), $\omega=8$ (red-dashed),
and $\omega=10$ (black-dashed). [Right] The behavior of the $g_{rr}$
component of the induced world volume metric with $L=1,r_0=0$ 
(black-solid), $r_0=3$ (red-dashed), and $r_0=4$ (blue-dashed).}
\label{fig:D3}
\end{center}
\end{figure}

Take rotating solutions to (\ref{D3eq}) of the form:

\begin{eqnarray}
\label{rotsold3}
\psi(r,t)&=&\omega t+f(r),\;\;\;\;\ f(r)=3r_0\int_{r_0}^{r}{\frac{dr}{r^2}
\sqrt{\frac{r^2-L^4\,\overline{\omega}^2}{r^2-r_0^2}}}.
\end{eqnarray}
Here we set $\overline{\omega}=\omega/3$ and note that
when $\omega=0$, our solution (\ref{rotsold3}) integrates to 
that of probe D3-brane in the BKS model,  \cite{Ben-Ami:2013lca}, 
reviewed in Sec.\,2, with the probe wrapping  $adS_2\times S^2$
in $adS_5\times T^{1,1}$ (see Eq.\,(\ref{BKSSol.D3})). The solution
(\ref{rotsold3}) is parameterized by $(\omega,r_0)$ and describes
probe D3-brane motion, with angular velocity $\omega$ about the
transverse $S^3\subset T^{1,1}$, starting and ending up at the 
boundary. The probe descends from the UV boundary at infinity to
the  minimal extension $r_0$ in the IR where it bends back up the
boundary. We also note that inspection of (\ref{rotsold3})
illustrates, in the limit $r\rightarrow\text{large}$, that 
the behavior of $df(r)/dr=f_r(r)$ does not depend on $\omega$
(see Fig.\,\ref{fig:D3}). This illustrates that in the large
radii limit the solution $f(r)$ in (\ref{rotsold3}) gives the
world volume field $\psi(r)$ of the BKS model  (see Sec.\,2)
with the boundary values $\psi_{\pm}$ in the asymptotic UV limit,
$r\rightarrow\infty$ (see also  Fig.\,\ref{fig:D3}). But, we note
that in the (other) IR limit, i.e., when $r\rightarrow\text{small}$, the
behavior of $f_r(r)$ does depend on $\omega$. Inspection of 
(\ref{rotsold3}) illustrates that in the IR the behavior of $f_r(r)$
with $\omega>0$ is comparable to that of with $\omega=0$, 
only if certain $\omega>0$ are chosen (see Fig.\,\ref{fig:D3}). 
This illustrates that in the small radii limit in the IR the behavior of
$f_r(r)$ (here) does compare to that of $\psi^{\prime}(r)=d\phi/dr$
in the BKS model (see Sec.\,2),--with $\psi^{\prime}(r)\rightarrow\infty$
in the IR limit $r\rightarrow r_0$--, consistent with U-like embedding, only
for certain $\omega>0$.

To derive the induced metric on the D3-brane, we put the rotating
solution (\ref{rotsold3}) into the background metric (\ref{10DKWmet})
and obtain:
\begin{eqnarray}
\label{indD31}
dS_{D3}^2&=&-\frac{1}{L^2}(r^2-L^4\overline{\omega}^2)dt^2+
\frac{L^2}{r^2}\bigg[\frac{r^4-r_0^2 L^4\overline{\omega}^2}
{r^2(r^2-r_0^2)}\bigg]dr^2+\frac{2L^2r_0\overline{\omega}}{r^2}
\sqrt{\frac{r^2-L^4\overline{\omega}^2}{r^2-r_0^2}}drdt.\notag\\
\end{eqnarray}

To eliminate the cross-term in (\ref{indD31}), we consider a coordinate transformation:

\begin{equation}
\tau=t-L^4\overline{\omega} r_0\int{\frac{dr}{r^2 \sqrt{(r^2-L^4\overline{\omega}^2)
(r^2-r_0^2)}}}.
\end{equation}
The induced world volume metric on the rotating probe D3-brane then reads:
\begin{equation}
\label{indD3}
dS_{D3}^2=-\frac{1}{L^2}(r^2-L^4\overline{\omega}^2)d\tau^2+\frac{L^2}{r^2-r_0^2}dr^2.
\end{equation}
The induced world volume metric (\ref{indD3}) is not given by the black hole geometry
with thermal horizon  solving $-g_{\tau\tau}=g^{rr}\big|_{r=r_H}=0$. Inspection of (\ref{indD3})
may seem to indicate though that for a very specific, degenerate choice of parameters,
$r_0=L^2\overline{\omega}$, the induced world volume metric (\ref{indD3}) is that of 
the BTZ black hole metric minus the angular coordinate. But, when the embedding is fixed
U-like ($r_0>0$), this degeneracy is inconsistent with the continuity of the `would be' world 
volume horizon: The `would be' world volume horizon, $r_H=L^2\overline{\omega}$, has 
to grow continuously with increasing the angular velocity continuously, i.e., from zero, 
$\omega=0$, to any positive definite values, $\omega>0$. This implies that when 
$\omega=0$, the embedding is V-like, and only then U-like, when $\omega>0$. Thus, 
in this example, there is no continuous world volume horizon consistent with the U-like
embedding of the probe. We therefore conclude that the induced world volume metric
on the U-like embedded probe D3-brane wrapping the $adS_2\times S^2$ spacetime
(in the BKS model)  and rotating about the transverse $S^3\subset T^{1,1}$ does not
describe the world volume black hole geometry with thermal horizon and Hawking 
temperature of expected features.\footnote{This is in contrast with the conformal 
$adS_5\times S^5$ solution where the induced world volume metric on the rotating
probe D3-brane describes the world volume black hole geometry with thermal horizon,
 $r_H=\omega$, and Hawking temperature,  $T_{H}=\frac{\sqrt{3}\omega}{2\pi}$,
of expected features \cite{Das:2010yw}.}

\section{Discussion}

In this work, we studied the induced world volume metrics and Hawking 
temperatures of all type IIB rotating probe flavor $Dp$-branes $(p=3,5,7)$
in the holographic BKS models. Furthermore, we studied the related 
energy--stress tensors and energy flow of the rotating probes in these
holographic BKS models. By gauge/gravity duality, the Hawking temperatures
of such U-like embedded rotating probes in the gravity dual correspond to
the  temperatures of flavors at finite R--charge in the (d)CFT -- the gauge 
theory conformal and chiral flavor symmetry breakdown, and the energy flow 
from the thermal probes into the system, to the energy dissipation from the
flavor sector into the gauge theory conformal and chiral flavor symmetry 
breakdown. Such non-equilibrium finite-spin  systems and their energy flow
have been previously studied in the literature in the  Kuperstein--Sonnenschein
holographic model, including U-like embedded spacetime filling probe flavor
D7-branes with spin in the KW gravity  dual of CFT  with the conformal and 
chiral flavor symmetry breakdown. The aim of this study was to extend and
generalize such former  studies to more generic holographic BKS models, 
including all U-like embedded  type IIB probe flavor branes in the KW gravity
dual of (d)CFT with conformal and chiral flavor symmetry breakdown. The 
motivation of our study was to exemplify novel  non-equilibrium systems in
holographic models dual to defect gauge theories, including dCFTs with 
spontaneous breakdown of the conformal and chiral flavor symmetry.

We started our analysis by reviewing certain aspects of our previous 
case-study of U-like embedded spacetime filling probe D7-branes 
wrapping $adS_3\times S^3$ in $adS_5\times T^{1,1}$ and rotating
about the trasverse $S^2\subset T^{1,1}$ in the KW gravity dual of 
CFT with spontaneous conformal and chiral flavor symmetry breakdown. 
We showed that when the embedding is U-like, i.e., when the embedding
parameter,--given by the minimal extension of the probe--, is positive 
definite and additional spin, is turned on, the induced world volume metric
on the probe admits thermal horizons and Hawking temperatures of expected
features. We found that the world volume horizon forms about the minimal 
extension of the probe and grows continuously with increasing the angular
velocity, as expected.  We found that the related world volume Hawking 
temperature increases continuously with increasing the angular velocity, 
as expected. We also found that by decreasing/increasing the minimal
extension, the temperature scale increases/decreases dramatically,
while the temperature behavior remains unchanged. We therefore found
that varying the minimal extension sets large hierarchies of temperature
scales, while leaves the temperature behavior unchanged. We found, 
however, that the behavior of the world volume temperature can change
dramatically in the presence of the world volume gauge field. We found
that when the world volume electric field is turned on, the world volume
temperature admits two distinct branches. We found that there is one
branch where the temperature increases and another where it decreases
with growing horizon size, corresponding to `large' and `small' black holes, 
respectively. Nonetheless, we found this behavior of the world volume 
Hawking temperature parameter dependent -- depending on the size
of the minimal extension of the  probe. We found that when the minimal 
extension is increased to relatively larger values, the temperature changes
again dramatically -- only increasing with growing horizon size and therefore
including only one branch -- corresponding only to `large' black holes. 
By gauge/gravity duality, we thus found, in this example, that when the
IR scale of conformal and chiral flavor symmetry breakdown is positive definite
and the flavor sector gets, in addition, R--charged, the flavor sector of the CFT
becomes thermal. By gauge/gravity duality, we also found, in this example, that
the temperature of flavors with finite R--charge and density chemical potential
is dramatically controlled by the IR scale of conformal and chiral flavor symmetry
breakdown -- giving the VEV deformations of the dual gauge theory. We then 
noted that by taking into account the backreaction of the above solution to the
KW bulk gravity dual of CFT, one naturally expects the U-like embedded rotating
D7-brane to form a mini black hole in the KW gravity dual of CFT with spontaneous
breakdown of the conformal and chiral flavor symmetry, in the probe limit. Hence, 
we found the U-like embedded rotating probe D7-brane describing a thermal object
in the dual CFT with spontaneous conformal and chiral symmetry breakdown.
In this example, the system was dual to $\mathcal{N}=1$ gauge theory (CFT)
coupled to a spacetime filling flavored quark subject to an external electric field.
Since the CFT itself was at zero temperature while the flavored quark was at 
finite temperature, we found, in this example, that such system describes 
non-equilibrium steady state in the CFT -- the gauge theory conformal and chiral
flavor  symmetry breakdown. However, by computing the energy--stress tensor of 
the rotating probe flavor brane, we then showed that the energy from the 
probe will eventually dissipate into the bulk KW, dual to energy dissipation
from the flavor sector into the CFT with spontaneous conformal and chiral
symmetry breakdown. We first found that,  independent from the parameters
choice, at the minimal extension, at the IR scale of conformal and chiral flavor
symmetry breakdown, the energy density blows up and thereby showed the
backreaction in the IR is non-negligible.  We then showed that when the minimal
extension is positive definite and spin is turned on, the energy flux is  
non-vanishing and found the energy can flow from the brane into the bulk, 
forming, with the large backreaction, a black hole in the bulk KW, independent 
from the electric field. We also argued how this external injection of energy
may be understood in our stationary solutions. We considered UV and IR cut
offs in our rotating D7-brane system and noted from the energy--stress
tensor that the incoming energy from the UV equals the outgoing energy
from the IR where the large backreaction forms a black hole intaking the 
injected energy. By gauge/gravity duality, we thus found, in this example,
the energy dissipation from the flavor sector into the CFT with spontaneous
conformal and chiral flavor symmetry breakdown, independent from the 
electric field.

We then continued our analysis by considering other examples U-like 
embedded type IIB probe branes with spin in the KW gravity dual of dCFT
with (defect) flavors at finite R-charge and spontaneous breakdown of
the conformal and chiral flavor symmetry. These included the example
of probe D3-brane and the two possible examples of probe D5-branes 
wrapping non-trivial cycles in the KW gravity dual. 

In the first example, we considered the U-like embedded  probe D5-brane
wrapping the $adS_3\times S^3$ in $adS_5\times T^{1,1}$ and rotating
about the transverse space $S^2\subset T^{1,1}$ in the KW gravity dual
of dCFT with spontaneous conformal and chiral flavor symmetry breakdown. 
In this case, we found that the induced world volume metric on the rotating
probe describes the black hole geometry despite the absence of black holes
in the bulk. We showed that when the embedding is U-like, i.e., when the 
embedding parameter,--given by the minimal extension of the probe--, is 
positive definite and additional spin is turned on,  the induced world volume
metric on the probe admits thermal horizons and  Hawking temperatures of
expected features. We found that the world volume horizon forms about the
minimal extension of the probe and grows continuously with increasing the
angular velocity, as expected. We found that the world volume Hawking 
temperature increases continuously with increasing the angular velocity, 
as expected. We also found that by decreasing/increasing the minimal 
extension, the temperature scale increases/decreases dramatically, while
the temperature behavior remains unchanged. We therefore found that
varying the minimal extension sets large hierarchies of temperature scales, 
while the temperature behavior remains unchanged. We found, however, 
that the behavior of the world volume temperature can change dramatically
in the presence of the world volume gauge electric field. We found that when
the world volume electric field is turned on, the world volume temperature
admits two distinct branches. 
We found that there is one branch where the temperature increases and 
another where it decreases with growing horizon size, corresponding to 
`large' and `small' black holes, respectively. We found this behavior of the
 world volume Hawking temperature parameter independent -- independent
from the size of the minimal extension of the probe. We found that when 
the minimal extension is increased to relatively larger values, the temperature
remains unchanged -- still including the two distinct branches -- corresponding
to `large'  and `small' black holes, respectively. By gauge/gravity duality,
we thus found, in this example, that when the IR scale of conformal and chiral
flavor symmetry breakdown is positive definite and the (defect) flavor sector
gets, in addition, R--charged, the (defect) flavor sector of the dCFT becomes
thermal. By gauge/gravity duality, we also found, in this example, that the
temperature of (defect) flavors with finite R--charge and density chemical 
potential is set by the baryon number density number and is almost independent
from the IR scale of conformal and chiral flavor symmetry breakdown -- giving the
VEV deformations of the dual gauge theory. We then noted that by taking into 
account the backreaction of the above solution to the KW bulk gravity dual of
dCFT, one naturally expects the U-like embedded D5-brane to form a mini black
hole in the KW gravity dual of dCFT with spontaneous breakdown of the conformal
and chiral flavor symmetry, in the probe limit. Hence, we found the U-like embedded
rotating probe D5-brane describing a thermal object in the dual dCFT with 
spontaneous conformal and chiral symmetry breakdown. In this example, the 
system was dual to $\mathcal{N}=1$ gauge theory coupled to a defect flavor
(dCFT). Since the gauge theort itself was at zero temperature while the defect
flavor was at finite temperature, we found, in this example, that such system
describes non-equilibrium steady state in the dCFT --  the defect gauge theory
 conformal and chiral flavor symmetry breakdown. However, by computing the
energy--stress tensor of the probe, we then showed that the energy from the
defect flavor sector will eventually dissipate into the gauge theory with spontaneous
conformal and chiral symmetry breakdown. We first found that, independent from
the parameters choice, at the minimal extension, at the IR scale of conformal and
chiral flavor symmetry breakdown, the energy density blows up and thereby showed
the backreaction in the IR is non-negligible. We then showed that when the minimal
extension is positive definite and spin is turned on, the energy flux is non-vanishing
and found the energy can flow from the brane into the bulk,  forming, with the large
backreaction, a localised black hole in the bulk, independent from the electric field. 
We also argued how this  external injection of energy may be understood in our 
stationary solutions. We  considered UV and IR cut offs in our  rotating D5-brane
system and noted from the energy--stress tensor that  the incoming energy from
the UV equals the  outgoing energy from the IR where the large backreaction forms
a black hole intaking the injected energy. By gauge/gravity duality, we thus found,
in this  example, the energy dissipation from the defect (flavor) sector into the gauge
theory with spontaneous conformal and chiral flavor symmetry breakdown, independent
from the electric field. 

In the second example, we considered the U-like embedded probe D5-brane
wrapping the $adS_4\times S^2$ in $adS_5\times T^{1,1}$ and rotating
about the transverse internal space  $S^3\subset T^{1,1}$ in the KW gravity
dual of dCFT with spontaneous conformal and chiral flavor symmetry breakdown. 
On contrary with the first example, in this case, we found that the induced world
volume metric on the probe D5-brane is not given by the black hole geometry, 
$-g_{\tau\tau}=g^{rr}|_{r=r_H}=0$, of expected features. We first noted that
in order to render the induced world volume metric of the form akin to the black
hole geometry, the parameter space has to be degenerate, i.e., the minimal 
extension has to equal the angular velocity, $r_0=\omega$. However, we then
found, in this example, that when the  embedding is U-like, $r_0>0$, the 
angular velocity, $\omega$, cannot increase continuously, as it should -- from
zero to positive definite values. We thus found that the degenerate choice of
parameters obstructs, in turn, the continuous growth of the `would be' world
volume horizon. We therefore found, in this example, that the induced world
volume metric on the rotating probe does not describe the black hole geometry
with thermal horizon and Hawking temperature of expected features.

In the third example, we considered the simplest example of U-like embedded
probe flavor brane with additional spin, dual to defect flavors with finite R-charge
and spontaneous breakdown of the conformal and chiral flavor symmetry. This
included the U-like embedded probe D3-brane wrapping the $adS_2\times S^2$
spacetime in $adS_5\times T^{1,1}$ and rotating about the transverse space
$S^3\subset T^{1,1}$ in the KW gravity dual of dCFT with spontaneous 
conformal and chiral flavor symmetry breakdown.  We found that the induced 
world volume metric on the probe  D3-brane is not given by the black hole 
geometry, $-g_{\tau\tau}=g^{rr}|_{r=r_H}=0$, of expected features. We
first noted that in order to render the induced world volume metric of the
form akin to the black hole geometry, such as BTZ minus the angular coordinate, 
the parameter space has to be degenerate, i.e., the minimal extension has to
equal the angular velocity, $r_0=\omega$.  However, we then found, in this 
example, that when the embedding is U-like, $r_0>0$, the angular velocity, 
$\omega$, cannot increase continuously, as it should -- from zero to positive
definite values. We thus found that the degenerate choice of parameters
obstructs, in turn, the continuous growth of the `would be' world volume
horizon. We therefore found, in this example, that the induced world volume
metric on the rotating probe does not describe the world volume black hole
geometry with thermal horizon and Hawking temperature of expected features.

By comparing the world volume black hole solution of the U-like embedded
rotating probe D5-brane with that of the U-like embedded rotating probe
D7-brane in the KW gravity dual, we found similarities and differences in 
the scale and behavior of the world volume horizons and Hawking temperatures.
We found, in each case, that the world volume horizon forms about the minimal 
extension of  the probe with the world volume Hawking temperature increasing
continuously with increasing the angular velocity. We also found, in each case,
that the temperature scale changes dramatically with changing the minimal
extension, which therefore sets large hierarchies of temperature scales.
In the case of probe D7-branes, we found, however, that by varying the
minimal extension, the temperature scale admits much larger hierarchies
than in the case of probe D5-branes. In both cases, we found though that
by varying the minimal extension the temperature behavior remains
unchanged. In each case, we found, however, that when, in addition, the 
world volume electric field is turned on, the behavior of the world volume
Hawking temperature changes dramatically -- admitting two distinct branches,
corresponding to `large' and `small' black holes, respectively. However, we 
found then that by varying the minimal extension of the probe, the behavior
of the world volume Hawking  temperature of the rotating probe D7-brane
varies dramatically once again, in contrast with that of the rotating probe
D5-brane. We found that by increasing the minimal extension of the probe,
the  temperature behavior of the rotating probe D7-brane changes once 
again drastically,--admitting only one branch--, corresponding to `large'  
black holes only, while that of the rotating probe D5-brane remains 
unchanged,--admitting two distinct branches--, corresponding to `large'  
as well as `small' black holes, respectively, as before. For both probes,
we also found that by varying the world volume electric field, via varying
the baryon density number, the temperature scales admit small and 
comparable hierarchies. Nonetheless, we found that by changing, in
addition, the minimal extension, in the case of probe D7-branes, the
world volume temperature admits larger hierarchies of scales than in
the case of probe D5-branes. By gauge/gravity duality,  we thus found
that by varying the IR scale of conformal and chiral flavor symmetry 
breakdown, the temperature scale of flavors at finite R--charge
in the (d)CFT varies  dramatically and thereby admits large hierarchies, 
while the temperature behavior remains unchanged. By gauge/gravity
duality,  we found, however, that when, in addition, the external electric
field is turned on, the temperature behavior of flavors at finite
R--charge and density chemical potential in the (d)CFT changes
dramatically, while the temperature scale changes moderately and thereby
admits moderate hierarchies. By gauge/gravity duality,  we also found that
by varying, in addition, the IR scale of conformal and chiral flavor symmetry
breakdown, the temperature behavior of flavors at finite R--charge and 
baryon density chemical potential in the CFT, varies once again dramatically,
in contrast with the temperature of (defect) flavors at finite R--charge and 
density chemical potential in the dCFT. Furthermore, by comparing
the energy-stress tensors of the U-like embedded rotating probe D5-brane
with that of the U-like embedded rotating probe D7-brane, we found the 
scale and behavior of the energy-densities very similar. We found,
in each case, that the energy-density of the rotating probe always starts 
from its infinite value at the minimal extension where the backreaction is 
non-negligible, and then settles its minimum finite value near the minimal 
extension, from where it eventually increases to finitely larger values away
from the minimal extension. However, we found that by keeping the minimal
extension fixed while varying the other parameters, in the case of probe
D7-brane, the energy-density scale changes moderately away from the 
minimal extension, in contrast with the case of probe D5-brane where it
remains more or less unchanged. Nonetheless, for both probes, we
found that by varying the parameters of the solution, the energy-density
behavior remains unchanged. Moreover, in both cases,  we found that the
energy rotating probe will eventually dissipate into the bulk, forming, with
the large backreaction in the IR a (localised) black hole in the bulk. 
By gauge/gravity duality, we thus found, in both cases, that the energy 
from the (defect) flavor sector at finite R--charge and density chemical
potential will finally dissipate into the $\mathcal{N}=1$ KW gauge theory
with spontaneous conformal and chiral flavor symmetry breakdown.

We conclude that when the probe embedding is U-like, the induced world
volume metrics of only certain type IIB probe flavor branes with spin in the
KW gravity dual of $\mathcal{N}=1$ (d)CFT admit thermal horizons and
Hawking temperatures of expected features despite the absence of black 
holes in the bulk. We conclude, in  particular, that the world volume black hole
formation on U-like embedded type IIB probe flavor branes with spin in the KW
gravity dual of (d)CFT is non-trivial and depends on the world volume spacetime
codimension and topology of the non-trivial internal cycle wrapped by the probe.
We also conclude that when the probe embedding is U-like, the world volume 
Hawking temperature of the rotating black hole solution on type IIB probe flavor
branes in the KW gravity dual of (d)CFT is dramatically parameter dependent. 
We conclude, in particular, that the world volume temperature profile of the
U-like embedded probe spacetime filling flavor brane with spin in KW is freely
and dramatically controlled by the embedding parameter --  the minimal
extension of the probe, while the world volume temperature profile of the U-like
embedded probe (defect) flavor brane in KW is independent from the minimal
extension and set by the world volume electric field density number. 
By gauge/gravity duality, we therefore conclude that when the IR scale of
symmetry breakdown is positive definite and the flavor sector of the KW (d)CFT
with spontaneous breakdown of the conformal and chiral flavor symmetry gets, in
addition, R--charged,  flavor thermalization and non-equilibrium steady state
formation depends non-trivially on the type of flavors inserted to the gauge theory.
By gauge/gravity duality, we also conclude that the temperature profile of the
spacetime filling flavors in the KW CFT is freely and dramatically controlled by 
the IR scale of spontaneous conformal and chiral flavor symmetry breakdown -- 
giving the VEV deformations of the gauge theory, while the temperature profile
of (defect) flavors in the KW dCFT is independent from the IR scale and set by
the VEV of baryon number.  We conclude, however, that the energy of the rotating
thermal probes embedded U-like in the KW gravity dual will eventually dissipate into
the bulk,  forming, with the large backreaction in the IR, a  (localized) black hole in
the bulk. By gauge/gravity duality, we thus conclude that the energy form the
(defect)  flavor sector will finally dissipate into the $\mathcal{N}=1$ KW gauge
theory conformal and chiral flavor symmetry breakdown.

There are the following limitations to our work. In the examples we constructed, we
were unable to fully solve the $Dp$-brane $(p=3,5,7)$ equations of motion to determine
the explicit analytic form of the $Dp$-brane  world volume scalar fields, when conserved
angular motion and world volume gauge fields were turned on. Therefore, in our
examples, we were not able to provide full details about the probe $Dp$-brane solution
itself, when the angular velocity and world volume gauge fields were set non-vanishing.
Nevertheless, in order to derive the induced world volume metrics and compute the
induced world volume Hawking temperatures, we did not need to find the explicit
form of the $Dp$-brane world volume scalar fields. However, in all examples we constructed, we
deliberately chose ans\"atze of solutions and induced world volume metrics that did
reproduce those of the BKS models, when we turned off the angular
velocity and world volume gauge fields. Moreover, in all examples, we demonstrated
that, independent from the value of angular velocity and world volume gauge fields,
in the large radii limit, the radial derivatives of the world volume field in our
ans\"atze coincide with that of vanishing angular velocity and vanishing world volume
gauge fields. Thus, in  all examples, in the large radii limit, our ans\"atze solved
to the $Dp$-brane scalar fields with asymptotic UV boundary values of the 
BKS models.  Furthermore, in all our examples, we
demonstrated that, in the small radii limit, in the IR, for specific range of angular
velocities, the radial derivatives of the world volume field in our ans\"atze behave
as that of the BKS models, consistent with U-like 
embeddings. 

The other important issue that we did not discuss in this work is the
Goldstone boson of conformal symmetry breakdown at finite R-charge -- dual to
rotation. We first note that the existing models in the literature identify the Goldstone
in the pure case by using very specific brane embedding configurations replacing polar
with Cartesian coordinates. We then note that the brane field equations of interest here 
involve rotational symmetry and such are usually systematically solved in polar coordinates.
We thus find the identification of the Goldstone boson at finite R-charge in the embedding
configurations of the pure model  ambiguous and formidable task. We suspect, however, 
that this issue is technically more involved and may need alternative methods of inspection
that go beyond the scope of this paper.

The analysis of this work may be extended in two obvious ways. One obvious
extension is to study probe flavor brane thermalization in U-like embeddings in
more generic type IIB holographic models, such as the regular KS and singular
KT gravity duals of $\mathcal{N}=1$ non-conformal gauge theories including
confinement and chiral symmetry breaking, and RG flow, respectively. The other
obvious extension is to study probe flavor brane thermalization in U-like 
embeddings of type IIA holographic BKS models based on ABJM theory. We leave
these extensions for our future studies.

\end{document}